\documentclass[a4paper,11pt]{article}

\usepackage{jheppub_mod} 
\usepackage[T1]{fontenc} 

\usepackage{hyperref}
\usepackage{graphicx}
\usepackage{amsmath,amssymb,slashed}
\usepackage{booktabs,tabulary}
\usepackage{dsfont}
\usepackage{color}
\usepackage{multirow}
\usepackage{subcaption}

\newcommand{\Fpi}{F_\pi}
\newcommand{\mpi}{M_{\pi}}
\newcommand{\mpii}{M_{\pi^0}}
\newcommand{\Order}{\mathcal{O}}
\newcommand{\MeV}{\,\text{MeV}}
\newcommand{\GeV}{\,\text{GeV}}

\newcommand{\beq}{\begin{equation}}
\newcommand{\eeq}{\end{equation}}
\newcommand{\diff}{\text{d}}
\newcommand{\sis}{s_\text{is}}
\newcommand{\siv}{s_\text{iv}}
\newcommand{\sthr}{s_\text{thr}}
\newcommand{\sm}{s_\text{m}}
\newcommand{\grhop}{g_\text{eff}}
\newcommand{\Mrhop}{M_\text{eff}}
\newcommand{\PiO}{\pi^0}
\newcommand{\F}{\mathcal{F}}
\newcommand{\A}{\mathcal{A}}
\newcommand{\disc}{\text{disc}\,}
\renewcommand{\Re}{\text{Re}\,}
\renewcommand{\Im}{\text{Im}\,}
\newcommand{\mmu}{m_\mu}
\newcommand{\Nc}{N_c}
\newcommand{\ml}{m_\ell}
\newcommand{\Tr}{\text{Tr}}
\newcommand{\eps}{\epsilon}
\newcommand{\unity}{\mathds{1}}
\newcommand{\diag}{\text{diag}}

\allowdisplaybreaks[1]

\preprint{INT-PUB-18-042}
\title{Dispersion relation for hadronic light-by-light scattering: pion pole}

\author[a]{Martin Hoferichter,}
\author[b]{Bai-Long Hoid,}
\author[b]{Bastian Kubis,}
\author[c]{Stefan Leupold,}
\author[b]{and Sebastian~P.~Schneider}

\affiliation[a]{
Institute for Nuclear Theory, University of Washington, Seattle, WA 98195-1550, USA}

\affiliation[b]{
Helmholtz-Institut f\"ur Strahlen- und Kernphysik (Theorie) and \\
Bethe Center for Theoretical Physics, Universit\"at Bonn, 53115 Bonn, Germany}

\affiliation[c]{
Institutionen f\"or fysik och astronomi, Uppsala Universitet, Box 516, 75120 Uppsala, Sweden}

\emailAdd{mhofer@uw.edu}
\emailAdd{longbai@hiskp.uni-bonn.de}
\emailAdd{kubis@hiskp.uni-bonn.de}
\emailAdd{stefan.leupold@physics.uu.se}
\emailAdd{schneider@hiskp.uni-bonn.de}

\abstract{
The pion-pole contribution to hadronic light-by-light scattering in the anomalous magnetic moment of the muon $(g-2)_\mu$ is fully determined by the doubly-virtual pion transition form factor. Although this crucial input quantity is, in principle, directly accessible in experiment, a complete measurement covering
all kinematic regions relevant for $(g-2)_\mu$ is not realistic in the foreseeable future.
Here, we report in detail on a reconstruction from available data, both space- and time-like, using a dispersive representation that accounts for all the low-lying singularities, 
reproduces the correct high- and low-energy limits, and proves convenient for the evaluation of the $(g-2)_\mu$ loop integral. We concentrate on the systematics of the fit to $e^+e^-\to 3\pi$ data, which are key in constraining the isoscalar dependence, as well as the matching to the asymptotic limits. 
In particular, we provide a detailed account of the pion transition form factor at low energies in the time- and space-like region, including
the error estimates underlying our final result for the pion-pole contribution, $a_\mu^{\pi^0\text{-pole}}=62.6^{+3.0}_{-2.5}\times 10^{-11}$, 
and demonstrate how forthcoming singly-virtual measurements will further reduce its uncertainty. 
}

\begin{document} 
\maketitle

\newpage

\section{Introduction}
\label{sec:intro}

For decades the anomalous magnetic moment of the muon, $a_\mu=(g-2)_\mu/2$, has been one of the prime physical quantities both to test the Standard Model (SM) at quantum loop level, tracing back to the early milestone calculation performed in~\cite{Schwinger:1948iu}, and to monitor the signals coming from physics beyond the Standard Model (BSM). It can be experimentally measured to a very high precision, with the up-to-date value~\cite{Bennett:2006fi,Mohr:2015ccw}
\beq
a_{\mu}^{\text{exp}}= 116\, 592\, 089(63) \times 10^{-11},
\eeq 
revealing a tantalizing deviation of about $(3$--$4)\sigma$ from the SM prediction.\footnote{Recently, there have been hints for another deviation from the SM emerging in the anomalous magnetic moment of the electron, $(g-2)_e$, albeit presently only at the level of $2.5\sigma$~\cite{Hanneke:2008tm,Parker:2018,Davoudiasl:2018fbb,Crivellin:2018qmi}.} For this reason, an even more ambitious upgraded experiment at Fermilab~\cite{Grange:2015fou} and a complementary one at J-PARC~\cite{Saito:2012zz} are aiming at a four-fold improvement to achieve a precision of  $16\times 10^{-11}$ (see~\cite{Gorringe:2015cma} for a detailed comparison of the two approaches). Potential BSM contributions to $a_{\mu}$ notwithstanding, the current theoretical uncertainties of the SM contributions are required to be controlled more precisely in order to synchronize with the upcoming experimental precision. 

\begin{figure}[t]
	\centering
	\includegraphics[width=0.7\linewidth]{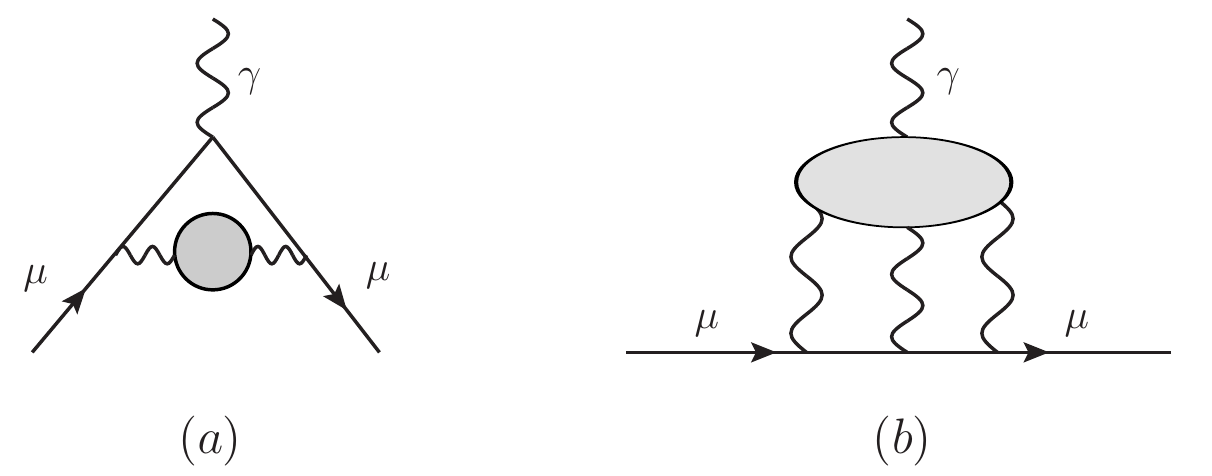}
	\caption{Diagrammatic representation of $(a)$ HVP and $(b)$ HLbL.}
	\label{fig:HCs}
\end{figure}

The dominant SM uncertainty arises from hadronic contributions~\cite{Jegerlehner:2009ry,Prades:2009tw,Benayoun:2014tra}, given that the uncertainty estimates of QED up to five loops~\cite{Aoyama:2012wk,Aoyama:2014sxa,Aoyama:2017uqe} (with analytical cross checks evaluated to four-loop order~\cite{Kurz:2015bia,Kurz:2016bau,Laporta:2017okg}) and electroweak contributions to two loops (including three-loop leading logarithms)~\cite{Czarnecki:2002nt,Gnendiger:2013pva} amount to $\lesssim 1\times 10^{-11}$. The first leading category, hadronic vacuum polarization (HVP) illustrated in diagram $(a)$ of Fig.~\ref{fig:HCs}, enters at $\Order(\alpha^2)$ in the expansion of the fine-structure constant, followed by the second hadronic light-by-light (HLbL) scattering category shown in diagram $(b)$ of Fig.~\ref{fig:HCs} at $\Order(\alpha^3)$. Higher-order insertions of HVP and HLbL scattering are already controlled sufficiently accurately~\cite{Calmet:1976kd,Kurz:2014wya,Colangelo:2014qya}. 
Despite the non-perturbative nature of these two contributions, it is possible to derive data-driven estimates based on dispersion relations. The HVP corrections can be related to the total cross section of $e^+e^-\to \text{hadrons}$~\cite{Bouchiat:1961,Brodsky:1967sr}. Therefore, its evaluation benefits from improved experimental measurements,  with most recent compilations~\cite{Jegerlehner:2017lbd,Davier:2017zfy,Keshavarzi:2018mgv,Colangelo:2018mtw} already providing uncertainties comparable to or less than HLbL. In contrast, current estimates of HLbL rely 
heavily on hadronic models~\cite{deRafael:1993za,Bijnens:1995cc,Bijnens:1995xf,Bijnens:2001cq,Hayakawa:1995ps,Hayakawa:1996ki,Hayakawa:1997rq,Knecht:2001qg,Knecht:2001qf,Blokland:2001pb,RamseyMusolf:2002cy,Melnikov:2003xd,Engel:2012xb,Masjuan:2012wy,Engel:2013kda,Roig:2014uja,Bijnens:2016hgx}, which despite being based on chiral symmetry or large-$N_c$ arguments\footnote{To ensure anomaly cancellation in the SM subtleties arise in the
large-$\Nc$ counting related to a rescaling of the quark charges. In consequence, the $\pi^0$- and $\eta_8$-pole contributions become suppressed
by two orders in $\Nc$ compared to their naive scaling, which strongly challenges the viability of the
large-$\Nc$ expansion as an organizing principle for HLbL scattering. This issue will be addressed below in App.~\ref{app:largeNc}.} and (partially) fulfilling constraints from perturbative QCD (pQCD) involve model uncertainties that are difficult to control. 
In this regard, a dispersive framework for the  evaluation of HLbL scattering based on the general principles of analyticity, unitarity, and crossing symmetry has been recently developed~\cite{Hoferichter:2013ama,Colangelo:2014dfa,Colangelo:2014pva,Colangelo:2015ama}, including first numerical results for two-pion contributions~\cite{Colangelo:2017qdm,Colangelo:2017fiz}.
Such a framework thus provides an alternative model-independent determination of HLbL scattering complementary to lattice QCD calculations~\cite{Blum:2014oka,Green:2015sra,Blum:2015gfa,Blum:2016lnc,Blum:2017cer}, attributing the contributions to on-shell form factors and scattering amplitudes that are, at least in principle, accessible experimentally. 

The single-meson poles constitute the simplest singularities of the HLbL tensor, whose residues are determined by the doubly-virtual transition form factors (TFFs). Therefore, the numerically dominant 
pion-pole contribution would be fully determined if the doubly-virtual pion TFF could be measured for all (relevant) space-like momenta. In the absence of such double-tag experiments for $e^+e^-\to e^+e^-\pi^0$, we dispersively reconstruct the pion TFF in light of the measurements of the $\pi^0\to\gamma\gamma$ decay width, the $e^+e^-\to 3\pi$ cross section, and the space-like singly-virtual form factor from $e^+e^-\to e^+e^-\pi^0$ again owing to the constraints from analyticity and unitarity. The resulting form factor representation
\beq
\label{eq:TFF_final}
F_{\pi^0\gamma^*\gamma^*}=F_{\pi^0\gamma^*\gamma^*}^\text{disp}+F_{\pi^0\gamma^*\gamma^*}^\text{eff}+F_{\pi^0\gamma^*\gamma^*}^\text{asym}
\eeq
takes into account all low-energy intermediate states by the first dispersive part, incorporates the normalization and space-like  high-energy data by the second (small) contribution from higher intermediate states, and implements the asymptotic constraints for arbitrary virtualities at $\Order(1/Q^2)$ via the last term. The pion-pole contribution is then evaluated based on this 
comprehensive dispersive determination of the pion TFF, completing previous efforts devoted to the data-driven determination of 
$a_\mu^{\pi^0\text{-pole}}$~\cite{Niecknig:2012sj,Schneider:2012ez,Hoferichter:2012pm,Hoferichter:2014vra,Hoferichter:2017ftn,Hoferichter:2018dmo} 
(see also~\cite{Adlarson:2012bi,Czerwinski:2012ry,Amaryan:2013eja,Bijnens:2014fya,Benayoun:2014tra,Adlarson:2014hka,Leupold:2018mgr}).

The paper is formatted as follows. The (unambiguous) definition of the pion-pole contribution to $a_{\mu}$ in the dispersive approach to HLbL scattering is recalled in Sect.~\ref{sec:pion_pole}, 
in terms of the on-shell pion TFF. Section~\ref{sec:disp_rel} is devoted to the dispersive reconstruction of the TFF based on its isospin decomposition and unitarity relation, the fits to the $e^+e^-\to3\pi$ cross section, and the double-spectral representation of the form factor. The decomposition~\eqref{eq:TFF_final} gives rise to various energy scales that are discussed in Sect.~\ref{sec:scales}. The asymptotic constraints dictated by pQCD are discussed in Sect.~\ref{sec:asymptotics}. The numerical results for the form factor in  both time-like and space-like regions as well as the pion-pole contribution to $a_{\mu}$ including a detailed discussion of its uncertainty estimates are presented in Sect.~\ref{sec:numerics}. Conclusions are drawn in Sect.~\ref{sec:sum} and additional supplementary material is collected in the appendices.

\section{Pion-pole contribution to $\boldsymbol{a_\mu}$}
\label{sec:pion_pole}

In order to evaluate the HLbL scattering contribution to the muon $(g-2)_\mu$, we define the full fourth-rank HLbL tensor $\Pi_{\mu\nu\lambda\sigma}$ following~\cite{Colangelo:2015ama},
\beq
\Pi_{\mu\nu\lambda\sigma}(q_1, q_2,q_3)=-i \int \diff^{4}x\, \diff^{4}y\, \diff^{4}z\,e^{-i(q_1\cdot x +q_2\cdot y+q_3\cdot z)} \langle 0|T\,\{ j_\mu (x)j_\nu (y)j_\lambda (z)j_\sigma (0)\}|0\rangle,
\eeq
where 
\beq
\label{eq:defemcurr}
j_\mu(x) = \frac{2}{3}(\bar u \gamma_\mu u)(x)-\frac{1}{3}(\bar d \gamma_\mu d)(x)-\frac{1}{3}(\bar s \gamma_\mu s)(x)
\eeq
denotes the electromagnetic currents carried by the light quarks and $q_i$ are the four-momenta of the photons. The leading-order HLbL contribution is then obtained by the projection technique~\cite{Brodsky:1966mv}:
\begin{align}
a_{\mu}^{\textrm{HLbL}} &= -\frac{e^6}{48m_\mu}
\int\frac{\diff^4 q_1}{(2\pi)^4} \int \frac{\diff^4 q_2}{(2\pi)^4}
\,\frac{1}{q_1^2 q_2^2 (q_1 + q_2)^2}\bigg[\frac{\partial}{\partial k^{\rho}}\Pi_{\mu\nu\lambda\sigma}(q_1, q_2,k-q_1-q_2)\bigg]_{k=0} \notag \\
&\times \textrm{Tr}\, \left\{(\slashed{p}+m_\mu)[\gamma^\rho,\gamma^\sigma](\slashed{p}+m_\mu)\gamma^\mu\frac{1}{\slashed{p}+\slashed{q}_1-m_\mu}\gamma^\lambda\frac{1}{\slashed{p}-\slashed{q}_2-m_\mu}\gamma^\nu\right\}, 
\end{align}
where $p$ is the four-momentum of the muon and $q_1 + q_2+q_3=0$. 

\begin{figure}[t] 
	\centering
	\includegraphics[width=\linewidth]{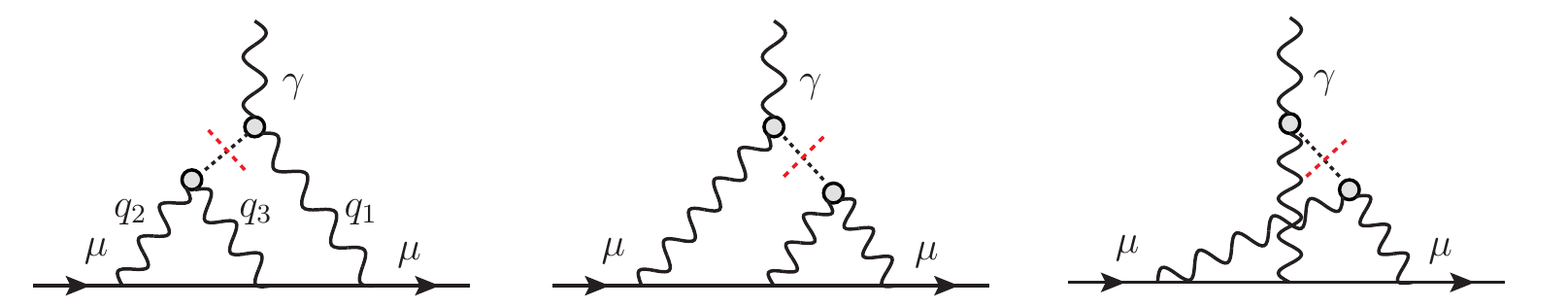} 
	\caption{The pion-pole contribution to HLbL scattering of the muon $(g-2)_\mu$.}
	\label{fig:pion_pole}
\end{figure}

 Diagrammatically, the pion-pole contribution can be attributed to the one-particle reducible piece of the HLbL tensor arising from a single pion propagator. There are three Feynman diagrams shown in Fig.~\ref{fig:pion_pole}, where the momenta are indicated in the hadronic subgraph. 
 
 After projection onto the muon anomaly, we obtain the result~\cite{Knecht:2001qf}
\begin{align}
\label{eq:tw}
a_{\mu}^{\pi^0\text{-pole}} &= - e^6 
\int\frac{\diff^4 q_1}{(2\pi)^4} \int \frac{\diff^4 q_2}{(2\pi)^4}
\,\frac{1}{q_1^2 q_2^2 (q_1 + q_2)^2[(p+q_1)^2-m_\mu^2][(p - q_2)^2 - m_\mu^2]} \notag \\
&\times \bigg[ 
\frac{F_{\pi^0\gamma^*\gamma^*}(q_1^2, (q_1 + q_2)^2)\, F_{\pi^0\gamma^*\gamma^*}( q_2^2, 0)}{q_2^2-\mpii^2}\,\hat T_1(q_1,q_2;p) \notag \\ 
&+\frac{F_{\pi^0\gamma^*\gamma^*}(q_1^2, q_2^2)\, F_{\pi^0\gamma^*\gamma^*}((q_1 + q_2)^2, 0)}{(q_1 + q_2)^2-\mpii^2} \,\hat T_2(q_1,q_2;p) \bigg],  
\end{align}
where $p^2=m_\mu^2$, $F_{\pi^0\gamma^*\gamma^*}$ is the on-shell pion TFF, and the integral kernels $\hat T_1$ and $\hat T_2$ are shown in App.~\ref{app:IK}. The first and second diagram give identical contributions collected in $\hat T_1$, while the third diagram leads to the term containing $\hat T_2$.
Critically, this diagrammatic derivation happens to coincide with its dispersive definition, obtained by carefully isolating the respective residues in the HLbL tensor~\cite{Colangelo:2014dfa,Colangelo:2015ama}.

After performing Wick rotations for the two-loop integrals, five out of six angular integrations can be carried out for arbitrary form factors resorting to Gegenbauer-polynomial techniques, 
which leads to a three-dimensional integral representation for the pion-pole contribution~\cite{Jegerlehner:2009ry},
\begin{align}
\label{eq:pion_pole}
a_\mu^{\pi^0\text{-pole}} & =\Big(\frac{\alpha}{\pi}\Big)^3\int_0^\infty \diff Q_1 \int_0^\infty \diff Q_2
\int_{-1}^{1} \diff \tau \notag\\
&\times \Big[w_1(Q_1,Q_2,\tau)\,F_{\pi^0\gamma^*\gamma^*}(-Q_1^2, -Q_3^2)\,F_{\pi^0\gamma^*\gamma^*}(-Q_2^2,0)  \notag\\
& + w_2(Q_1,Q_2,\tau)\,F_{\pi^0\gamma^*\gamma^*}(-Q_1^2, -Q_2^2)\, F_{\pi^0\gamma^*\gamma^*}(-Q_3^2,0)\Big],
\end{align}
where $Q_{1/2}^2=-q_{1/2}^2$, $Q_3^2 = Q_1^2 + 2 Q_1 Q_2 \tau + Q_2^2$, and $\tau= \cos \theta$, with $\theta$ the remaining angle between the Euclidean four-momenta $Q_1$ and $Q_2$.
The weight functions appearing in~\eqref{eq:pion_pole} are given by 
\begin{align}
\label{eq:km}
w_1(Q_1,Q_2,\tau)  & = -\frac{2\pi}{3}
\sqrt{1-\tau^2} \, \frac{Q_1^3 Q_2^3}{Q_2^2 +\mpii^2} \,
T_1(Q_1,Q_2,\tau), 
\notag\\ 
w_2(Q_1,Q_2,\tau)  & = -\frac{2\pi}{3}
\sqrt{1-\tau^2} \, \frac{Q_1^3 Q_2^3}{Q_3^2 +\mpii^2} \,
T_2(Q_1,Q_2,\tau),
\end{align}
where the kernel functions $T_1$ and $T_2$ are reproduced in App.~\ref{app:IK}. 

The relation~\eqref{eq:pion_pole} constitutes a special case of the master formula for the complete HLbL contribution to $a_\mu$~\cite{Colangelo:2015ama,Colangelo:2017fiz}, obtained by decomposing the HLbL tensor into scalar basis functions according to the general recipe established in~\cite{Bardeen:1969aw,Tarrach:1975tu} that ensure the absence of kinematic singularities and zeros, critical for 
the applicability of a dispersive representation. In the end, twelve combinations of these scalar functions $\bar \Pi_i$ enter the master formula 
\beq
\label{eq:masterformula}
a_\mu^\text{HLbL}= \frac{2 \alpha^3}{3 \pi^2} \int_0^\infty \diff Q_1 \int_0^\infty \diff Q_2 \int_{-1}^1 \diff\tau \sqrt{1-\tau^2} Q_1^3 Q_2^3 \sum_{i=1}^{12} \bar T_i(Q_1,Q_2,\tau) \bar \Pi_i(Q_1,Q_2,\tau),
\eeq
in which the pion pole only contributes to $\bar \Pi_1$ and $\bar \Pi_2$ 
\begin{align}
\label{Pi_pipole}
\bar \Pi_1^{\pi^0\text{-pole}}(Q_1,Q_2,\tau)&=-\frac{F_{\pi^0\gamma^*\gamma^*}(-Q_1^2,-Q_2^2)F_{\pi^0\gamma^*\gamma^*}(-Q_3^2,0)}{Q_3^2+\mpii^2}, \notag\\ 
\bar \Pi_2^{\pi^0\text{-pole}}(Q_1,Q_2,\tau)&=-\frac{F_{\pi^0\gamma^*\gamma^*}(-Q_1^2,-Q_3^2)F_{\pi^0\gamma^*\gamma^*}(-Q_2^2,0)}{Q_2^2+\mpii^2},
\end{align}
reproducing the equivalent representation~\eqref{eq:pion_pole} with $\bar T_1=T_2$ and $\bar T_2=T_1$.  

If dispersion relations are not derived for the HLbL tensor but for the Pauli form factor directly~\cite{Pauk:2014rfa},
this equivalence has so far only been confirmed for a vector-meson-dominance (VMD) form factor, and in general it is not guaranteed that dispersion relations for different quantities lead to the same notion of the pion pole. Moreover, in model calculations different definitions have been employed in the past,
including off-shell pions~\cite{Bartos:2001pg,Dorokhov:2008pw,Nyffeler:2009tw,Hong:2009zw,Cappiello:2010uy,Goecke:2010if,Dorokhov:2011zf,Kampf:2011ty,Greynat:2012ww,Dorokhov:2012qa,Roig:2014uja} and a variant introducing a constant form factor at one vertex~\cite{Melnikov:2003xd}. 
However, these ambiguities are specific to each particular model and do not occur in the dispersive approach to the HLbL tensor.
Once an organizing principle in terms of its singularities is accepted, 
the pion-pole contribution as given by the master formula~\eqref{eq:pion_pole} and~\eqref{eq:masterformula} follows unambiguously. 
In consequence, the most recent phenomenological evaluations~\cite{Nyffeler:2016gnb,Masjuan:2017tvw,Guevara:2018rhj}  and lattice QCD calculation~\cite{Gerardin:2016cqj} of the pion-pole contribution have adopted this dispersive definition. 

\begin{figure}[t]
	\centering
	\includegraphics[width=0.49\linewidth]{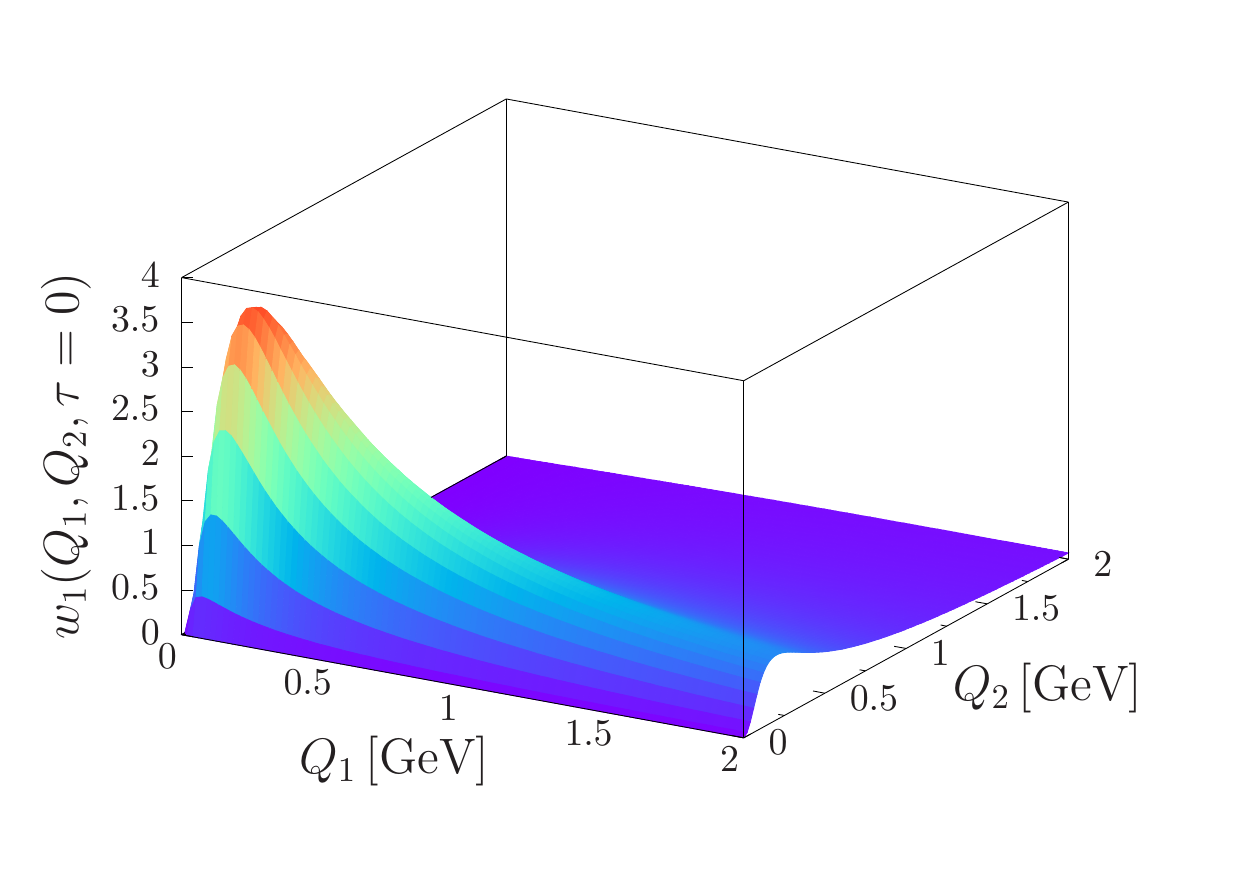}
	\includegraphics[width=0.49\linewidth]{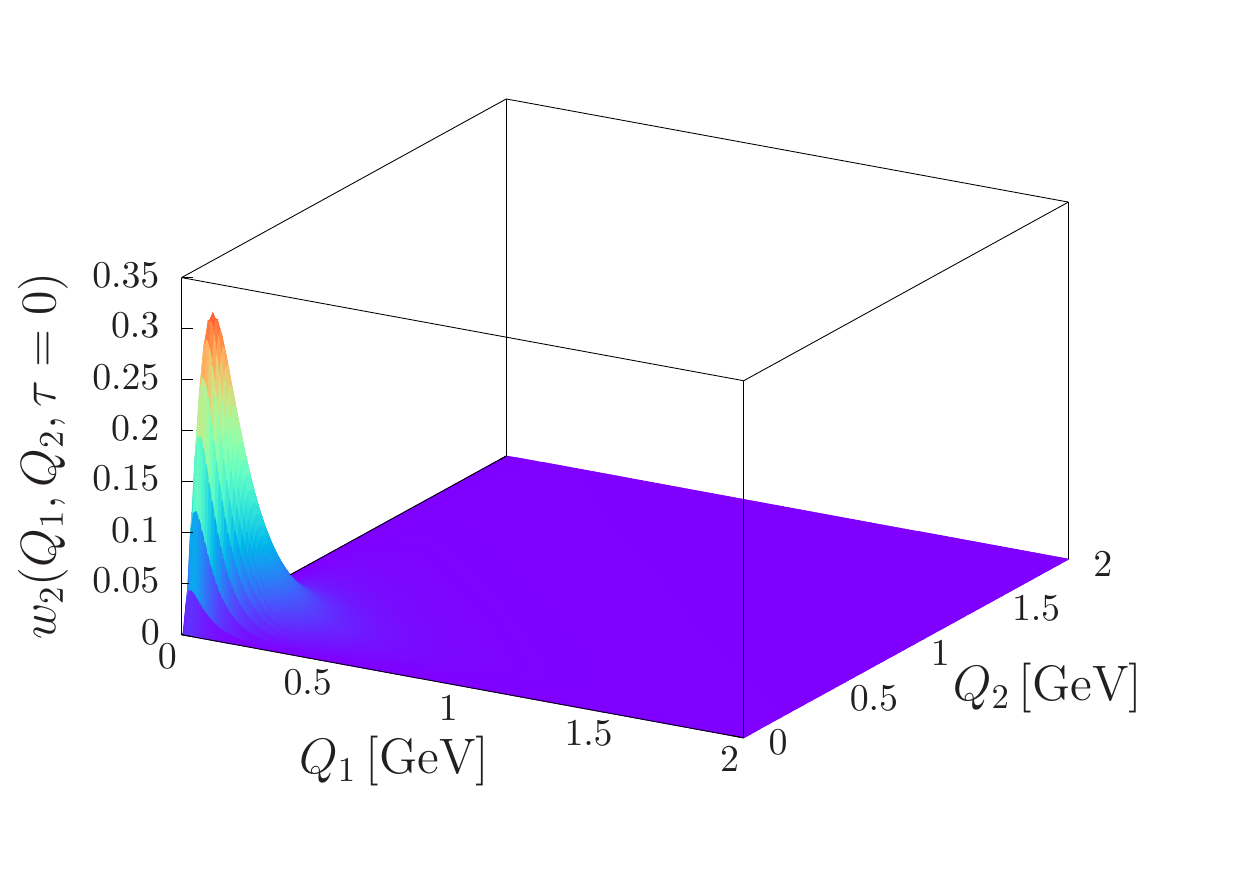}
	\caption{The weight functions $w_1(Q_1,Q_2,\tau)$ (left diagram) and $w_2(Q_1,Q_2,\tau)$ (right diagram) as  functions of $Q_1$ and $Q_2$ for $\tau=0,\, \theta=90^\circ$.}
	\label{fig:wf}
\end{figure}

The properties of the weight functions $w_1$ and $w_2$ have been studied extensively in~\cite{Nyffeler:2016gnb}. We briefly summarize their main features to gain some intuition for the evaluation of the multi-dimensional integral in the master formula~\eqref{eq:pion_pole}.
$w_1(Q_1,Q_2,\tau)$ and $w_2(Q_1,Q_2,\tau)$ are dimensionless, $w_2(Q_1,Q_2,\tau)$ is symmetric under $Q_1\leftrightarrow Q_2$, and both tend to zero for $Q_i\to0$ and $\tau\to\pm1$. Asymptotically, they behave according to
\begin{align}
\label{wi_asym}
\lim_{Q_1\to\infty}w_1(Q_1,Q_2,\tau)&\to \frac{1}{Q_1},\qquad \lim_{Q_2\to\infty}w_1(Q_1,Q_2,\tau)\to \frac{1}{Q_2^2},\notag\\
\lim_{Q_i\to\infty}w_2(Q_1,Q_2,\tau)&\to \frac{1}{Q_i^3},
\end{align}
hence assuring the convergence of the three-dimensional integral~\eqref{eq:pion_pole} for a form factor approaching zero at large momenta. In fact, the contribution from $w_2$ even converges for a pointlike form factor. 
To better understand the divergence structure of the integral, it is instructive to consider 
the leading order in chiral perturbation theory (ChPT).  
Since this corresponds to a pointlike form factor, the loop integral diverges, demanding a counter term that cannot be determined independently by other means but $a_\mu^{\pi^0\text{-pole}}$ itself. However, as pointed out in~\cite{RamseyMusolf:2002cy,Knecht:2001qg}, the chiral analysis does predict the logarithmically enhanced pieces, in a parameter-free way for the double logarithm and in terms of a low-energy constant (LEC) related to $P\to\ell^+\ell^-$ decays ($P=\pi^0,\eta$, $\ell=e,\mu$) for the single logarithm~\cite{Savage:1992ac,Ametller:1993we,Vasko:2011pi,Husek:2014tna}. 
In the dispersive approach, this relation to pseudoscalar dilepton decays is accounted for automatically in terms of the TFFs, see App.~\ref{app:EFT}, as a matter of fact more accurately without any need to rely on the chiral expansion. This relation between the TFF and pseudoscalar decays is well-established in the literature~\cite{Knecht:1999gb,Dorokhov:2007bd,Dorokhov:2008cd,Dorokhov:2009xs,Masjuan:2015lca,Husek:2015wta,Masjuan:2015cjl,Weil:2017knt}, and indeed the representation for the TFF derived here for $(g-2)_\mu$ should prove valuable for an improved prediction for the $\pi^0\to e^+e^-$ decay as well.

Finally, $w_1(Q_1,Q_2,\tau)$ and $w_2(Q_1,Q_2,\tau)$ are plotted as functions of $Q_1$ and $Q_2$ for $\tau=0$ ($\theta=90^\circ$) in Fig.~\ref{fig:wf}. 
It can been seen that the maximum peaks appear in the momenta range below $0.2\GeV$ for both $w_1(Q_1,Q_2,\tau)$ and $w_2(Q_1,Q_2,\tau)$. In line with the asymptotic behavior~\eqref{wi_asym} we find that $w_2(Q_1,Q_2,\tau)$ is roughly an order of magnitude smaller than $w_1(Q_1,Q_2,\tau)$ for the same values of $\tau$ and falls off faster compared to $w_1(Q_1,Q_2,\tau)$ after reaching the maximum peak.
In summary, the peaks of the weight functions $w_1(Q_1,Q_2,\tau)$ and $w_2(Q_1,Q_2,\tau)$ are concentrated in the momentum range $Q_i\leq0.5\GeV$ so that the most prevailing contribution in the master formula~\eqref{eq:pion_pole} arises from the low-energy region. Moreover, this is exactly the region where the pion TFF can be precisely determined in our dispersive framework, hence providing a possibility to model-independently evaluate the dominant pion-pole contribution with well-controlled uncertainties. Accordingly, we now turn to the dispersive determination of the pion TFF itself.

\section{Dispersion relations for the pion transition form factor}
\label{sec:disp_rel}

\subsection{Definition and low-energy properties}
\label{subsec:def_LEP}

The pion TFF is defined by the QCD vertex function
\beq
\label{eq:defpiTFF}
i\int \diff^4x \, e^{iq_1\cdot x}  \langle 0|T \, \{j_\mu(x)j_\nu(0)\}|\pi^0(q_1+q_2)\rangle 
 = -\epsilon_{\mu\nu\alpha\beta} \, q_1^\alpha \, q_2^\beta \, F_{\pi^0\gamma^*\gamma^*}(q_1^2,q_2^2),
\eeq
where $j_\mu$ are the light quark currents defined in~\eqref{eq:defemcurr} and $\eps^{0123}=+1$.\footnote{Note that the definition of $j_\mu$ in~\cite{Schneider:2012ez,Hoferichter:2012pm,Hoferichter:2014vra,Hoferichter:2017ftn} differs from~\eqref{eq:defemcurr} by a factor $e$. For $(g-2)_\mu$, however, the standard convention separates all factors of $e$ upfront, which leads to the normalization given in~\eqref{eq:LET}.} It describes the interaction between an on-shell neutral pion ($(q_1+q_2)^2=\mpii^2$) and two off-shell photons with four-momenta $q_1$ and $q_2$. The normalization of the form factor for real photons is dictated by the Adler--Bell--Jackiw anomaly~\cite{Adler:1969gk,Bell:1969ts,Bardeen:1969md}, 
\beq
\label{eq:LET}
F_{\pi^0\gamma^*\gamma^*}(0,0)=\frac{1}{4\pi^2\Fpi} \equiv F_{\pi\gamma\gamma},
\eeq
where $\Fpi=92.28(9)\MeV$~\cite{Olive:2016xmw} is the pion decay constant. It is related to the neutral pion decay width into two photons by $F_{\pi^0\gamma^*\gamma^*}^2(0,0)=4\,\Gamma(\pi^0 \to\gamma\gamma)/(\pi\alpha^2\mpii^3)$, which has been tested up to $1.4\%$ in a Primakoff measurement of the $\pi^0 \to \gamma\gamma$ decay width~\cite{Larin:2010kq}
(chiral and radiative corrections have been worked out in~\cite{Bijnens:1988kx,Goity:2002nn,Ananthanarayan:2002kj,Kampf:2009tk}). 
We will use the chiral tree-level prediction~\eqref{eq:LET} including the quark-mass renormalization of $\Fpi$, together with its $1.4\%$ uncertainty, as the central value and uncertainty estimate 
for the normalization of the TFF. The updated PrimEx-II experiment is expected to achieve a precision of $0.85\%$~\cite{Gasparian:2016oyl,Larin:2018}, so that, very likely, the dominant source of uncertainty might soon be of systematic nature in understanding the emerging tension with the chiral $2$-loop prediction~\cite{Kampf:2009tk}.   

\begin{figure}[t]
	\centering
	\includegraphics[width=0.4\linewidth]{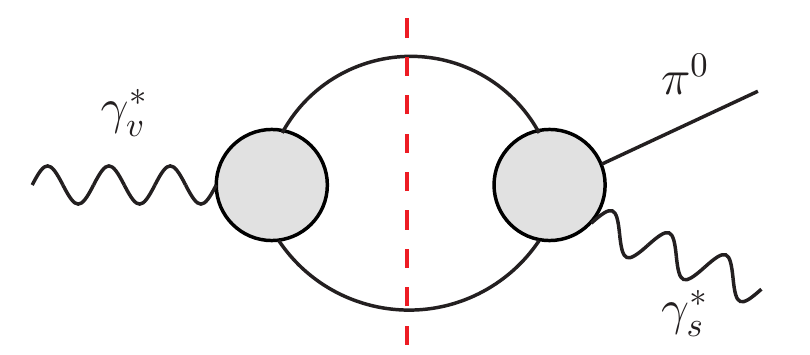}	
	\caption{Two-body unitarity relation for $\gamma^*_v\to\gamma^*_s \PiO$. The gray blobs represent the pion vector form factor and the $\gamma^*_s\to3\pi$ amplitude, respectively, and the solid lines pion intermediate states.}
	\label{fig:tbu}
\end{figure}

In a dispersive approach, the pion TFF is reconstructed from the most important lowest-lying singularities in the unitarity relation.\footnote{In general, we restrict our attention to purely hadronic states, i.e.\ neglect radiative processes/corrections, which is justified by the smallness of the electromagnetic 
coupling constant. An exception is the energy range of the $\omega$ meson due to its eight-percent branching to 
$\pi^0\gamma$~\cite{Olive:2016xmw}. This coupling of the three-pion states to $\pi^0\gamma$ is taken into account, see~\eqref{eq:allthewidths}.}  Assuming exact isospin symmetry, one of the photons in the $\pi^0 \gamma^* \gamma^*$ vertex must be an isovector ($I=1$) state and the other an isoscalar ($I=0$). Therefore, the form factor can be decomposed into definite-isospin virtualities as
\beq
\label{eq:isod}
F_{\pi^0\gamma^*\gamma^*}(q_1^2, q_2^2)=F_{vs}(q_1^2,q_2^2)+F_{vs}(q_2^2,q_1^2),
\eeq
where the isovector and isoscalar virtualities are labeled by the indices $v$ and $s$. At low energies, the unitarity relation for $\gamma^*_v\to\gamma^*_s \pi^0$ is dominated by the $\gamma^*_v\to\pi^+\pi^-\to\gamma^*_s \pi^0$ process as shown in Fig.~\ref{fig:tbu}. 
Consequently, the building blocks in the sub-diagrams are the pion vector form factor and the $\gamma^*_s\to3\pi$ amplitude. 

The pion vector form factor is described by two differently subtracted variants of the Omn\`es representation~\cite{Omnes:1958hv}. First, it is parameterized by  
\beq
F_\pi^V(s)=\big(1+\alpha_Vs\big)\Omega(s), \qquad
\Omega(s) = \exp\bigg\{\frac{s}{\pi}\int_{4M_\pi^2}^{\infty}\diff s'\frac{\delta(s')}{s'(s'-s)}\bigg\},
\eeq
where $\Omega(s)$ is the Omn\`{e}s function~\cite{Omnes:1958hv},
and three different $\pi\pi$ $P$-wave phase-shift inputs are used for $\delta(s)$: Bern and Madrid phases~\cite{Caprini:2011ky,GarciaMartin:2011cn}, respectively, are based on analyses of Roy- and Roy-like equations of $\pi\pi$ scattering. In addition, we consider an extension of~\cite{Caprini:2011ky} including the $\rho'(1450)$ and $\rho'^\prime (1700)$ resonances in an elastic approximation~\cite{Schneider:2012ez},
fit to the pion vector form factor as measured in $\tau$ decays~\cite{Fujikawa:2008ma}, in order to estimate the impact of inelasticities on the $\pi\pi$ input.
The coefficient $\alpha_V\sim(1$--$10)\times10^{-2}\GeV^{-2}$ is again obtained from a fit to~\cite{Fujikawa:2008ma} up to $1.0\GeV$ for Bern and Madrid phases and the full range for the third variant. The polynomial is set to a constant above $1.0\GeV$ ($1.9\GeV$ for the third phase) to attain a better high-energy behavior. Second, a twice-subtracted version as in~\cite{Guo:2008nc,Hoferichter:2014vra} is used below $1.3\GeV$ (below $1.9\GeV$ for the third phase),
\beq
F_\pi^V(s)=\exp\Bigg\{\frac{\langle r^2\rangle_\pi^V}{6}s+\frac{s^2}{\pi}\int_{4M_\pi^2}^{\infty}\diff s'\frac{\delta(s')}{s'^2(s'-s)}\Bigg\},
\eeq
with a fit radius $\langle r^2\rangle_\pi^V\sim0.436\,\text{fm}^2$ covering the data up to $1.0\GeV$. It is smoothly guided to the once-subtracted representation at $1.9\GeV$ by
adjusting the radius to the value that follows from the once-subtracted version by means of a sum rule, 
$\langle r^2_{\text{sum}}\rangle_\pi^V\sim0.420\,\text{fm}^2$.
The difference between both variants of $F_\pi^V$ enters the dispersive uncertainty for subsequently calculated quantities.

\begin{figure}[t]
 \begin{minipage}{0.495\linewidth}
 	\centering
	\includegraphics[width=0.8\linewidth]{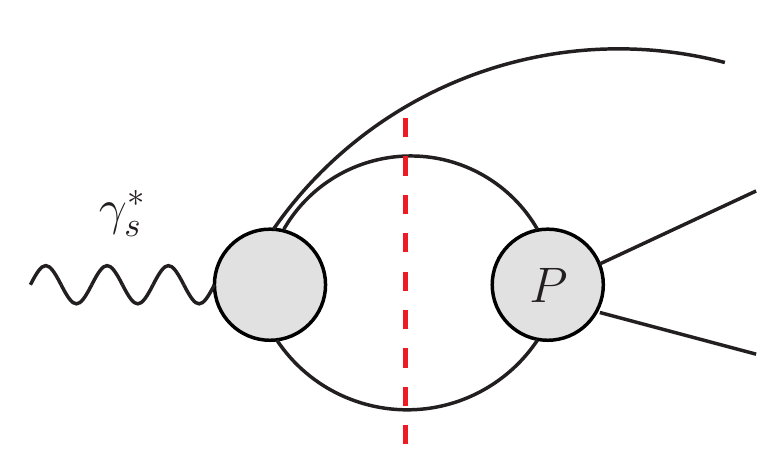}
\end{minipage}
\begin{minipage}{0.495\linewidth}
	 \centering
\includegraphics[width=\linewidth]{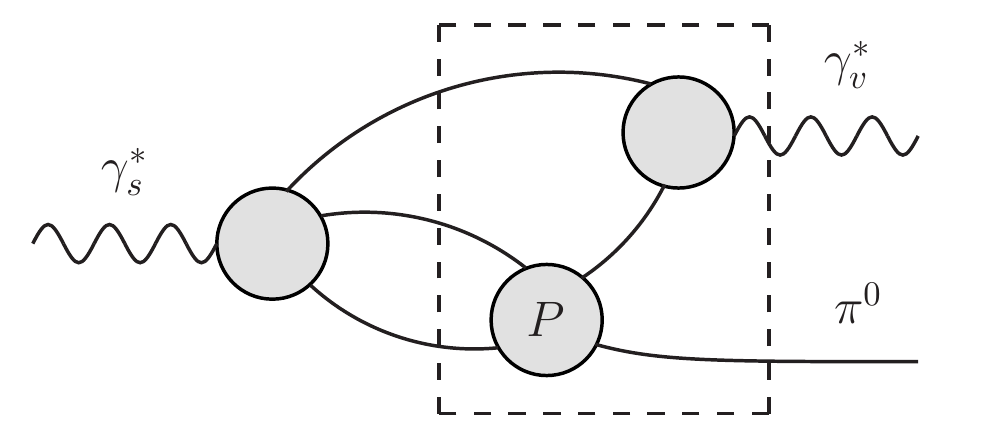}
\end{minipage}
	\caption{Two-body unitarity relation for the $\gamma^*_s\to3\pi$ amplitude (left) and the approximation for three-body unitarity in $\gamma^*_s\to\gamma^*_v\PiO$ that follows from the two-body rescattering (right). The part of the diagram in the dashed box can be viewed as a special case of the full $\pi^+\pi^-\PiO\to\gamma^*_v\PiO$ amplitude. The gray blob labeled $P$ refers to 
	the $P$-wave $\pi\pi$ scattering amplitude.}
	\label{fig:thbu}
\end{figure}

Turning to the $\gamma^*_s\to3\pi$ amplitude, its two-body unitarity relation is shown in the left diagram of Fig.~\ref{fig:thbu}. 
It involves the final-state interactions between pion pairs, which can be resummed in terms of the $P$-wave phase shift in the dispersive framework. However, it possesses a more complex analytic structure as a three-body decay process, which will be discussed in detail in Sect.~\ref{subsec:3pi}. While the full three-body unitarity $\gamma^*_s\to\pi^+\pi^-\PiO\to\gamma^*_v\PiO$ governing the unitarity relation for $\gamma^*_s\to\gamma^*_v\PiO$ cannot be implemented exactly in our approach, the $\pi\pi$ rescattering in the two-body unitarity relation for $\gamma_s^*\to3\pi$ already generates the leading topologies containing three-pion cuts for $\gamma^*_s\to\gamma^*_v\PiO$ as presented in the right diagram of Fig.~\ref{fig:thbu}, approximating the left-hand cut structure in $3\pi\to\gamma^*_v\PiO$ by pion-pole terms~\cite{Hoferichter:2014vra}.

\subsection[Parameterization of $e^+e^-\to 3\pi$]{Parameterization of $\boldsymbol{e^+e^-\to 3\pi}$}
\label{subsec:3pi}

\subsubsection[The $\gamma^*_s\to3\pi$ formalism]{The $\boldsymbol{\gamma^*_s\to3\pi}$ formalism}
\label{subsubsec:3piformalism}

We define the following matrix element in terms of the scalar function $\F(s,t,u; q^2)$ for the investigation of the $\gamma_s^*(q)\to\pi^+(p_+)\pi^-(p_-)\PiO(p_0)$ amplitude
\beq
\langle 0|j_\mu(0)|\pi^+(p_+)\pi^-(p_-)\PiO(p_0)\rangle =-\epsilon_{\mu\nu\rho\sigma}\, p_+^{\,\nu}p_-^{\,\rho} p_0^{\,\sigma}\F(s,t,u;q^2),
\eeq
with $q=p_++p_-+p_0$. The Mandelstam variables are chosen as $s=(q-p_0)^2$, $t=(q-p_+)^2$, and $u=(q-p_-)^2$, which fulfill $s+t+u=3\mpi^2+q^2$.

At low energy, the Wess--Zumino--Witten anomaly~\cite{Wess:1971yu,Witten:1983tw} provides a normalization for $\F$ in the chiral limit~\cite{Adler:1971nq,Terentev:1971cso,Aviv:1971hq}, which reads
\beq
\F(0,0,0; 0)=\frac{1}{4\pi^2 F^3_\pi}\equiv F_{3\pi}.
\eeq
So far, it has been tested only at the $10\%$ level both in the extraction from Primakoff measurements~\cite{Antipov:1986tp} and from the reaction $\pi^- e^-\to \pi^- e^-\PiO$~\cite{Giller:2005uy}. 
Therefore, a dispersive framework was proposed in~\cite{Hoferichter:2012pm,Hoferichter:2017ftn} to extract the chiral anomaly from the $\gamma\pi\to\pi\pi$ cross section up to $1\GeV$, 
using forthcoming data on $\gamma\pi^-\to \pi^-\pi^0$ taken in the COMPASS Primakoff program~\cite{Seyfried}.

 The partial-wave expansion of $\F$ in the $s$-channel reads~\cite{Jacob:1959at}
\beq
\F(s,t,u;q^2) =\sum_{l\;{\text{odd}}}f_l(s,q^2)P'_l(z_s),
\eeq
where only partial waves with odd angular momenta contribute and $z_s=\cos\theta_s$ is the cosine of the scattering angle in the $s$-channel. $P'_l(z_s)$ denotes the derivatives of the Legendre polynomials so that the dominant $P$-wave is projected out by 
\beq
f_1(s,q^2) = \frac{3}{4}\int_{-1}^1\diff z_s\,(1-z_s^2)\, \F(s,t,u;q^2).
\eeq

Neglecting discontinuities of $F$- and higher partial waves,\footnote{The effect of $F$-waves has been studied for $\omega\to 3\pi$~\cite{Niecknig:2012sj} and $\gamma\pi\to\pi\pi$~\cite{Hoferichter:2017ftn}, demonstrating that the non-zero contributions that arise in the vicinity of the $\rho_3(1690)$ resonance can be safely  ignored in the present context.} $\F$ can be decomposed into single-variable functions based on the reconstruction theorem~\cite{Stern:1993rg,Knecht:1995tr},
\beq
\label{eq:ee3piAmp}
\F(s,t,u; q^2)=\F(s, q^2)+\F(t, q^2)+\F(u, q^2).
\eeq
$\F(s, q^2)$ is related to the $l=1$ partial wave according to 
\begin{align}
	\label{eq:hat}
	f_1(s,q^2) &= \F(s,q^2) + \hat \F(s,q^2),\notag\\ \hat\F(s,q^2)& =  \frac{3}{2}\int_{-1}^1\diff z_s\,(1-z_s^2) \,\F\big(t(s,q^2,z_s),q^2\big),
\end{align} 
where
\beq
t(s,q^2,z_s) = \frac12 \, (3\mpi^2+q^2-s) + \frac12 \,\sigma_\pi(s)\,\lambda^{1/2}(q^2,\mpi^2,s)\, z_s,
\eeq
with $\sigma_\pi(s)=\sqrt{1-4\mpi^2/s}$ and $\lambda(x,y,z)=x^2+y^2+z^2-2(xy+yz+xz)$ the K\"all\'en function. $\hat \F(s,q^2)$ contains the left-hand-cut contribution to the partial wave $f_1(s,q^2)$ arising from the crossed-channel singularities. Furthermore, the angular integration in $\hat \F(s,q^2)$ imposes a complex analytic structure in the decay region $q^2>9\mpi^2$, which is explained in detail in~\cite{Niecknig:2012sj}. The discontinuity equation for $\F(s,q^2)$ reads
\beq
\label{F-dis}
\disc {\F}(s, q^2)=2i\,\big(\F(s, q^2)+\hat \F(s, q^2)\big)\,\theta(s-4M_\pi^2)\sin \delta(s)\, e^{-i\delta(s)},
\eeq
whose solution is given by a once-subtracted dispersive representation~\cite{Niecknig:2012sj}: 
\beq
\label{eq:solution}
\F(s, q^2)=\Omega(s)\bigg\{ a(q^2) + \frac{s}{\pi}\int_{4M_\pi^2}^{\infty} \diff s' \frac{\hat\F(s', q^2)\sin \delta(s')} {s'(s'-s)|\Omega(s')|} \bigg\}.
\eeq

The numerical calculation of the integral equation~\eqref{eq:solution} relies on the iterative solution of Khuri--Treiman (KT) equations~\cite{Khuri:1960zz} based on the observation that $\hat\F$ is linear in $\F$. In practice, we solve~\eqref{eq:solution} for $a(q^2)\to1$ (and a finite cutoff $\Lambda_{3\pi}$ above which we assume the asymptotic behavior $\hat\F(s, q^2)\sim 1/s$) and restore the full dependence as an overall normalization of the iterative solution.
For $q^2=M^2_{\omega/\phi}$, the solutions of~\eqref{eq:solution} 
have been used to describe the vector-meson decays $\omega/\phi\to3\pi$~\cite{Niecknig:2012sj}, where the subtraction constants $a$ are fixed from the partial decay widths of $\omega/\phi\to3\pi$. In the present case, $a(q^2)$, as a function of $q^2$, contains the information about the coupling of the isoscalar photon to $3\pi$ states. Therefore, $a(q^2)$ was determined from $e^+e^-\to3\pi$ cross section data in~\cite{Hoferichter:2014vra}, assuming that three-body unitarity for $\gamma^*_s\to3\pi$ is dominated by the narrow resonances $\omega$ and $\phi$. 

In this work, we further improve the parameterization of $a(q^2)$ by introducing a conformal polynomial to account for the effects from inelastic channels. 
In detail, we employ a once-subtracted representation with the addition of a conformal-polynomial term $C_p(q^2)$,
\beq
\label{eq:a-par}
a(q^2)=\alpha_A+\frac{q^2}{\pi}\int_{s_\text{thr}}^{\infty} \diff s'\frac{\Im\A (s')}{s'(s'-q^2)}+C_p(q^2),
\eeq
in which the function $\A$ is given by the sum of Breit--Wigner parameterizations
\beq
\A(q^2)=\sum_{V}\frac{c_V}{M_V^2-q^2-i\sqrt{q^2}\, \varGamma_V(q^2)},
\eeq
where $V$ represents $\omega$ and $\phi$ and as well as $\omega'(1420)$ and $\omega''(1650)$ as the description of the $e^+e^-\to3\pi$ cross section extends to $1.8\GeV$. The energy-dependent widths $\varGamma_{\omega/\phi}(q^2)$ of the $\omega/\phi$ mesons derive from their  main decay channels according to
\begin{align}
	\label{eq:allthewidths}
	\varGamma_\omega(q^2) &= \frac{\gamma_{\omega\to3\pi}(q^2)}{\gamma_{\omega\to3\pi}(M_\omega^2)}\varGamma_{\omega\to3\pi}
	+ \frac{\gamma_{\omega\to\pi^0\gamma}(q^2)}{\gamma_{\omega\to\pi^0\gamma}(M_\omega^2)}\varGamma_{\omega\to\pi^0\gamma}, \notag\\
	\varGamma_\phi(q^2) &= \frac{\gamma_{\phi\to3\pi}(q^2)}{\gamma_{\phi\to3\pi}(M_\phi^2)}\varGamma_{\phi\to3\pi} 
	+
	\sum \limits_{K=K^+,K^0}
	\frac{\gamma_{\phi\to K\bar K}(q^2)}{\gamma_{\phi\to K\bar K}(M_\phi^2)}\varGamma_{\phi\to K\bar K}, 
\end{align}
with $\varGamma_i$ the measured partial decay width for the decay $i$ and the energy-dependent coefficients 
\beq
\gamma_{\omega\to\pi^0\gamma}(q^2) = \frac{(q^2-\mpi^2)^3}{(q^2)^{3/2}},\qquad
\gamma_{\phi\to K\bar K}(q^2) = \frac{(q^2-4M_K^2)^{3/2}}{q^2}.
\eeq
The phase space $\gamma_{\omega/\phi\to3\pi}(q^2)$ is calculated as described in~\cite{Niecknig:2012sj}. These main channels amount to about $98\%$ of the $\omega$ and $\phi$ total widths, while the missing $2\%$ are remedied by rescaling all partial widths accordingly. We also considered adding the leading missing channels $\omega\to\pi^+\pi^-$ and $\phi\to\eta\gamma$ explicitly to the parameterization, but this yields an almost identical effect compared to the simple rescaling of the partial widths. For the $\omega'$ and $\omega''$  excited-state resonances, with masses and widths taken from~\cite{Olive:2016xmw}, we assume a $100\%$ branching ratio to $3\pi$. Due to the $\pi^0\gamma$ channel, the integration starts at $\sthr=\mpii^2$.
The subtraction constant $\alpha_A$ in equation~\eqref{eq:a-par} is fixed by the chiral anomaly at the real-photon point for $\gamma^*_s\to3\pi$ (corrected by quark-mass renormalization)~\cite{Bijnens:1989ff,Hoferichter:2012pm},
\beq
\label{eq:alphaF3pi}
\alpha_A= \frac{F_{3\pi}}{3}\times 1.066(10).
\eeq

Finally, the new conformal-polynomial term in~\eqref{eq:a-par} is given by
\beq
\label{Cp}
C_p(q^2)=\sum_{i=1}^p c_i\big(z(q^2)^i-z(0)^i\big), \qquad z(q^2)=\frac{\sqrt{s_\text{inel}-s_1}-\sqrt{s_\text{inel}-q^2}}{\sqrt{s_\text{inel}-s_1}+\sqrt{s_\text{inel}-q^2}},
\eeq
where the inelastic threshold $s_\text{inel}$ is chosen at $1\GeV^2$ motivated by the nearby $K\bar K$ threshold and the second parameter is fixed at $s_1=-1\GeV^2$. The degree $p$ of the conformal polynomial
is larger than the actual number of free parameters for the following reasons.
First, the $S$-wave cusp must be eliminated because of the $P$-wave nature of the photon. Second, $a(q^2)$ is constructed in such a way that the sum rule for the subtraction constant $\alpha_A$ is exactly fulfilled,
\beq
\label{eq:a_SR}
\alpha_A=\frac{1}{\pi}\int_{s_\text{thr}}^{\infty} \diff s'\frac{\Im{a}(s')}{s'}
=\frac{1}{\pi}\int_{s_\text{thr}}^{\infty} \diff s'\frac{\Im\A(s')}{s'}+\frac{1}{\pi}\int_{s_\text{inel}}^{\infty} \diff s'\frac{\Im{C_p}(s')}{s'},
\eeq
which induces another constraint on the coefficients $c_i$ in~\eqref{Cp}. Third, the integration in~\eqref{eq:a_SR} extends to infinity to fulfill the sum rule exactly, but in practice 
an isoscalar integration cutoff $\sis$ needs to be introduced, both for the double-spectral representation of the TFF that we will derive below
to satisfy the asymptotic constraints from pQCD 
and because the description of the $e^+e^-\to3\pi$ data based on KT equations cannot be justified to arbitrarily high energies.
In practice, we take $\sis=(1.8\GeV)^2$, so that, to ensure the validity of~\eqref{eq:a_SR}, the imaginary part of the conformal polynomial has to decrease sufficiently fast.
For that reason, we constrain the $c_i$ further to cancel the leading asymptotic behavior for $q^2\to\infty$. For a degree $p$ and $n$ constraints on the asymptotic behavior
the imaginary part behaves as $q^{-(2n+1)}$ and $p-n-2$ free parameters remain. 
We find that the low-energy $e^+e^-\to3\pi$ data can be well described with two free parameters
for $n=3$--$5$ and three free parameters for $n=6$, with small deviations starting around $1.6\GeV$. The representation for $a(q^2)$ constructed in this manner not only results in an improved description of the data, in particular above the $\phi$ resonance, but also guarantees the internal consistency of the different representations for the TFF when generalizing the single dispersion relation~\eqref{eq:un_DR} to the double-spectral representation~\eqref{eq:dsr_disp},
see Sect.~\ref{subsec:ds_rep}.

\subsubsection[Fit results for $e^+e^-\to3\pi$]{Fit results for $\boldsymbol{e^+e^-\to3\pi}$}
\label{subsubsec:3pifits}

We determine the normalization $a(q^2)$ by fitting the residues $c_V$ and the coefficients of the conformal polynomial $c_i$  to the $e^+e^-\to3\pi$ data. To this end, 
the relation between the $e^+e^-\to3\pi$ cross section (neglecting the electron mass) and the $\gamma^*_s \to \pi^+ \pi^-\pi^0$ amplitude~\eqref{eq:ee3piAmp} is given by 
\beq
\label{eq:epemcross1}
\sigma_{e^+ e^- \to 3\pi}(q^2) = \alpha^2\int_{s_\text{min}}^{s_\text{max}} \diff s \int_{t_\text{min}}^{t_\text{max}} \diff t \,
\frac{(s-4\mpi^2)\,\lambda(q^2,\mpi^2,s)\sin^2\theta_s}{768 \, \pi \, q^6}  \, |\F(s,t,u;q^2)|^2, 
\eeq
where the integration boundaries are
\begin{align}
s_\text{min} &= 4 M_\pi^2, \qquad\qquad \,s_\text{max} = \big(\sqrt{q^2}-M_\pi \big)^2, \notag \\ 
t_\text{min/max}&= (E_-^*+E_0^*)^2-\bigg( \sqrt{E_-^{*2}-M_\pi^2} \pm  \sqrt{E_0^{*2}-M_\pi^2} \bigg)^2,
\end{align}
with
\beq
E_-^*=\frac{\sqrt{s}}{2},\qquad E_0^*=\frac{q^2-s-M_\pi^2}{2\sqrt{s}}.
\eeq

As detailed in~\cite{Hoferichter:2014vra}, the most comprehensive single data sets of the $e^+ e^- \to 3\pi$ cross section at low and high energies are provided by SND~\cite{Achasov:2002ud,Achasov:2003ir} and BaBar~\cite{Aubert:2004kj}, respectively, so that the combined SND+BaBar data set yields the dominant constraint for the entire energy region below $1.8\GeV$, with negligible differences when fitting to the full data
base instead (see the fits in~\cite{Hoferichter:2014vra} to the data compilation from~\cite{Hagiwara:2011af}). The uncertainty estimates for the fits are generated based on the following variations: $\F(s,q^2)$ is calculated using the three different $\pi\pi$ phase shifts introduced in Sect.~\ref{subsec:def_LEP} in the context of the pion vector form factor. Additionally, the cutoff $\Lambda_{3\pi}$  in the integral equation~\eqref{eq:solution} above which the asymptotic behavior is assumed is varied from $1.8$ to $2.5\GeV$. 

\begin{figure}[t]
 	\centering
 	\includegraphics[width=0.8\linewidth]{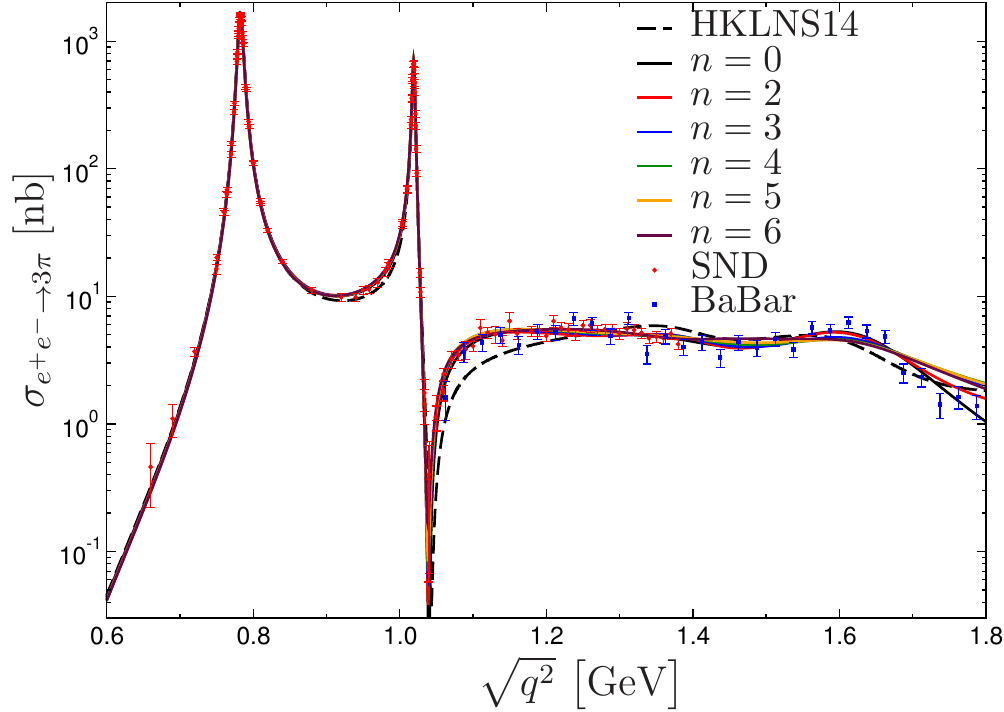}
 	\caption{Fits to the $e^+e^-\to 3\pi$ cross section from SND~\cite{Achasov:2002ud,Achasov:2003ir} and BaBar~\cite{Aubert:2004kj} with the different variants of the conformal polynomial labeled by $n$, the phase shift from~\cite{Caprini:2011ky}, and  $\Lambda_{3\pi}=2.5\GeV$,  in comparison to~\cite{Hoferichter:2014vra} (HKLNS14).}
 	\label{fig:fit3pi}
 \end{figure}

The $e^+ e^- \to 3\pi$ cross sections for different values of $n$ fit to the SND+BaBar data sets below $1.8\GeV$ using the phase shift from~\cite{Caprini:2011ky} and a cutoff $\Lambda_{3\pi}=2.5\GeV$ are shown in Fig.~\ref{fig:fit3pi}. 
It can be clearly seen that the fit results are substantially improved above the $\phi$ peak by introducing the conformal polynomial in comparison to the results obtained in~\cite{Hoferichter:2014vra}. The uncertainty bands for individual $n$ are not included in the plot as the curves would be hard to distinguish otherwise especially below $1.6\GeV$. The differences in the reduced $\chi^2$, see
 Table~\ref{tab:3pifit} for the explicit fit results for the different phase shifts and cutoffs $\Lambda_{3\pi}$, are almost exclusively generated by the high-energy end of the fit range, thus indicating that indeed our KT description starts to break down around $1.8\GeV$. The low-energy data, however, are described with a reduced $\chi^2/\text{dof}\sim 1$. 
 
\begin{table}[t]
	\small
	\centering
	\renewcommand{\arraystretch}{1.3}
	\begin{tabular}{lcccc}
		\toprule
		&$n=3$ & $n=4$ & $n=5$  & 	$n=6$ \\
		\midrule
		$c_\omega \ [\text{GeV}^{-1}]$& $2.87\ldots2.90$  & $2.85\ldots2.88$ & $2.84\ldots2.87$  & $2.83\ldots2.86$\\	
		
		$c_\phi \ [\text{GeV}^{-1}]$ & $-(0.400\ldots0.412)$ & $-(0.400\ldots0.414)$ & $-(0.400\ldots0.414)$ & $-(0.400\ldots0.413)$\\	
		
		$c_{\omega'} \ [\text{GeV}^{-1}]$ &$-(0.24\ldots0.52)$&  $-(0.14\ldots0.39)$ &  $-(0.040\ldots0.33)$ & $-0.15\ldots0.14$  \\	
		
		$c_{\omega''} \ [\text{GeV}^{-1}]$ &$-(0.80\ldots1.16)$ &$-(0.60\ldots0.94)$& $-(0.49\ldots0.90)$ & $-(0.45\ldots0.78)$\\
		
		$c_1 \ [\text{GeV}^{-3}]$ & $-(1.56\ldots 1.79)$ & $ -(1.75\ldots1.96)$& $-(1.81\ldots 2.08)$ & $-(2.00\ldots 2.24)$ \\
		
		$c_2 \ [\text{GeV}^{-3}]$ & $-(1.05\ldots 1.16)$ & $-(1.28\ldots1.40)$ & $-(1.44\ldots 1.50)$  & $-(1.67\ldots 1.73)$ \\
		
		$c_3 \ [\text{GeV}^{-3}]$ & ---& --- & --- & $-0.05\ldots0.12$  \\
		
		$\chi^2/\text{dof}$ & $1.37\ldots 1.70$ &  $1.58\ldots 2.03$ & $1.68\ldots 2.15$  & $1.71\ldots 2.16$\\
		\bottomrule
		\renewcommand{\arraystretch}{1.0}
	\end{tabular}
	\caption{Fit parameters and reduced $\chi^2$ for the $e^+e^-\to3\pi$ fits to SND+BaBar~\cite{Achasov:2002ud,Achasov:2003ir,Aubert:2004kj} using different versions of the conformal polynomial with asymptotic behavior $q^{-(2n+1)}$. The ranges indicate the variation found for the different $\pi\pi$ phase shifts and values of $\Lambda_{3\pi}$.} 
	\label{tab:3pifit}
\end{table}

\subsection{Double-spectral representation}
\label{subsec:ds_rep}

The previous discussion of the isospin decomposition~\eqref{eq:isod} and the crucial building blocks in the unitarity relation for the pion TFF, the pion vector form factor $F_\pi^V(s)$ and the $\gamma_s^*\to3\pi$ $P$-wave amplitude $f_1(s,q^2)$, defines the quantities that enter 
a once-subtracted dispersion relation in the isovector virtuality (for fixed isoscalar virtuality)~\cite{Hoferichter:2012pm},
\beq
\label{eq:os_DR}
F_{vs}(q_1^2,q_2^2)=F_{vs}(0,q_2^2)+\frac{q_1^2}{12\pi^2}\int^\infty_{4\mpi^2}\diff x\frac{q_{\pi}^3(x)\big(F_\pi^{V}(x)\big)^*f_1(x,q_2^2)}{x^{3/2}(x-q_1^2)},
\eeq
with $q_\pi(s)=\sqrt{s/4-M_\pi^2}$. For $q_2^2=M_{\omega/\phi}^2$, the representation~\eqref{eq:os_DR} has been used to describe the $\omega/\phi\to\pi^0\gamma^*$ TFFs, where the sum rule for the subtraction function
\beq
\label{eq:os_DR_SR}
F_{vs}(0,q_2^2) = \frac{1}{12\pi^2}\int^\infty_{4\mpi^2}\diff x\frac{q_{\pi}^3(x)}{x^{3/2}}\big(F_\pi^{V}(x)\big)^*f_1(x,q_2^2)
\eeq
is related to the real-photon decays~\cite{Schneider:2012ez}. For $q_2^2=0$,~\eqref{eq:os_DR} yields the isovector part of the singly-virtual pion TFF, 
\beq
\label{eq:os_DR_sv}
F_{vs}(q_1^2,0)=F_{vs}(0,0)+\frac{q_1^2}{12\pi^2}\int^\infty_{4\mpi^2}\diff x\frac{q_{\pi}^3(x)\big(F_\pi^{V}(x)\big)^*f_1(x,0)}{x^{3/2}(x-q_1^2)},
\eeq
where the sum rule $F_{vs}(0,0)=F_{\pi\gamma\gamma}/2$ is typically saturated at the $90\%$ level~\cite{Hoferichter:2012pm,Hoferichter:2014vra}. 

For the $(g-2)_\mu$ application~\eqref{eq:pion_pole} we need a representation of the space-like doubly-virtual form factor that can be evaluated at arbitrarily high energies, matching smoothly onto 
the asymptotic behavior expected from pQCD, see Sect.~\ref{sec:asymptotics}.
In this regard, the once-subtracted representation is disfavored because it approaches a constant for large virtualities, contradicting the pQCD scaling, unless 
the sum rule for the subtraction constant is  fulfilled exactly. In practice, however, the uncertainties in the input always generate
variants of the form factor that behave as a constant at high energies, and such a constant
form factor does not lead to a convergent $(g-2)_\mu$ integral.
Therefore, we start from an unsubtracted dispersion relation~\cite{Hoferichter:2014vra}
\beq
\label{eq:un_DR}
F_{vs}(q_1^2,q_2^2)=
\frac{1}{12\pi^2}\int^{\infty}_{4\mpi^2}\diff x\frac{q_{\pi}^3(x)\big(F_\pi^{V}(x)\big)^{*}f_1(x,q_2^2)}{x^{1/2}(x-q_1^2)},
\eeq
despite the expected $10\%$ violation of the sum rule for the normalization $F_{\pi\gamma\gamma}/2$. To remedy this shortcoming, we introduce
an isovector integration cutoff $\siv$ and add an effective pole collecting the contributions from higher intermediate states and high-energy contributions in the $2\pi$ and $3\pi$ channels, see Sect.~\ref{sec:asymptotics} for details. In this manner, the representation~\eqref{eq:un_DR}, in principle, already determines the general doubly-virtual form factor. 
However, to find a representation that facilitates the evaluation in the entire space-like region
we derive a more compact double-spectral representation that makes the analyticity of the form factor $F_{\pi^0\gamma^*\gamma^*}(q_1^2,q_2^2)$ in both of its arguments $q_1^2$ and $q_2^2$ explicit,
\beq
\label{eq:dsr}
F_{\pi^0\gamma^*\gamma^*}(q_1^2,q_2^2)=\frac{1}{\pi^2}\int_0^{\infty} \diff x \int_0^{\infty} \diff y \frac{\rho(x,y)}{(x-q_1^2)(y-q_2^2)},
\eeq
 where $\rho(x,y)$ is the double-spectral density that we aim to reconstruct from the low-lying hadronic intermediate states. Accordingly, the single dispersion relation~\eqref{eq:un_DR} is elevated to the double-spectral form by performing yet another dispersion relation in the isoscalar variable,
 \beq
  F_{vs}(-Q_1^2,q_2^2)=\frac{1}{\pi}\int_{\sthr}^{\sis} \diff y\frac{\Im F_{vs}(-Q_1^2,y)}{y-q_2^2}=\frac{1}{12\pi^2}\int^{\siv}_{4\mpi^2}\diff x\frac{q_{\pi}^3(x)\big(F_\pi^{V}(x)\big)^{*}f_1(x,q_2^2)}{x^{1/2}\big(x+Q_1^2\big)},
 \eeq
 where $\sis$ is the isoscalar integration cutoff and the threshold $\sthr=\mpii^2$ is the same as in~\eqref{eq:a_SR}. This leads to a double-spectral representation of the form factor,
 \begin{align}
 \label{eq:dsr_disp}
 F_{\pi^0\gamma^*\gamma^*}^\text{disp}(-Q_1^2,-Q_2^2)&=
 \frac{1}{\pi^2} \int_{4M_\pi^2}^{\siv} \diff x \int_{\sthr}^{\sis}\diff y \frac{ \rho^\text{disp}(x,y)}{\big(x+Q_1^2\big)\big(y+Q_2^2\big)} +\big(Q_1\leftrightarrow Q_2\big),\notag\\
 \rho^\text{disp}(x,y)&=\frac{q_\pi^3(x)}{12\pi\sqrt{x}}\Im \Big[\big(F_\pi^{V}(x)\big)^*f_1(x,y)\Big],
 \end{align}
 to describe the low-energy properties, which can be applied to space-like doubly-virtual  kinematics.  The nonzero imaginary part of $F_{vs}(-Q_1^2,q^2)$ is attributed to three-body unitarity in the isoscalar virtuality, both the three-pion cuts which result in the deviation of the phase of $f_1(s,q^2)$ from the phase of $F_\pi^V(s)$ in the decay region $q^2>9\mpi^2$~\cite{Schneider:2012ez} and the complex nature of $a(q^2)$ as well.  
 In fact, the complicated analytic structure of the partial wave $f_1(s,q^2)$ itself might make it seem surprising that the TFF fulfills a dispersive representation
 as simple as~\eqref{eq:dsr}, see App.~\ref{sec:anom_thr} for a more detailed discussion.

 Formally, the equivalence of the single dispersion relation~\eqref{eq:un_DR} and the double-spectral representation~\eqref{eq:dsr_disp} for $F_{vs}(q_1^2,q_2^2)$ implies a sum rule
 \beq
 \label{sum_rule}
 \big(F_\pi^{V}(s)\big)^*f_1(s,q_2^2)= \frac{1}{\pi}\int_{s_\text{thr}}^{\sis} \diff y\frac{\Im \Big[\big(F_\pi^{V}(s)\big)^*f_1(s,y)\Big]}{y-q_2^2}, 
 \eeq
 which, once finite cutoffs are applied, requires that the singularities be concentrated in the low-energy region to ensure overall consistency, precisely the motivation for constraining 
 the high-energy behavior of the imaginary part of $a(q^2)$ accordingly. In this context, due to the pseudothreshold singularities located at $s=(\sqrt{q_2^2}-\mpi)^2$~\cite{Schneider:2012ez}, it becomes more convenient to consider the integrated quantities instead, which is why we do not pursue the sum rule~\eqref{sum_rule} itself any further.

\section{Relevant scales for the transition between low and high energies}
\label{sec:scales}

Having presented the construction of the dispersive representation of the low-energy properties of the pion TFF, we first wish to offer a qualitative understanding of the relevant \textit{scales} that show up in its subsequent quantitative completion at higher energies. 
To this end, we will use phenomenologically successful models.  
The following reasoning is meant to be of qualitative use to help understanding the characteristic mass or energy scales that we find later in the model-independent final calculations. 

For the calculation of the hadronic quantum fluctuations in the magnetic moment of the muon, the latter's mass provides
a scale somewhat smaller than the masses of pions; hence it is clear that the low-energy sector of QCD plays the 
most important role for these quantum fluctuations.
Yet, concerning the pion-pole contribution, it turns out that only a proper high-energy behavior of 
the pion TFF guarantees the convergence of the corresponding integrals. Thus, pure low-energy information is not enough for a 
quantitatively reliable determination of the pion-pole contribution. Fortunately, pQCD provides some input for
the asymptotic behavior of the pion TFF~\cite{Lepage:1979zb,Lepage:1980fj,Brodsky:1981rp}. 
Also from a practical point of view, an interpolation between 
the low-energy region and the asymptotic behavior is more constraining and therefore more accurate than a pure extrapolation.
The question related to relevant scales is then: where is the effective onset $\sm$ of the asymptotic region? 

The central piece of our framework is the dispersive representation of the pion TFF: at low energies, the virtual photons couple 
dominantly to two- and three-pion states.
Below about $1\GeV$, these two- and three-pion states essentially behave elastically. Their rescattering 
is quantitatively under control by the dispersive framework 
developed in~\cite{Hoferichter:2012pm,Niecknig:2012sj,Schneider:2012ez,Hoferichter:2014vra,Hoferichter:2017ftn}.
We use  the phrase ``low-energy region'' to characterize the regime dominated by elastic reactions.
Above $1\GeV$, new channels, i.e.\ inelasticities become important. For instance, in the isovector channel, 
the two-pion states (and the virtual photon) couple to 
four-pion states~\cite{TheBaBar:2017vzo,Colangelo:2001df,Ecker:2002cw,GarciaMartin:2011cn,Caprini:2011ky,Niecknig:2012sj}. 
Although the threshold for four pions lies significantly below $1\GeV$, both the smallness of four-pion phase space near threshold
and the derivative couplings of the pions 
demanded by chiral symmetry effectively delay the onset of the importance of the four-pion states
to the $\pi\omega$ threshold.  
In the isoscalar channel, the three-pion states (and the virtual photon) couple to kaon pairs;
this is particularly significant in the energy region 
of the $\phi$ meson, which has sizable branching fractions to kaon pairs and to three pions~\cite{Olive:2016xmw}. 
Of course, these are only examples: at higher energies, more and more channels come into play.

From a technical point of view, it is much more 
challenging to deal with the coupled-channel dynamics above $1\GeV$. 
On the other hand, it should be clear that for our purposes 
a less detailed knowledge of the regime beyond the low-energy region is acceptable. 
We have to expect an effective scale $\Mrhop$ of the higher-lying inelasticities, 
i.e.\ the effective scale of the physics 
not covered by two- and three-pion states and their respective elastic rescattering, to reside 
at an energy larger than $1\GeV$; but we shall argue now that it cannot be far away from it either. 

The pion TFF is a part of the $PVV$ 
three-point correlator, where $P$/$V$ denotes a quark current with pseudoscalar/vector quantum 
numbers. With the standard Lehmann--Symanzik--Zimmermann procedure, one can map out the pion-pole contribution to the $PVV$ correlator; see, e.g., \cite{Husek:2015wta}. The crucial point is that the whole $PVV$ correlator would vanish if chiral symmetry were not 
broken~\cite{Knecht:2001xc}. On the other hand, chiral symmetry breaking is a long-distance, low-energy phenomenon. 
Quantitatively, it is characterized by the scale $4\pi F_\pi \approx 1\GeV$~\cite{Manohar:1983md}.
Thus, the pion TFF as part of the $PVV$ correlator cannot be influenced too
much from high-lying inelasticities, and we expect $\Mrhop$ more or less close to $1\GeV$. 

This reasoning is not entirely independent of the question concerning the onset $\sm$ of the asymptotic region; 
yet, it is not the same question. The asymptotic region concerns large \textit{space-like} momenta where one can apply 
pQCD and the operator product expansion (OPE), 
while the higher-lying inelasticities concern the \textit{time-like} input for a dispersive representation. 
To relate the frameworks of OPE and dispersion theory, we use the 
QCD sum rule method~\cite{Shifman:1978by,Novikov:1983jt,Reinders:1984sr,Narison:1989aq,Leinweber:1995fn}, 
to be more specific: the light-cone QCD sum rules (LCSRs)~\cite{Chernyak:1983ej,Khodjamirian:1997tk,Agaev:2010aq,Mikhailov:2016klg}. 
The details of this analysis with the aim of an estimate for $\sm$ are provided in App.~\ref{sec:LV-app}. 
In the QCD sum rule language, $\sm$ coincides with the duality 
threshold. It enters as a free parameter that must be determined by comparison to data. 
For the case at hand, we compare to the singly-virtual pion TFF. By construction, the duality threshold must lie above 
the low-energy regime that is parameterized explicitly by hadronic resonances in the sum rule method, yet the analysis of 
App.~\ref{sec:LV-app} reveals that the duality threshold cannot lie significantly higher either. Figure~\ref{fig:LV}
in App.~\ref{sec:LV-app} shows that the best agreement with the data on the singly-virtual pion TFF is achieved by 
low values of $\sm$, again not much larger than $1\GeV^2$.

\section{Matching to the asymptotic behavior}
\label{sec:asymptotics}

The dispersive double-spectral density of~\eqref{eq:dsr_disp} incorporates all the low-lying singularities in the $2\pi$ and $3\pi$ channels, but does not account for higher intermediate states nor the correct matching to pQCD. Therefore, we now develop the explicit form of the effective and asymptotic contributions in~\eqref{eq:TFF_final}, considering both 
leading-order (LO) and next-to-leading-order (NLO) pQCD dynamics as well as an effective pole in order to impose the correct normalization $F_{\pi\gamma\gamma}$ and incorporate the constraints from space-like singly-virtual data.

\subsection{Leading-order perturbative QCD}
\label{subsec:LOpQCD}

 If both momenta $q_1^2$ and  $q_2^2$ are large (and have the same sign), the $T$-product of the electromagnetic currents $j_\mu$ in~\eqref{eq:defpiTFF} can be expanded along the light cone $x^2=0$. The lowest-order and leading-twist expansion of the TFF reads~\cite{Lepage:1979zb,Lepage:1980fj,Brodsky:1981rp}
\beq
\label{eq:pQCD}
F_{\pi^0\gamma^*\gamma^*}(q_1^2,q_2^2)=-\frac{2\Fpi}{3}\int_0^1\diff u\frac{\phi_\pi(u)}{u q_1^2+(1-u) q_2^2} +\Order\big(q_i^{-4}\big),
\eeq
where powers of asymptotic momenta are denoted by $q_i$. The twist-two pion distribution amplitude can be expanded in terms of Gegenbauer polynomials $C_{2n}^{3/2}$ as
\beq
\phi_\pi(u,\mu)=6u(1-u)\bigg[1+\sum_{n=1}^{\infty}a_{2n}(\mu)C_{2n}^{3/2}(2u-1)\bigg],
\eeq
which provides a universal asymptotic distribution amplitude $\phi_\pi(u)=6u(1-u)$ at large factorization scale $\mu \to \infty$ as the logarithmically $\mu$-dependent coefficients $a_{2n}$ tend to zero.
Since at low scales the non-perturbative coefficients $a_{2n}$ are largely unknown,  we will use the asymptotic distribution amplitude $\phi_\pi(u)$ in the following analysis, ignoring the higher-order terms $n\geq 1$ as well as higher-twist corrections. 

Introducing an asymmetry parameter $\omega=(q_1^2-q_2^2)/(q_1^2+q_2^2)$, the leading expression~\eqref{eq:pQCD} can be changed into the form
 \beq
 \label{eq:pQCD_2}
F_{\pi^0\gamma^*\gamma^*}(q_1^2,q_2^2)=-\frac{4\Fpi}{3}\frac{f(\omega)}{q_1^2+q_2^2} +\Order\big(q_i^{-4}\big),
\eeq
where
 \beq
 \label{eq:f_omega}
f(\omega)=\int_0^1\diff u\frac{\phi_\pi(u)}{u (1-\omega )+(1-u) (1+\omega )}.
\eeq
Specifically, this implies the OPE limit~\cite{Nesterenko:1982dn,Novikov:1983jt} for the diagonal form factor ($\omega=0$),
\beq
\label{eq:OPE_limit}
F_{\pi^0\gamma^*\gamma^*}(-Q^2,-Q^2)=\frac{2\Fpi}{3Q^2}+\Order\big(Q^{-4}\big).
\eeq
In addition, formal evaluation at $\omega=\pm1$ produces
\beq
\label{eq:BL_limit}
F_{\pi^0\gamma^*\gamma^*}(-Q^2,0)=F_{\pi^0\gamma^*\gamma^*}(0,-Q^2)=\frac{2\Fpi}{Q^2}+\Order\big(Q^{-4}\big),
\eeq
usually referred to as the Brodsky--Lepage (BL) limit of the singly-virtual form factor. However, the OPE expansion justifies~\eqref{eq:pQCD}
only for $|\omega|<1/2$~\cite{Gorsky:1987,Manohar:1990hu}, otherwise its derivation cannot be considered rigorous. Apart from these two frequently studied conventional limits,~\eqref{eq:pQCD_2} also predicts the asymptotic behavior for arbitrary virtualities $q_1^2$ and $q_2^2$ by~\eqref{eq:f_omega}. Hence, our representation will be matched to $f(\omega)$ to fully take into account the entire domain of space-like virtualities, instead of just two particular limits~\eqref{eq:OPE_limit} and~\eqref{eq:BL_limit}. Beyond the leading expansion~\eqref{eq:pQCD}, calculations including $\alpha_s$ corrections~\cite{delAguila:1981nk,Braaten:1982yp}, higher terms in the Gegenbauer-polynomial expansion of $\phi_\pi(u)$~\cite{Chernyak:1981zz,Chernyak:1983ej} within QCD sum rules~\cite{Khodjamirian:1997tk,Agaev:2010aq,Mikhailov:2016klg,Radyushkin:1996tb}, Dyson--Schwinger equations~\cite{Raya:2015gva,Eichmann:2017wil}, and Regge theory~\cite{RuizArriola:2006jge,Arriola:2010aq,Gorchtein:2011vf} could be considered, but a consistent treatment of all subleading corrections
becomes very complicated with little numerical impact on $(g-2)_\mu$. As an explicit example we will consider $\alpha_s$ corrections in Sect.~\ref{subsec:NLOpQCD}. 

At LO, we implement the pQCD constraints as follows. First, it has been observed that~\eqref{eq:pQCD} can be transformed into a dispersion relation by a simple change of variables $u\to x/(x-q_2^2)$ for space-like virtuality $q_2^2$~\cite{Khodjamirian:1997tk},
\beq
F_{\pi^0\gamma^*\gamma^*}(q_1^2,q_2^2)
=\frac{1}{\pi}\int_0^\infty \diff x \frac{\Im F_{\pi^0\gamma^*\gamma^*}(x,q_2^2)}{x-q_1^2},
\eeq
with
\beq
\Im F_{\pi^0\gamma^*\gamma^*}(x,q_2^2)=\frac{2\pi \Fpi}{3(x-q_2^2)}\phi_\pi\Big(\frac{x}{x-q_2^2}\Big).
\eeq
Furthermore, we find that identifying the discontinuities in the second variable $q_2^2$ leads to a new double-spectral representation for the asymptotic expression:
\beq
\label{double_spectral_asym}
F_{\pi^0\gamma^*\gamma^*}(q_1^2,q_2^2)=\frac{1}{\pi^2}\int_0^\infty  \diff x \int_0^\infty\diff y\frac{\rho^\text{asym}(x,y)}{(x -q_1^2)(y- q_2^2)},
\eeq
where
\beq
\label{eq:dsd}
\rho^\text{asym}(x,y)=-2\pi^2\Fpi xy\delta''(x-y)
\eeq
is a double-spectral density proportional to $xy$ and concentrated along the diagonal direction $x=y$ because of the second derivative of the delta function. 
Note that the singular nature of $\rho^\text{asym}(x,y)$ along the diagonal direction is a rather general feature not restricted to the asymptotic distribution amplitude $\phi_\pi(u)$. 
For instance,  a constant pion distribution amplitude $\phi_\pi(u)=1$ produces a double spectral density $(2\pi^2\Fpi/3)\delta(x-y)$ proposed in the context of QCD sum rules~\cite{Novikov:1983jt}.
 
The double-spectral form of the pQCD expression~\eqref{double_spectral_asym} then suggests to decompose the TFF in terms of the   
different integration regions
\begin{align}
\label{eq:decom}
F_{\pi^0\gamma^*\gamma^*}(q_1^2,q_2^2&)=\frac{1}{\pi^2}\int_0^{\sm} \diff x \int_0^{\sm} \diff y \frac{\rho(x,y)}{(x-q_1^2)(y-q_2^2)}
+\frac{1}{\pi^2}\int_{\sm}^\infty \diff x \int_{\sm}^\infty \diff y \frac{\rho(x,y)}{(x-q_1^2)(y-q_2^2)}\notag\\
&+\frac{1}{\pi^2}\int_0^{\sm} \diff x \int_{\sm}^\infty \diff y\, \frac{\rho(x,y)}{(x-q_1^2)(y-q_2^2)}
+\frac{1}{\pi^2}\int_{\sm}^\infty \diff x\int_0^{\sm} \diff y\, \frac{\rho(x,y)}{(x-q_1^2)(y-q_2^2)},
\end{align}
where $\sm$ is a continuum threshold introduced to separate the different regions, see the discussion in Sect.~\ref{sec:scales}. On the one hand, the low-energy input to the double-spectral density has been derived in~\eqref{eq:dsr_disp}. On the other, the spectral density in the doubly-asymptotic region can be identified with $\rho^\text{asym}(x,y)$ in~\eqref{eq:dsd}.
The spectral densities in the third and fourth mixed low- and high-energy regions are not well constrained, e.g.\ the asymptotic spectral density $\rho^\text{asym}(x,y)$ applied in these regions 
simply vanishes. Given that the contribution from the doubly-asymptotic region alone can provide the correct asymptotic behavior and that both the BL limit as well as the available data 
can be described with a combination of the low-energy dispersive contribution and an effective pole,
we will discard the contributions from the mixed regions altogether assuming that the effective pole sufficiently takes care of them. In the end, this defines the asymptotic contribution
\begin{align}
\label{eq:asym}
F_{\pi^0\gamma^*\gamma^*}^\text{asym}(q_1^2,q_2^2)
&= 2\Fpi\int_{\sm}^\infty \diff x \frac{q_1^2q_2^2}{(x-q_1^2)^2(x-q_2^2)^2}\notag\\
&=\frac{2\Fpi q_1^2q_2^2}{(q_1^2-q_2^2)^2}\bigg(\frac{1}{\sm-q_1^2}+\frac{1}{\sm-q_2^2}+\frac{2}{q_1^2-q_2^2}\log\frac{\sm-q_1^2}{\sm-q_2^2}\bigg),
\end{align}
which reproduces the limit defined by~\eqref{eq:pQCD} for non-vanishing virtualities.

We remark that an asymptotic contribution of the form~\eqref{eq:asym} could also be used to impose the correct asymptotic behavior on a hadronic model. For instance, for a VMD-inspired model one could write
\beq
\label{VMD_modified}
F_{\pi^0\gamma^*\gamma^*}^\text{VMD}(q_1^2,q_2^2)=F_{\pi\gamma\gamma}\bigg(\frac{(1-\eps)\,M_1^4}{(M_1^2-q_1^2)(M_1^2-q_2^2)}
+\frac{\eps\,M_2^4}{(M_2^2-q_1^2)(M_2^2-q_2^2)}\bigg)+F_{\pi^0\gamma^*\gamma^*}^\text{asym}(q_1^2,q_2^2),
\eeq
which amounts to a simplified model for our full representation~\eqref{eq:TFF_final}.
By construction, all asymptotic limits for non-vanishing virtualities are correct, while the strict BL limit~\eqref{eq:BL_limit} emerges for $(1-\eps)M_1^2+\eps M_2^2=8\pi^2\Fpi^2$.
We tried to describe our full result using~\eqref{VMD_modified} as an approximation, treating either $M_1$, $M_2$, and $\eps$, or, in addition, $\sm$ as free fit parameters. Such an ansatz seems to work reasonably well, with 
systematic errors introduced at the level of $a_\mu$ around $0.5\times 10^{-11}$, but of course cannot replace the full calculation.

\subsection{Next-to-leading-order perturbative QCD}
\label{subsec:NLOpQCD}

Higher orders in pQCD beyond the leading result~\cite{Lepage:1979zb,Lepage:1980fj,Brodsky:1981rp}
have been derived in~\cite{Braaten:1982yp}. Adapted to our notation, the corresponding correction can be expressed as
\begin{align}
\label{F_alpha_s}
 F_{\pi^0\gamma^*\gamma^*}(q_1^2,q_2^2)&=-\frac{2\Fpi}{3}\int_0^1\diff u\frac{\phi_\pi(u)}{u q_1^2+(1-u) q_2^2}\bigg(1+\frac{C_F\alpha_s(\mu_s^2)}{2\pi} f(u,-q_1^2,-q_2^2,-\mu^2)\bigg),\notag\\
 f(u,q_1^2,q_2^2,\mu^2)&=-\frac{9}{2}+\frac{L_{12}(L_{12}-2)}{2}\bigg(1-\frac{q_1^2q_2^2}{(q_1^2-q_2^2)^2u(1-u)}\bigg)+\frac{3}{2}L_{12}\notag\\
 &-\frac{q_1^2}{2(q_1^2-q_2^2)}\bigg(1-\frac{q_2^2}{(q_1^2-q_2^2)(1-u)}\bigg)L_1(L_1-2)
 +\frac{q_2^2}{2(q_1^2-q_2^2)u}(L_{12}-L_2)\notag\\
 &+\frac{q_2^2}{2(q_1^2-q_2^2)}\bigg(1+\frac{q_1^2}{(q_1^2-q_2^2)u}\bigg)L_2(L_2-2)
 -\frac{q_1^2}{2(q_1^2-q_2^2)(1-u)}(L_{12}-L_1),\notag\\
 L_i&=\log\frac{q_i^2}{\mu^2},\qquad L_{12}=\log\frac{u q_1^2+(1-u)q_2^2}{\mu^2}, \qquad C_F=\frac{N_c^2-1}{2N_c}=\frac{4}{3}.
\end{align}
In the singly-virtual limit we obtain
\beq
F_{\pi^0\gamma^*\gamma^*}(-Q^2,0)=\frac{2\Fpi}{Q^2}\bigg(1-\frac{5}{2}\frac{C_F\alpha_s(-Q^2)}{2\pi}\bigg)
=\frac{2\Fpi}{Q^2}\bigg(1-\frac{5}{3}\frac{\alpha_s(-Q^2)}{\pi}\bigg),
\eeq
in agreement with the result stated in~\cite{Braaten:1982yp}. Similarly, evaluation in the doubly-virtual limit produces
\beq
F_{\pi^0\gamma^*\gamma^*}(-Q^2,-Q^2)=\frac{2\Fpi}{3Q^2}\bigg(1-\frac{3}{2}\frac{C_F\alpha_s(-2Q^2)}{2\pi}\bigg)
=\frac{2\Fpi}{3Q^2}\bigg(1-\frac{\alpha_s(-2Q^2)}{\pi}\bigg).
\eeq
In each case, we have set $\mu_s^2=q_1^2+q_2^2$~\cite{Braaten:1982yp}. 
As a powerful check on~\eqref{F_alpha_s} the dependence on $\mu$ cancels also for general virtualities if the asymptotic form of the distribution amplitude is employed. Subleading terms in the Gegenbauer-polynomial expansion of the pion distribution amplitude again depend on $\mu$, which compensates the $\mu$ dependence within the non-asymptotic $\alpha_s$ corrections.

For the asymptotic contribution to the pion TFF we seek corrections to
\beq
F_{\pi^0\gamma^*\gamma^*}^\text{asym}(q_1^2,q_2^2)=2\Fpi\int_{\sm}^\infty\diff x\frac{q_1^2q_2^2}{(x-q_1^2)^2(x-q_2^2)^2}.
\eeq
Since the corresponding double-spectral function is peaked at $x=y$, the canonical choice of scale should be
\begin{align}
F_{\pi^0\gamma^*\gamma^*}^\text{asym}(q_1^2,q_2^2)&=2\Fpi\int_{\sm}^\infty\diff x\frac{q_1^2q_2^2}{(x-q_1^2)^2(x-q_2^2)^2}
\bigg(1+\frac{2}{3\pi}\alpha_s(-x)\delta(q_1^2,q_2^2,-x)\bigg),\notag\\
\delta(q_1^2,q_2^2,\mu^2)&=\frac{\int_0^1\diff u \frac{\phi_\pi(u)}{u q_1^2+(1-u) q_2^2}f(u,-q_1^2,-q_2^2,-\mu^2)}{\int_0^1\diff u \frac{\phi_\pi(u)}{u q_1^2+(1-u) q_2^2}},
\end{align}
and we have checked that for $Q^2$ values of practical importance this estimate yields corrections close to the naive expectation $-\alpha_s(-2Q^2)/\pi\sim -10\%$ from the doubly-virtual limit.
In the end, the uncertainty in the choice of matching scale $\sm$ in the LO contribution safely encompasses such corrections.

\subsection{Constraints from singly-virtual data}
\label{subsec:sv_data}

As the next step, we present the conceptual ideas how to incorporate high-energy TFF data in our representation~\eqref{eq:TFF_final}. 
The final results of the corresponding fits will be provided in Sect.~\ref{sec:numerics} together with all other results for the pion TFF in various kinematic regimes.

Despite the absence of doubly-virtual measurements of the TFF thus far, there is ample experimental information for space-like singly-virtual kinematics~\cite{Behrend:1990sr,Gronberg:1997fj,Aubert:2009mc,Uehara:2012ag}. 
These data sets cover primarily large virtualities and thus provide the opportunity to probe the high-energy behavior of the singly-virtual form factor 
beyond the low-energy region $\lesssim 1\GeV$, the latter being most relevant for $a_\mu$. Most high-energy data in fact corroborate the BL 
limit $\lim_{Q^2\to\infty}Q^2F_{\pi^0\gamma^*\gamma^*}(-Q^2,0)=2\Fpi$ with $f(|\omega|=1)=3/2$ despite the questionable convergence at $|\omega|=1$, in contrast to a naive continuation of the OPE $f(|\omega|=1)=1$ or $f(|\omega|=1)=5/2$ obtained form the Chernyak--Zhitnitsky distribution amplitude~\cite{Chernyak:1981zz,Chernyak:1983ej}. 
Potential deviations from the BL limit were suggested by the BaBar experiment~\cite{Aubert:2009mc}, where the measured form factor exceeded the BL limit by as much as $50\%$ at $Q^2>10\GeV^2$,
but the latest Belle measurement~\cite{Uehara:2012ag} did not find any evidence for such a rapid growth at high $Q^2$. We will assign sufficiently broad uncertainty bands that cover both scenarios, 
so that our final result for $a_\mu^{\pi^0\text{-pole}}$ will not depend on any prejudice either way.

Our representation evaluated for singly-virtual asymptotics receives contributions from the low-energy dispersive part~\eqref{eq:dsr_disp}, while the pQCD term~\eqref{eq:asym} vanishes.  
In practice, the low-energy representation~\eqref{eq:dsr_disp} already fulfills the BL limit at a level around $55\%$, so  that only the remainder needs to be generated by higher intermediate states
as well as high-energy contributions to the $2\pi$ and $3\pi$ channels. 
This can be conveniently achieved by an effective pole in the double-spectral density, which amounts to an extra term
\beq
\label{eff_pole}
F_{\pi^0\gamma^*\gamma^*}^\text{eff}(q_1^2,q_2^2)=\frac{\grhop}{4\pi^2\Fpi}\frac{\Mrhop^4}{(\Mrhop^2-q_1^2)(\Mrhop^2-q_2^2)},
\eeq
where the coupling $\grhop$ is determined by imposing the sum rule for $F_{\pi\gamma\gamma}$ and the mass parameter $\Mrhop$ is fit to the space-like singly-virtual data~\cite{Behrend:1990sr,Gronberg:1997fj,Aubert:2009mc,Uehara:2012ag}. The resulting parameters $\grhop$ and $\Mrhop$ are found to be around $10\%$ and $(1.5$--$2)\GeV$ respectively, in agreement with the assumption that an effective pole subsumes the contributions from higher intermediate states.
As pointed out in the discussions of the pion phase shift ($2\pi$ states) and of the fit to the cross section for $e^+ e^- \to 3\pi$ ($3\pi$ states), our dispersive representation includes some part of the spectral strength of the energy region $(1$--$2)\GeV$. Naively, one might then expect that the complementary part covered by the effective pole of~\eqref{eff_pole} should lead to a value of $\Mrhop$ significantly higher up in energy. However, as pointed out in Sect.~\ref{sec:scales}, there cannot be much spectral strength at very high energies contributing to the pion TFF. Phrased differently, the range found for $\Mrhop$ is completely reasonable and a better description of the region above $1\GeV$ would merely lead to a smaller value of $\grhop$ instead of a higher value of $\Mrhop$.

In view of the tension of the BaBar data~\cite{Aubert:2009mc} both with the BL limit and the other data sets we need to specify how we treat the corresponding systematic uncertainty in our fits.
First, we observe that, while otherwise the results are very stable with respect to the lower threshold $Q^2_\text{min}$
above which data are fit, including the BaBar data induces a strong sensitivity on $Q^2_\text{min}$, and the $\chi^2$ deteriorates appreciably if $Q^2_\text{min}$ is increased.
For this reason, we define the central value of our analysis by the fit to all data sets excluding BaBar, with $Q^2_\text{min}=5\GeV^2$, which leads to an asymptotic value almost exactly at the BL limit.
To estimate the systematic uncertainties, we perform fits with $Q^2_\text{min}=(5\text{--}10)\GeV^2$, with and without the BaBar data, and for each fit consider a $3\sigma$ error band. 
The envelope of all these fits corresponds to an uncertainty band $^{+20}_{-10}\%$ around the central value, where the asymmetric error reflects the fact that the BaBar data imply 
a systematic shift in the upward direction. 
In this way, we assign a very generous error band to the space-like fits, in such a way that the systematic uncertainties are safely covered by the corresponding error estimate in our final result.
Moreover, since only data above $5\GeV^2$ are included in the fit, the low-energy region remains a prediction, effectively improving the asymptotic behavior of the result from~\cite{Hoferichter:2014vra} by the matching to the pQCD constraints.

\section{Numerical results}
\label{sec:numerics}

In this section we present the numerical outcome of our analysis. First of all, the singly-virtual pion TFF in the time-like region is predicted and the resulting $e^+e^-\to \pi^0\gamma$ 
cross section is compared to the corresponding experimental results. Second, the space-like doubly-virtual form factor is discussed, in particular along the singly-virtual and the diagonal direction, and the asymptotic behavior in the entire domain of space-like kinematics is further confronted with the predictions from pQCD. 
Last, the pion-pole contribution to $a_\mu$ is calculated along with comprehensive uncertainty estimates, each of which will be related to the various experimental input quantities.

\subsection[Time-like form factor and $e^+e^-\to \pi^0\gamma$]{Time-like form factor and $\boldsymbol{e^+e^-\to \pi^0\gamma}$}
\label{subsec:timelike}

According to~\eqref{eq:os_DR_SR} and~\eqref{eq:os_DR_sv}, the time-like singly-virtual TFF obeys a once-subtracted dispersion relation: 
\beq
\label{eq:os_TFF_tl}
F_{\pi^0\gamma^*\gamma^*}(q^2,0)=F_{\pi\gamma\gamma}
+\frac{1}{12\pi^2}\int_{4M_\pi^2}^\infty \diff x\frac{q_\pi^3(x)\big(F_\pi^{V}(x)\big)^* }{x^{3/2}} \bigg\{f_1(x,q^2)-f_1(x,0) 
+\frac{q^2}{x-q^2}f_1(x,0)\bigg\},
\eeq
where the normalization at the real photon point $q^2=0$ is fixed to the chiral anomaly using again the sum rule~\eqref{eq:os_DR_SR}. For the studies in~\cite{Hoferichter:2014vra}, the isoscalar contribution corresponding to the first two terms in the integrand of~\eqref{eq:os_TFF_tl} was calculated using the previously determined partial wave $f_1(s,q^2)$, where an asymptotic continuation $\sim 1/x$ was assumed above the isovector integration cutoff $\siv$. 
The last term, the isovector piece, was determined using a finite matching point of $1.2\GeV$~\cite{Hoferichter:2012pm}.
Here, we will consider an update of this once-subtracted analysis based on the new parameterization for $a(q^2)$, including the conformal polynomial and the new isovector part, where $\siv$ is chosen as a strict integration cutoff for both isoscalar and isovector contributions in line with the dispersive representation~\eqref{eq:dsr_disp}. 
At the same time, the double-spectral representation~\eqref{eq:dsr} provides an unsubtracted form of the time-like TFF 
\beq
\label{eq:un_TFF_tl}
F_{\pi^0\gamma^*\gamma^*}(q^2,0)=F_{\pi^0\gamma^*\gamma^*}^\text{disp}(q^2,0)+F_{\pi^0\gamma^*\gamma^*}^\text{eff}(q^2,0),
\eeq
where the determination of the parameters $\grhop$ and $\Mrhop$ in the effective pole is described in Sect.~\ref{subsec:sv_data}.

The relation between the $e^+e^- \to \pi^0 \gamma$ cross section and the pion TFF reads (neglecting the mass of the electron for simplicity)
\beq
\label{eq:cross-sec-pig}
\sigma_{e^+e^- \to \pi^0 \gamma}(q^2) = \alpha^2\frac{(q^2-M_{\pi^0}^2)^3\,\pi}{ 6\, q^6} \, |F_{\pi^0\gamma^*\gamma^*}(q^2,0)|^2.
\eeq
We emphasize that our predictions of the time-like form factor and thus the cross section are entirely based on the dispersive framework with the input quantities described in the previous sections: the anomalies $F_{\pi\gamma\gamma}$ and $F_{3\pi}$, the  $\pi\pi$ $P$-wave phase shift, the pion vector form factor, and the $e^+ e^- \to 3\pi$ cross section data. 

\begin{figure}[t]
	\centering
	\includegraphics[width=0.9\linewidth]{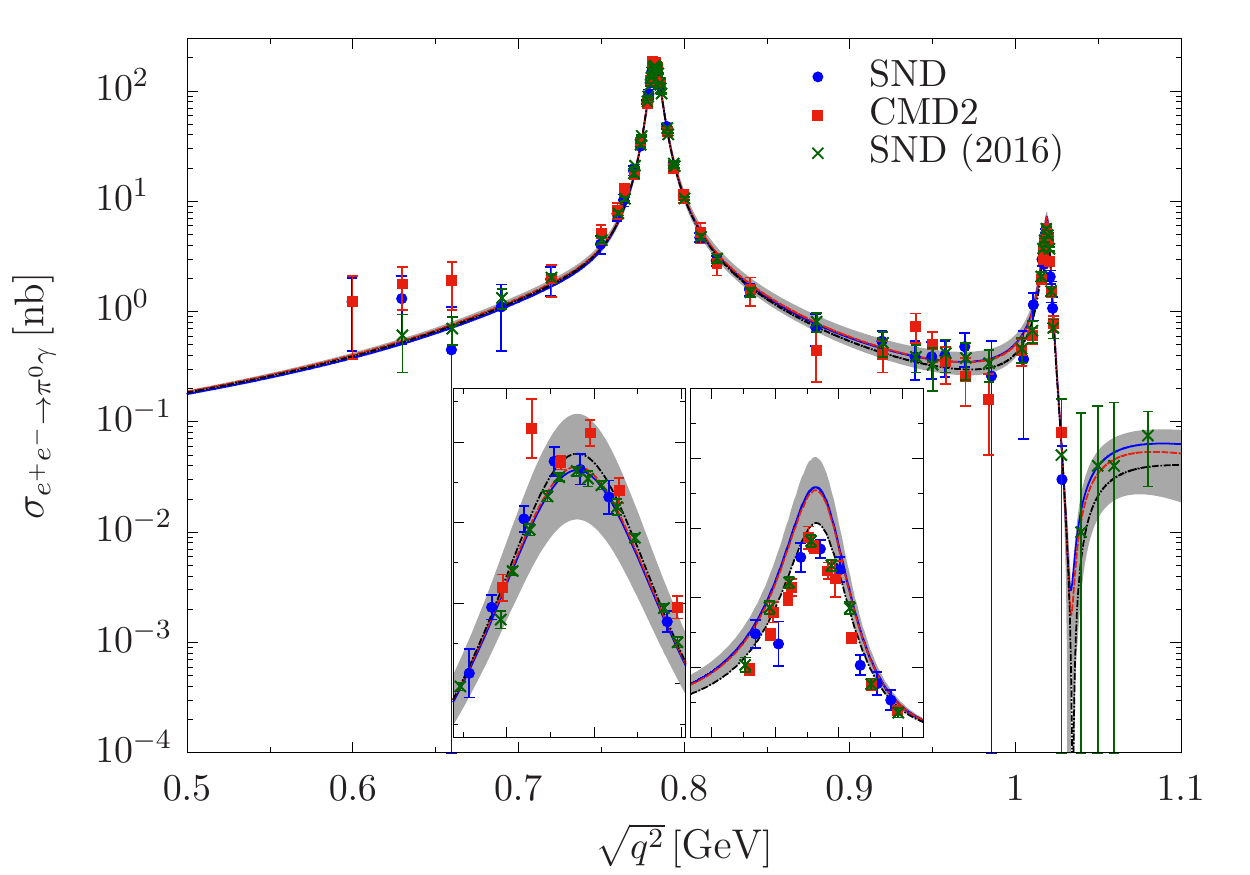}
    \caption{The $e^+e^-\to\pi^0\gamma$ cross section calculated from the once-subtracted TFF (blue solid line), the unsubtracted TFF (red dashed line), and~\cite{Hoferichter:2014vra} (black dot-dashed line), compared to the data of SND~\cite{Achasov:2000zd,Achasov:2003ed}, CMD2~\cite{Akhmetshin:2004gw}, and SND (2016)~\cite{Achasov:2016bfr}. The inserts show the same plot around the $\omega$ and $\phi$ peaks, respectively. The gray band indicates our uncertainty estimate for the unsubtracted TFF.} 
	\label{fig:pi0gamma}
\end{figure}

The resulting $e^+e^-\to\pi^0\gamma$ cross section predicted from the once-subtracted and the unsubtracted TFFs based on the new parameterization of $a(q^2)$ are compared to the previous analysis~\cite{Hoferichter:2014vra} in Fig.~\ref{fig:pi0gamma}. In addition to the $e^+e^-\to\pi^0\gamma$ cross section measurements~\cite{Achasov:2000zd,Achasov:2003ed,Akhmetshin:2004gw} already included in~\cite{Hoferichter:2014vra}, we also take into account the most accurate new data determined from the full data sample of the SND experiment~\cite{Achasov:2016bfr}. 
The mean values of our cross section are obtained averaging over the variations of the input quantities, $n$ from $3$--$6$ in the conformal polynomial of $a(q^2)$, 
and also the change of the integration cutoffs $\Lambda_{3\pi}$ and $\sqrt{\siv}$ in the range $(1.8$--$2.5)\GeV$.  
The band corresponding to the theoretical uncertainties $\sigma_\text{th}$, defined as the maximum deviations of all the variations from the average cross section, are only shown for the unsubtracted TFF in Fig.~\ref{fig:pi0gamma},
since otherwise the individual bands could hardly be differentiated.
These results are fully consistent with~\cite{Hoferichter:2014vra}, which is not immediately guaranteed for the unsubtracted version~\eqref{eq:un_TFF_tl} given that the effective
pole introduced to enforce the correct normalization implies a finite range of validity, the effects of which could potentially affect the low-energy region  in particular
for low masses $\Mrhop$. 

We further calculate the reduced $\chi^2$ corresponding to these results in the case of the different experimental data sets~\cite{Achasov:2000zd,Achasov:2003ed,Akhmetshin:2004gw,Achasov:2016bfr} for a more quantitative assessment of our description. The reduced $\chi^2/\text{dof}$ calculated below $1\GeV$ and $1.1\GeV$ is shown in Table~\ref{tab:pigamma}, together with a modified variant
\beq
\label{eq:modchisq}
\tilde\chi^2=\sum_{i=1}^N\frac{\big(y_i-y_\text{th}(q_i)\big)^2}{\sigma_i^2+\sigma_\text{th}^2(q_i)},
\eeq
where $q_i=\sqrt{q_i^2}$ and the difference between experiment and theory $y_i-y_\text{th}(q_i)$ is weighted by the combined uncertainty $\sqrt{\sigma_i^2+\sigma_\text{th}^2(q_i)}$. 
We observe very good agreement between the once- and unsubtracted TFFs, while, as expected, differences to~\cite{Hoferichter:2014vra} arise from the new parameterization of $a(q^2)$.
Below $1\GeV$, the $\chi^2$ deteriorates for the previously studied data sets from SND~\cite{Achasov:2000zd,Achasov:2003ed} and CMD2~\cite{Akhmetshin:2004gw}, 
but for the new SND data~\cite{Achasov:2016bfr}
the situation is reversed, here the new $a(q^2)$ leads to a better description. The difference can be traced back largely to the $\omega$ peak, see insert in Fig.~\ref{fig:pi0gamma}, 
where now the strength of the resonance is predicted almost perfectly, both for the subtracted and unsubtracted variants. In fact, the slight difference in the $\chi^2$ originates almost exclusively
from data outside the $\omega$ region.

\begin{table}[t]
	\centering
	\renewcommand{\arraystretch}{1.3}
	\begin{tabular}{lccccccc}
        \toprule 
		 &&SND & CMD2 & SND (2016)  \\
		\midrule	
		\multirow{2}{*}{once-subtracted TFF} 	&$\chi^2/\text{dof}$& $1.16$ [$2.76$] & $2.64$ [$12.7$] & $1.91$ [$4.73$]\\
		&$\tilde\chi^2/\text{dof}$& $0.43$ [$0.73$] & $1.10$ [$1.85$] & $0.42$ [$0.68$]\\	\midrule
		\multirow{2}{*}{unsubtracted TFF} &$\chi^2/\text{dof}$& $1.07$ [$2.51$] & $2.34$ [$11.5$] & $1.51$ [$4.04$]  \\		
		&$\tilde\chi^2/\text{dof}$& $0.36$ [$0.62$] & $0.95$ [$1.45$] & $0.29$ [$0.50$] \\ 	\midrule
		\multirow{2}{*}{HKLNS14} &$\chi^2/\text{dof}$& $0.90$ [$1.08$] & $1.82$ [$3.35$] & $2.15$ [$2.01$] \\		
		&$\tilde\chi^2/\text{dof}$ & $0.54$ [$0.62$] & $1.18$ [$1.39$]  & $0.68$ [$0.65$] \\
	    \bottomrule
		\renewcommand{\arraystretch}{1.0}
	\end{tabular}
	\caption{Reduced $\chi^2$ and $\tilde\chi^2$ for the $e^+e^-\to\pi^0\gamma$ cross section determined from the once-subtracted and the unsubtracted TFFs and from~\cite{Hoferichter:2014vra} (HKLNS14), compared to SND~\cite{Achasov:2000zd,Achasov:2003ed}, CMD2~\cite{Akhmetshin:2004gw}, and SND (2016)~\cite{Achasov:2016bfr} below $1\GeV$ [below $1.1\GeV$].} 
	\label{tab:pigamma}
\end{table}

Including the $\phi$ region, i.e.\ all data below $1.1\GeV$, we find that the slight mismatch at the resonance peak already observed in~\cite{Hoferichter:2014vra} is compounded, and accordingly the $\chi^2$ deteriorates appreciably when extending the energy region beyond $1\GeV$. This indicates that, most likely, the inelastic effects in $a(q^2)$ fit to the $3\pi$ channel, including imaginary parts that open around the $K\bar K$ threshold, cannot describe the same energy region in the $e^+e^-\to\pi^0\gamma$ spectrum, reflecting the fact that these inelastic effects do not have to affect 
the $3\pi$ and $\pi^0\gamma$ channels in the same way.
Accordingly, the marked improvement in the $3\pi$ channel just above the $\phi$ resonance comes at the expense of a mismatch in $\pi^0\gamma$. Phrased differently, the coefficients in 
the conformal polynomial if fit to $e^+e^-\to\pi^0\gamma$ instead of $3\pi$ would change, likely restoring agreement in the $\phi$ region. In addition, 
a quantitative description above $1\GeV$ would at some point be distorted by  
the influence of the effective poles in the unsubtracted TFF~\eqref{eq:un_TFF_tl}, so that the once-subtracted variant would become more appropriate for that purpose.
While it is therefore not unexpected that the $\chi^2$ of the central values increases in the $\phi$ region, we remark that when including the uncertainty estimates, 
see $\tilde \chi^2$ in Table~\ref{tab:pigamma}, the description hardly deteriorates and in the case of the new SND data and the unsubtracted TFF even slightly improves.
This demonstrates that the gradual breakdown of the predictive power of our formalism in the time-like region around the $\phi$ resonance is largely captured by our uncertainty estimates.

In this work, we are most interested in the space-like TFF as it enters in $(g-2)_\mu$, and the improved description of $3\pi$ was constructed in such a way as to better control the analytic continuation
to the space-like region. In principle, one could imagine fitting a similar representation of $a(q^2)$ to $e^+e^-\to \pi^0\gamma$ data alone and calculating the analytic continuation of the TFF based on the 
conformal parameters obtained in this fit. However, we conclude that the uncertainties in both the theoretical description and the data base are not competitive with a direct fit to $e^+e^-\to 3\pi$, 
which therefore provides the most reliable prediction of the space-like TFF. On the experimental side this conclusion is illustrated by the fact that the different data sets favor different theoretical predictions, see Table~\ref{tab:pigamma}, while on the theory side the complications become most apparent in the analytic continuation. For the application in $(g-2)_\mu$ the asymptotic behavior
requires an unsubtracted dispersion relation, but the effective pole would render precisely that variant unsuitable for a fit to the whole $e^+e^-\to\pi^0\gamma$ spectrum, as would be required for a reliable analytic continuation to the space-like region.

\subsection{Space-like form factor}
\label{subsec:spacelike}

\begin{figure}[t]
	\centering
	\includegraphics[width=0.8\linewidth]{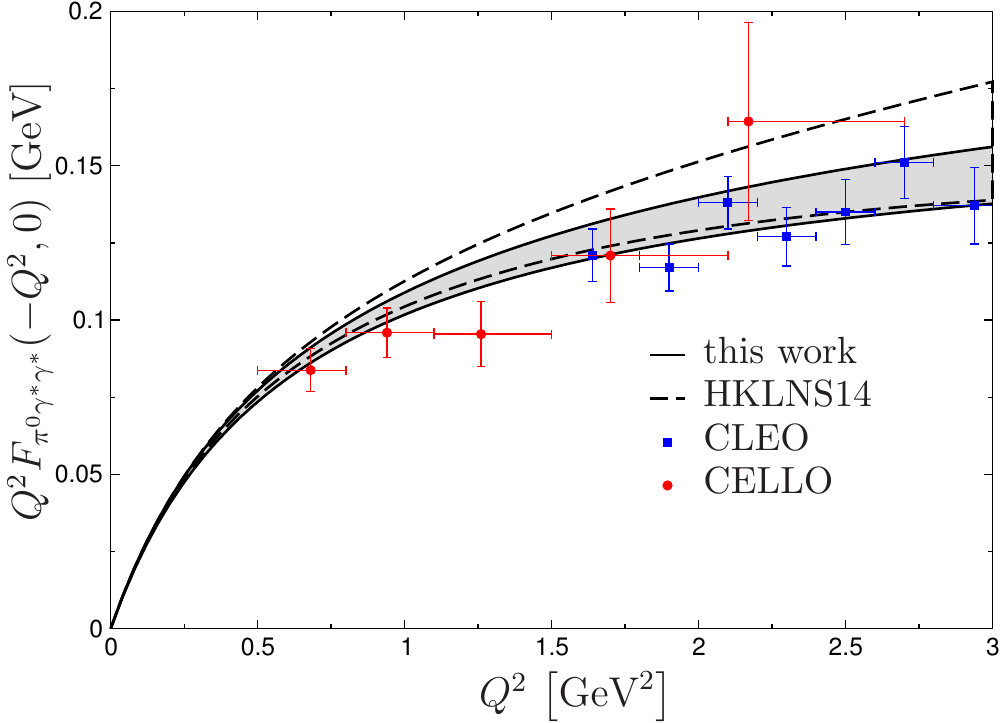}
	\caption{The singly-virtual form factors obtained in the current analysis (solid lines with gray uncertainty band) and from the once-subtracted representation~\cite{Hoferichter:2014vra}
	(HKLNS14, dashed lines) in the low-energy region, in comparison to CELLO~\cite{Behrend:1990sr} and CLEO~\cite{Gronberg:1997fj} data.}
	\label{fig:TFF_spacelike_low}
\end{figure}

After the discussion of the time-like TFF, we start the analysis of the space-like doubly-virtual TFF 
\beq
\label{eq:TFF_sl_final}
F_{\pi^0\gamma^*\gamma^*}(-Q_1^2,-Q_2^2)=F_{\pi^0\gamma^*\gamma^*}^\text{disp}(-Q_1^2,-Q_2^2)+F_{\pi^0\gamma^*\gamma^*}^\text{eff}(-Q_1^2,-Q_2^2)+F_{\pi^0\gamma^*\gamma^*}^\text{asym}(-Q_1^2,-Q_2^2)
\eeq
by first comparing our result for the singly-virtual TFF with the once-subtracted dispersive representation employed in~\cite{Hoferichter:2014vra},
\beq 
\label{eq:TFF_sl_os}
F_{\pi^0\gamma^*\gamma^*}(-Q^2,0) = F_{\pi\gamma\gamma} - \frac{Q^2}{\pi}\int_{s_\text{thr}}^\infty \diff s' 
\frac{\Im F_{\pi^0\gamma^*\gamma^*}(s',0)}{s'(s'+Q^2)}.
\eeq
For this purpose, the singly-virtual form factor at low energies up to $3\GeV^2$ is displayed in the form $Q^2F_{\pi^0\gamma^*\gamma^*}(-Q^2,0)$ as a function of $Q^2$ 
in Fig.~\ref{fig:TFF_spacelike_low}, together with the experimental data from CELLO~\cite{Behrend:1990sr} and CLEO~\cite{Gronberg:1997fj}, where the total uncertainties are obtained by adding the statistical and systematic errors in 
quadrature.\footnote{For the CELLO data, we directly take the uncertainties as given in~\cite{Behrend:1990sr} since systematic effects are not listed separately.} 
Our theoretical uncertainty of the singly-virtual form factor is estimated as the quadratic sum of the $\pm1.4\%$ $F_{\pi\gamma\gamma}$ normalization uncertainty varying $\grhop$, the dispersive uncertainty,
and the $^{+20}_{-10}\%$ BL uncertainty varying $\Mrhop$. Here, the dispersive error is defined as the maximum deviation from the central result found for different phase shifts and different pion vector form factors described in Sect.~\ref{subsec:def_LEP}, $n$ ranging from $3$--$6$ in the fit of $a(q^2)$ to the $e^+e^-\to3\pi$ cross section, and varying the integration cutoffs $\Lambda_{3\pi}$ and $\sqrt{\siv}$ between $(1.8$--$2.5)\GeV$. 
The resulting form factor depicted in solid lines is consistent with the available data and is close to the result obtained from the once-subtracted representation~\eqref{eq:TFF_sl_os} in dashed lines at low energies below $1\GeV^2$.  
At larger momenta the curves start to deviate, which is exactly expected from the matching of our representation to the correct high-energy behavior: the once-subtracted representation tends 
to show a linear behavior in the plot, whereas the unsubtracted form factor slowly converges to the BL limit.

\begin{figure}[t]
	\centering
	\includegraphics[width=0.8\linewidth]{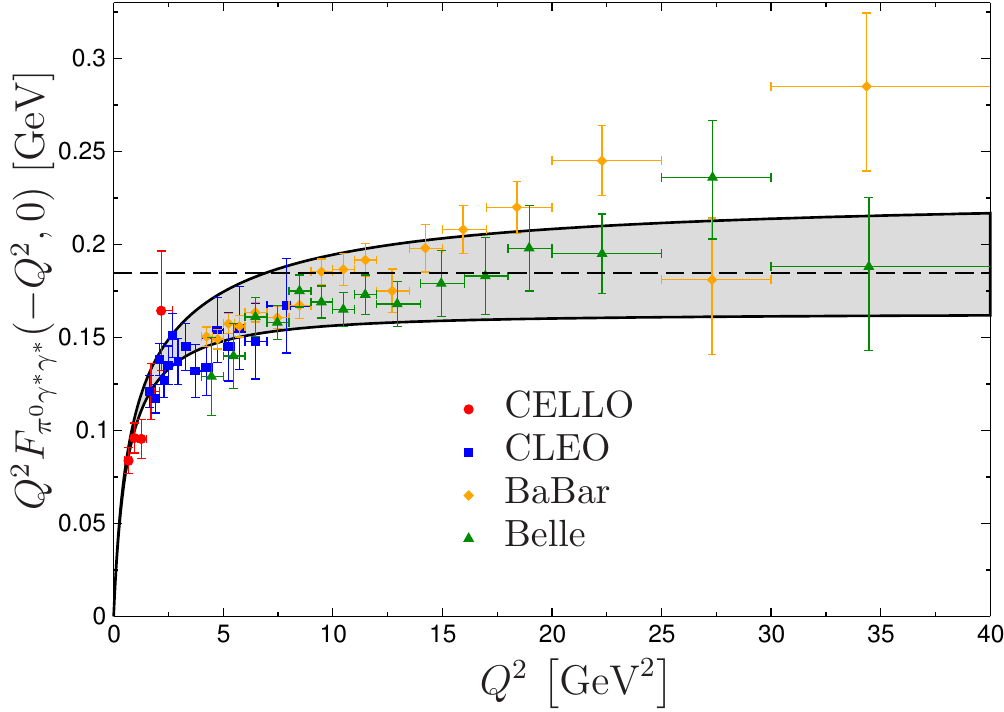}
	\caption{The singly-virtual form factor $Q^2F_{\pi^0\gamma^*\gamma^*}(-Q^2,0)$ as a function of $Q^2$, in comparison to the experimental data~\cite{Behrend:1990sr,Gronberg:1997fj,Aubert:2009mc,Uehara:2012ag}. The dashed horizontal line indicates the BL limit.}
	\label{fig:TFF_spacelike_asym}
\end{figure}

Next, we update the low-energy parameters characterizing the singly-virtual TFF, most notably its radius
\begin{align}
\label{eq:slope}
a_\pi&=\frac{\mpii^2}{F_{\pi\gamma\gamma}}\frac{\partial}{\partial q^2}F_{\pi^0\gamma^*\gamma^*}(q^2,0)\bigg|_{q^2=0}\notag\\
&=31.5(2)_{F_{\pi\gamma\gamma}}(8)_\text{disp}(3)_\text{BL}\times 10^{-3}=31.5(9)\times 10^{-3}.
\end{align}
The increased value compared to $a_\pi=30.7(6)\times 10^{-3}$~\cite{Hoferichter:2014vra} traces back to the matching to the asymptotic behavior and corresponds to the fact that our form factor is slightly smaller than the once-subtracted TFF~\eqref{eq:TFF_sl_os} as show in Fig.~\ref{fig:TFF_spacelike_low}. While fully consistent within uncertainties, the central value thus moves closer to the one
derived from Pad\'e approximants~\cite{Masjuan:2012wy}, $a_\pi=32.4(2.2)\times 10^{-3}$, and also to the current experimental average $a_\pi^{\text{exp}}=33.5(3.1)\times 10^{-3}$~\cite{Olive:2016xmw}, 
which is dominated by extractions from the Dalitz decay $\pi^0 \to e^+e^-\gamma$~\cite{TheNA62:2016fhr} (compare also~\cite{Adlarson:2016ykr}) and the space-like CELLO data~\cite{Behrend:1990sr}.
The dispersive approach continues to provide the most precise determination, due to the fact that other extractions are limited either by poor space-like data or the small kinematic region accessible in the Dalitz decay.

\begin{figure}[t]
	\centering
	\includegraphics[width=0.9\linewidth]{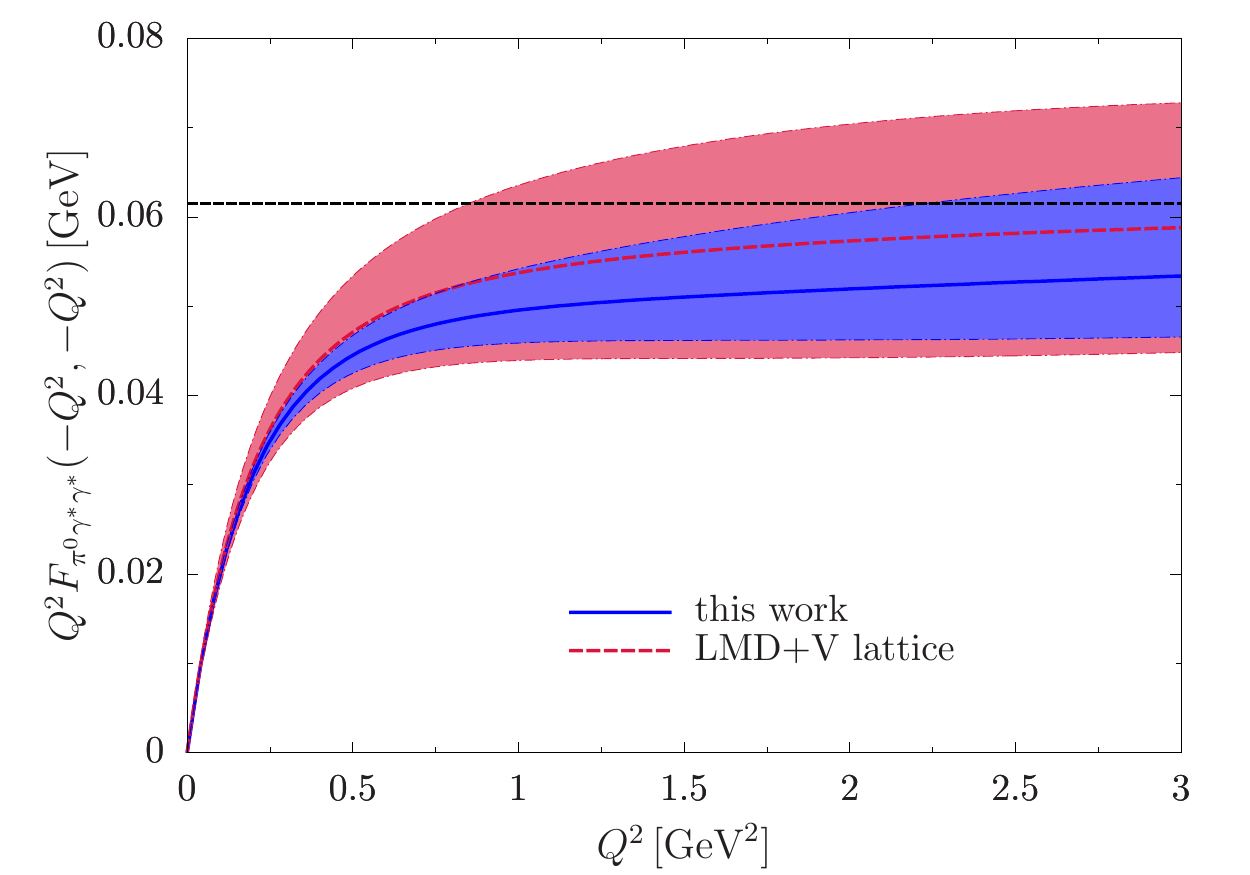}
	\caption{The diagonal form factor $Q^2F_{\pi^0\gamma^*\gamma^*}(-Q^2,-Q^2)$ versus $Q^2$ at low energies (blue solid line with uncertainty band), compared to the LMD+V model fit to the lattice data~\cite{Gerardin:2016cqj} (red dashed line with uncertainty band). The black dashed line shows the OPE limit.}
	\label{fig:TFF_diag_low_energy}
\end{figure} 

The next coefficient in the expansion around $q^2=0$ is evaluated as
\begin{align}
\label{eq:slope2}
b_\pi&=\frac{\mpii^4}{F_{\pi\gamma\gamma}}\frac{1}{2}\frac{\partial^2}{\partial (q^2)^2}F_{\pi^0\gamma^*\gamma^*}(q^2,0)\bigg|_{q^2=0}\notag\\
&=1.14(1)_{F_{\pi\gamma\gamma}}(4)_\text{disp}(1)_\text{BL}\times 10^{-3}=1.14(4)\times 10^{-3},
\end{align}
where the overall uncertainty is entirely dominated by the dispersive one as expected for a low-energy parameter. The larger dispersive uncertainty compared to the result $1.10(2)\times10^{-3}$ obtained in~\cite{Hoferichter:2014vra} partially originates from the fact that the uncertainty from the fits to the $e^+ e^- \to 3\pi$ cross section using different variants of the conformal polynomials in the parameterization~\eqref{eq:a-par} is included in the dispersive one. However, the total uncertainty is still appreciably smaller e.g.\ compared to $1.06(26)\times 10^{-3}$ from~\cite{Masjuan:2012wy}. 

\begin{figure}[t]
	\centering
    \includegraphics[width=0.9\linewidth]{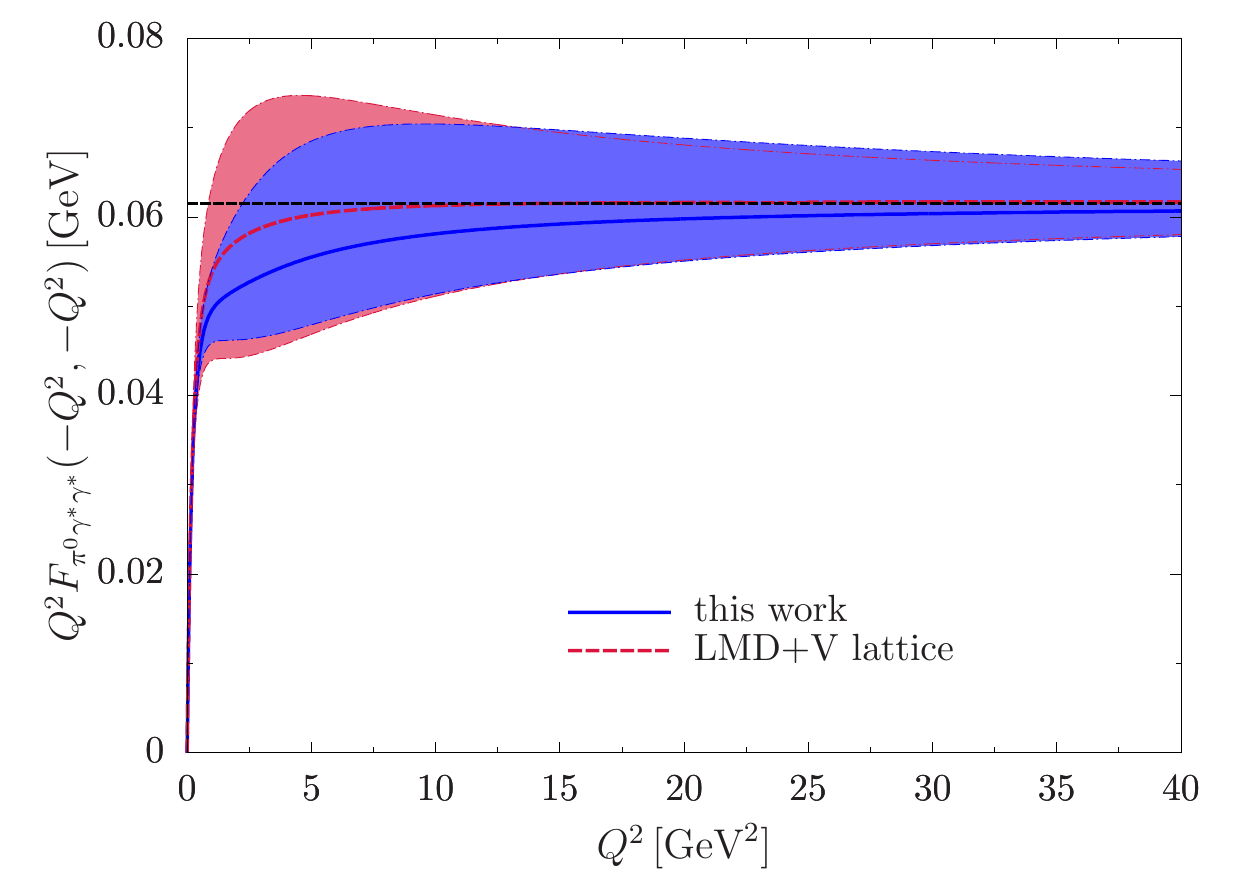}
	\caption{The diagonal form factor $Q^2F_{\pi^0\gamma^*\gamma^*}(-Q^2,-Q^2)$ (blue solid line with uncertainty band), in comparison to the LMD+V model fit to the lattice data~\cite{Gerardin:2016cqj} (red dashed line with uncertainty band). The OPE limit of the form factor is indicated by the black dashed line.}
	\label{fig:TFF_diag_asym}
\end{figure}

The asymptotic behavior of the singly-virtual TFF $Q^2F_{\pi^0\gamma^*\gamma^*}(-Q^2,0)$ at higher energies is shown in Fig.~\ref{fig:TFF_spacelike_asym}, along with the BaBar and Belle measurements~\cite{Aubert:2009mc,Uehara:2012ag} and the CELLO and CLEO data~\cite{Behrend:1990sr,Gronberg:1997fj} already included in Fig.~\ref{fig:TFF_spacelike_low}.\footnote{We include the $Q^2$-independent error components of the systematic errors into the total uncertainties of BaBar and Belle~\cite{Aubert:2009mc,Uehara:2012ag}.} 
We find that the central value of our result almost matches the BL prediction, slowly approaching this limit from below. 
Although even fits including the BaBar data and using an energy threshold of $10\GeV^2$ do not fully capture the rapid rise suggested by the BaBar data, our error band does cover all reasonably conceivable fit variants, see Sect.~\ref{subsec:sv_data}, which implies that the statistical significance of the last few BaBar data points does not suffice to drastically alter the fit results.

As the next step, we calculate the diagonal TFF $F_{\pi^0\gamma^*\gamma^*}(-Q^2,-Q^2)$ as another representative result for the doubly-virtual form factor. In the dispersive approach, the doubly-virtual diagonal form factor is completely determined by the singly-virtual inputs by virtue of its isospin structure. 
In particular, analyticity guarantees that the space-like form factor has to be a smooth function when matching to pQCD,  even though it receives contributions from three different terms in~\eqref{eq:TFF_sl_final}, including the asymptotic contribution~\eqref{eq:asym}. The uncertainty in this asymptotic piece is estimated by varying the threshold parameter $\sm$ in the range $1.7(3)\GeV^2$, which ensures a smooth matching and coincides with the typical range found with LCSRs~\cite{Chernyak:1983ej,Khodjamirian:1997tk,Agaev:2010aq,Mikhailov:2016klg}, see Sect.~\ref{sec:scales}. 
It is then added quadratically to the other three sources of uncertainty already discussed in the context of the singly-virtual form factor. 

  \begin{figure}[t]
 	\centering
 	\includegraphics[width=0.9\linewidth]{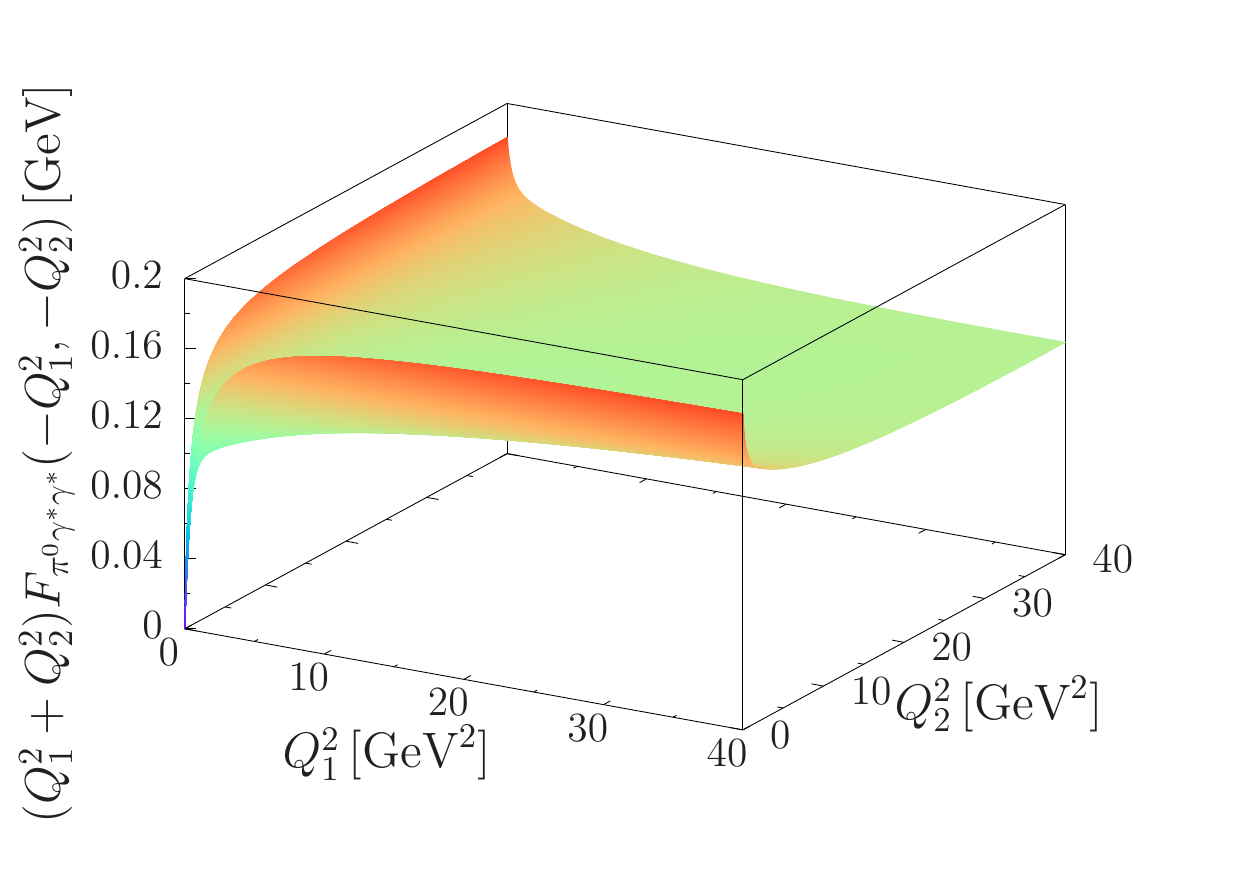}
 	\caption{Three-dimensional representation of $(Q_1^2+Q_2^2)F_{\pi^0\gamma^*\gamma^*}(-Q_1^2,-Q_2^2)$ as a function of the photon virtualities.}
 	\label{fig:TFF_dv}
 \end{figure} 

The asymptotic behavior of the diagonal form factor is known rigorously from the OPE, see~\eqref{eq:OPE_limit}. In the absence of experimental measurements, our result 
given in the form $Q^2F_{\pi^0\gamma^*\gamma^*}(-Q^2,-Q^2)$ in Fig.~\ref{fig:TFF_diag_low_energy} is compared to an LMD+V (lowest meson dominance + vector~\cite{Knecht:2001xc}) resonance model fit to lattice data extrapolated to the 
physical pion mass~\cite{Gerardin:2016cqj}. We find a slightly smaller diagonal form factor compared to the LMD+V model fit to lattice, otherwise observe consistency within the uncertainty bands.
Similarly, the results for $Q^2F_{\pi^0\gamma^*\gamma^*}(-Q^2,-Q^2)$  from our dispersive calculation and the lattice calculation of the TFF~\cite{Gerardin:2016cqj} at high energies up to $40\GeV^2$ are shown in Fig.~\ref{fig:TFF_diag_asym}, again in agreement within uncertainties.  
Our central value approaches the OPE limit from below, which indicates a negative subleading $\Order(1/Q^4)$ contribution as obtained in~\cite{Novikov:1983jt}. The total uncertainty at low energy is largely dominated by the one from the normalization $F_{\pi\gamma\gamma}$, but the uncertainties from the BL limit and the asymptotic contribution start to compete at higher energies. Accordingly, the uncertainty bands of both analyses shrink to the central results at higher energies since they are suppressed as subleading terms in $\Order(1/Q^2)$ and both analyses are matched correctly to the leading OPE limit~\eqref{eq:OPE_limit}.

\begin{figure}
	\centering
	\begin{subfigure}{.5\textwidth}
		\centering
			\includegraphics[width=\linewidth]{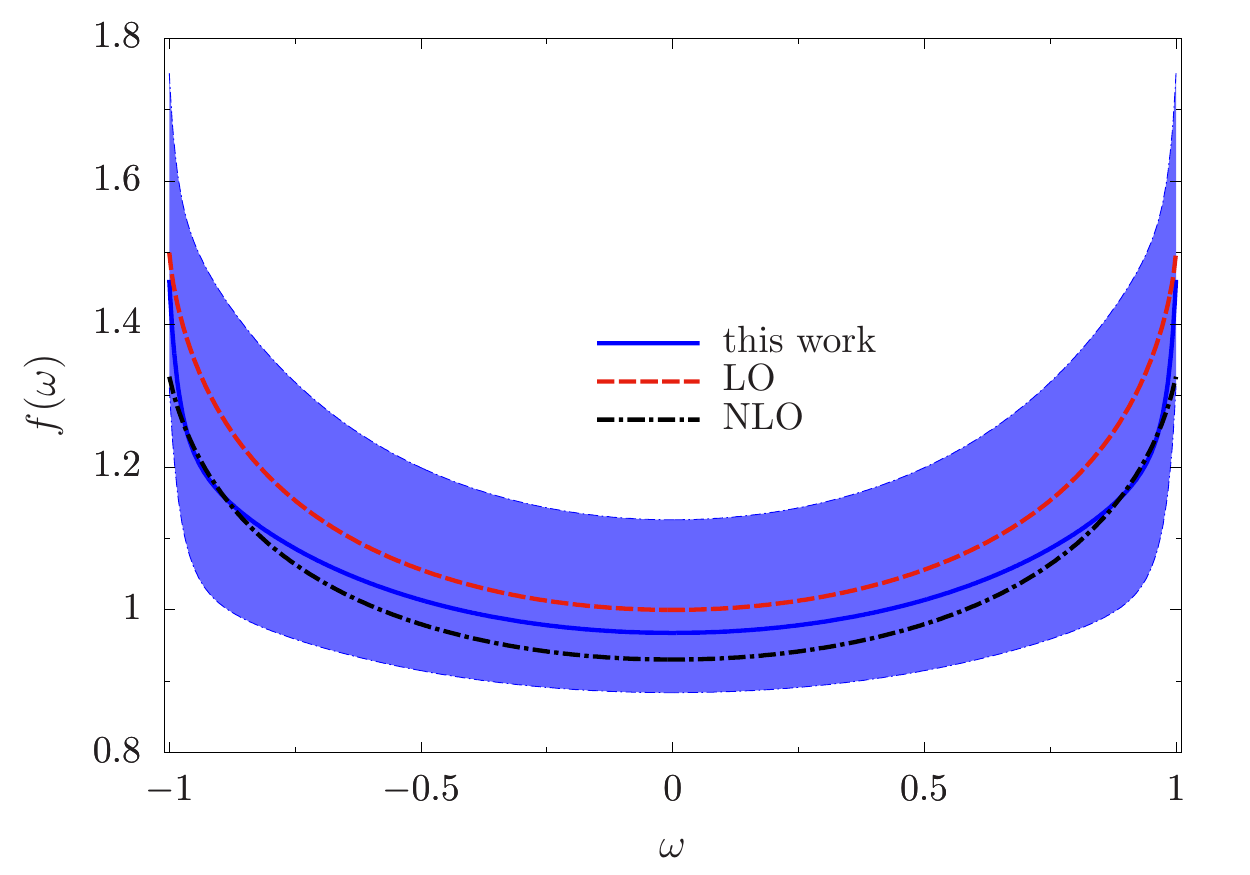}
	\end{subfigure}%
	\begin{subfigure}{.5\textwidth}
		\centering
			\includegraphics[width=\linewidth]{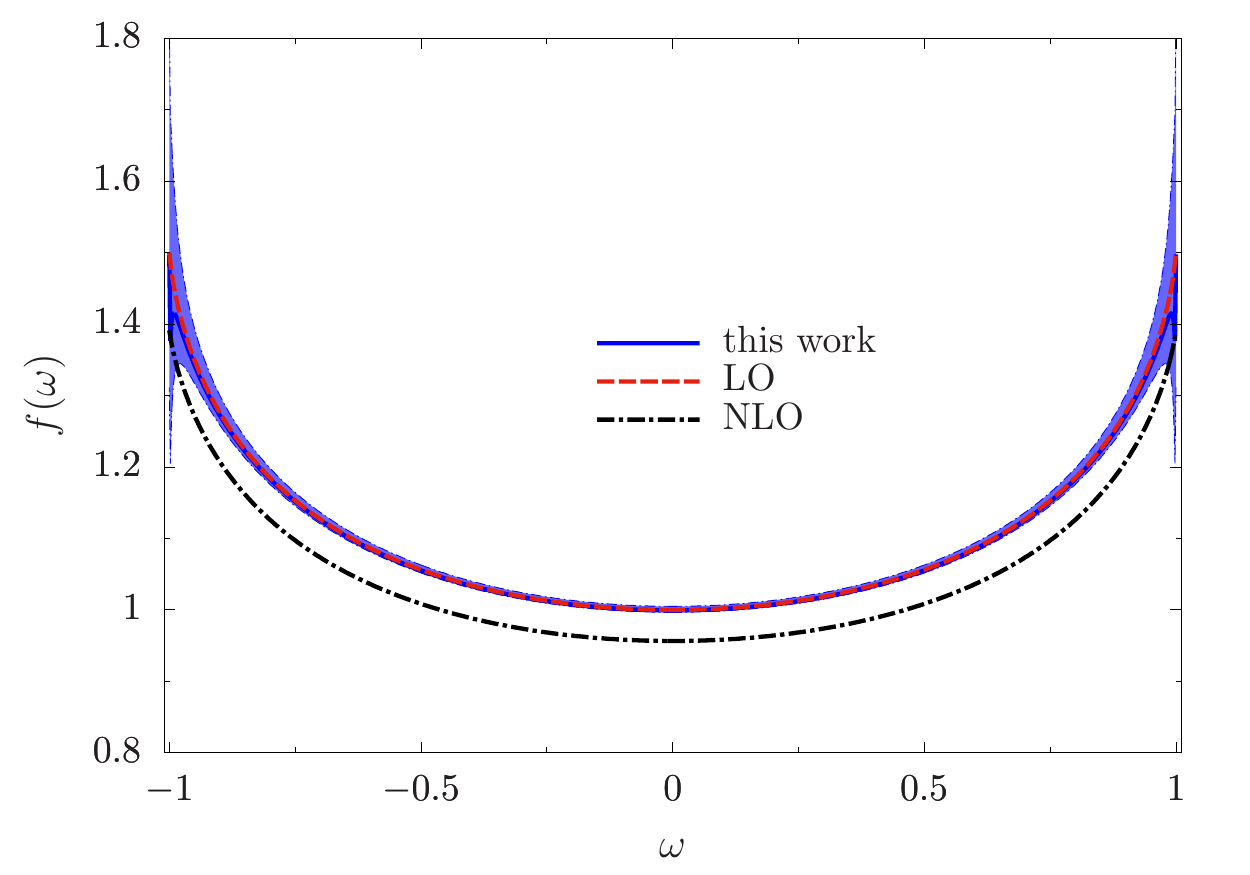}
	\end{subfigure}
	\caption{$f(\omega)$ obtained from the dispersive representation~\eqref{eq:TFF_sl_final} (blue solid line with uncertainty band) calculated at $Q_1^2+Q_2^2=35\GeV^2$ (left) and at $Q_1^2+Q_2^2=1.6\times 10^{3}\GeV^2$ (right), in comparison to $f(\omega)$ from the LO and NLO asymptotic pion distribution amplitudes $\phi_\pi(u)=6u(1-u)$ (red dashed line) and~\eqref{F_alpha_s} (black dot-dashed line).}
	\label{fig:f_omega}
\end{figure}
 
So far, we have shown the TFF for two special space-like kinematics to demonstrate consistency with experiment and lattice, respectively. 
However, the analysis is not complete since for $(g-2)_\mu$ we need the TFF as a function of two general photon virtualities. 
The full result, presented in the form $(Q_1^2+Q_2^2)F_{\pi^0\gamma^*\gamma^*}(-Q_1^2,-Q_2^2)$ as a function of $Q_1^2$ and $Q_2^2$, is depicted in Fig.~\ref{fig:TFF_dv}. 
The virtualities $Q_1^2$ and $Q_2^2$ cover broad ranges from low-energy to asymptotic regions of interest. The smooth transition and the correct high-energy behavior of the 
form factor in the entire kinematic domain are dictated by the analyticity of the form factor and the proper pQCD matching.\footnote{A data file containing the doubly-virtual TFF for space-like
kinematics, including the individual as well as the combined
uncertainties, is attached as ancillary material to this article.}

Finally, we compare the high-energy behavior of our dispersive representation~\eqref{eq:TFF_sl_final} to the predictions of the asymptotic behavior from pQCD by analyzing the function $f(\omega)$ 
defined in~\eqref{eq:pQCD_2} and~\eqref{eq:f_omega}. Its value encodes the asymptotic behavior of the TFF for arbitrary virtualities $Q_1^2$ and $Q_2^2$. $f(\omega)$ at the energy scale chosen as the highest accessible energy of the BaBar and Belle experiments~\cite{Aubert:2009mc,Uehara:2012ag} is illustrated in the left diagram of Fig.~\ref{fig:f_omega}. 
At these virtualities, our uncertainty band should safely  cover most of the modifications to $f(\omega)$ from higher terms in the Gegenbauer-polynomial expansion as well as other proposed modifications
of the pion distribution amplitude, as an example we show the $\alpha_s$ corrections~\eqref{F_alpha_s}. 
At very high energies, by construction, $f(\omega)$ obtained from the dispersive representation 
is nearly identical to the one obtained from the LO asymptotic pion distribution amplitude $\phi_\pi(u)=6u(1-u)$, therefore ensuring the correct high-energy behavior of the form factor (see right diagram). In this case, since $\alpha_s$ vanishes only logarithmically, the NLO curve is not covered anymore by our uncertainty band, but such large virtualities are irrelevant 
for the $(g-2)_\mu$ integral. Moreover, a complete NLO matching would actually be disadvantageous, given that, by chance, for the relevant energy range our central curve, although matched to the LO amplitude asymptotically, comes out closer to the NLO prediction (see left diagram).
We stress that Fig.~\ref{fig:f_omega} merely demonstrates to which extent the TFF has approached the pQCD limit for a particular choice of photon virtualities, it does not provide additional insights into the pion distribution amplitude beyond its asymptotic form.

\subsection[Consequences for $a_\mu$]{Consequences for $\boldsymbol{a_\mu}$}
\label{subsec:g-2}

We now turn to the main application of the detailed analysis of the space-like doubly-virtual TFF presented in the preceding section, 
the pion-pole contribution to $a_\mu$. Evaluating
the loop integrals in its definition~\eqref{eq:pion_pole} 
by means of the TFF representation~\eqref{eq:TFF_sl_final}, the final result reads
\begin{align}
	\label{eq:result_final}
	a_\mu^{\pi^0\text{-pole}}&=62.6(1.7)_{F_{\pi\gamma\gamma}}(1.1)_\text{disp}(^{2.2}_{1.4})_\text{BL}(0.5)_\text{asym}\times 10^{-11}\notag\\
	&=62.6^{+3.0}_{-2.5}\times 10^{-11}.
\end{align}
Here, the uncertainties from the numerical integration are negligible, in fact, we used both the standard variant~\eqref{eq:pion_pole} and a more symmetric parameterization of the integration region 
first suggested in~\cite{Eichmann:2015nra} and subsequently implemented in~\cite{Colangelo:2017qdm,Colangelo:2017fiz}. All uncertainties therefore derive from the TFF, with individual contributions estimated in close analogy to the previous sections. First, the central value is defined by the average over all variants of the dispersive formalism, i.e.\ $\pi\pi$ phase shifts, cutoff parameters, parameterizations of the pion vector form factor, and the conformal polynomial, with the uncertainty defined as the maximum deviation from this average. The normalization uncertainty 
then reflects the PrimEx result for the $\pi^0$ decay width corresponding to an uncertainty of $1.4\%$ in $F_{\pi\gamma\gamma}$, the BL error the uncertainty band from Fig.~\ref{fig:TFF_spacelike_asym}, and
the asymptotic error the impact of the variation of $\sm$ in~\eqref{eq:asym} according to $\sm=1.7(3)\GeV^2$. The quadratic sum of the four different sources of uncertainty defines our final estimate.
Note that while this strategy is completely analogous to the corresponding error estimates discussed before for the time- and space-like TFF, due to the fact that 
the TFF enters squared in the integral it is critical to perform this error estimate for each source individually at the level of $a_\mu$, using the total error band of the TFF instead would assume 
fully-correlated uncertainties and thereby overestimate the final error.

The decomposition~\eqref{eq:result_final} further suggests opportunities for future cross checks and improvements. First, the PrimEx-II measurement is expected to reduce the uncertainty in $F_{\pi\gamma\gamma}$
to $0.85\%$~\cite{Gasparian:2016oyl,Larin:2018}, which translates to normalization and total uncertainties of $1.0$ and $^{+2.7}_{-2.1}\times 10^{-11}$ in $a_\mu$, respectively.
Next, the dispersive uncertainties in particular in the low-energy space-like TFF could be cross-checked and potentially improved by upcoming data from BESIII~\cite{Redmer:2018gah},
while the $F_{3\pi}$ low-energy theorem, used to normalize $a(q^2)$ in~\eqref{eq:alphaF3pi}, is currently under study at COMPASS~\cite{Steffen:2018}.
A conclusive measurement of the asymptotic singly-virtual TFF at Belle~II~\cite{Abe:2010gxa,Kou:2018nap} would eliminate the systematic uncertainties from tensions between BaBar and Belle as well as the BL limit. 
In fact, simply taking the central fit to the full data base with $1\sigma$ uncertainties would formally reduce the BL error to $0.2\times 10^{-11}$ (with a central value of $63.1\times 10^{-11}$), which emphasizes the fact that our result, at the level of accuracy quoted in~\eqref{eq:result_final}, is insensitive to the tensions in the asymptotic behavior.  
Strictly speaking, all singly-virtual data on the space-like pion TFF~\cite{Behrend:1990sr,Gronberg:1997fj,Aubert:2009mc,Uehara:2012ag} result from doubly-virtual measurements extrapolated to the point where one photon is on-shell. With our doubly-virtual TFF~\eqref{eq:TFF_sl_final} at hand, agreement with data could be checked directly 
or our TFF could be used for the extrapolation. Also for this purpose, the values for radius~\eqref{eq:slope} and curvature~\eqref{eq:slope2} might prove useful.
Finally, absent doubly-virtual data it is not possible to reduce the pQCD uncertainties directly, but input from lattice QCD would allow one to further scrutinize this contribution. 

\begin{table}[t]
	\centering
	\renewcommand{\arraystretch}{1.3}
	\begin{tabular}{lcccr}
        \toprule 
        method & $a_\mu^{\pi^0\text{-pole}}$ & $a_\mu^{\pi^0\text{-``exchange''}}$ & $a_\mu^{\pi^0\text{-``const''}}$ & reference \\\midrule
        NJL model & & $81.8(16.5)$ & & \cite{Bartos:2001pg}\\
        LMD+V & & $72(12)$ & & \cite{Nyffeler:2009tw}\\
        holographic model & & $65.4(2.5)$ & & \cite{Cappiello:2010uy}\\
        Dyson--Schwinger equations & & $57.5(6.9)$ & & \cite{Goecke:2010if}\\
        nonlocal chiral quark model & & $50.1(3.7)$ & & \cite{Dorokhov:2011zf}\\
        resonance chiral theory & & $65.8(1.2)$ & & \cite{Kampf:2011ty}\\
        constituent chiral quark model & & $68(3)$ & & \cite{Greynat:2012ww}\\
        resonance chiral theory & & $66.6(2.1)$ & & \cite{Roig:2014uja}\\\midrule
        LMD+V & & & $78(10)$ & \cite{Melnikov:2003xd}\\\midrule
        ENJL, VMD & $59(9)$ & & & \cite{Bijnens:1995xf,Bijnens:2001cq}\\
        VMD & $57(6)$ & & & \cite{Hayakawa:1997rq}\\
        LMD+V & $58(10)$ & & & \cite{Knecht:2001qf}\\
        lattice QCD, LMD+V fit & $65.0(8.3)$ & & & \cite{Gerardin:2016cqj}\\
        rational approximants & $63.6(2.7)$ & & & \cite{Masjuan:2017tvw}\\
        resonance chiral theory & $58.1(9)$ & & & \cite{Guevara:2018rhj}\\
        dispersion relations & $62.6^{+3.0}_{-2.5}$ & & & this work, \cite{Hoferichter:2018dmo}\\
	 \bottomrule
	\renewcommand{\arraystretch}{1.0}
	\end{tabular}
	\caption{Comparison to previous results for $a_\mu^{\pi^0\text{-pole}}$. 
The uncertainties are reproduced as given in the respective publication, see main text for further discussion. For completeness, we also list works that calculate contributions involving an off-shell pion instead ($\pi^0$-``exchange'') or put one of the form factors to a constant ($\pi^0$-``const''), but stress that these results either 
	depend on the interpolator of the pion field or do not correspond to the dispersively defined pion pole, respectively, and therefore cannot be compared with the on-shell pion-pole contribution.} 
	\label{tab:amu}
\end{table}

Our central result~\eqref{eq:result_final} is compared to previous calculations in Table~\ref{tab:amu}. For completeness, we have also provided references that consider an off-shell 
pion-exchange contribution, but emphasize that these results are model-dependent, corresponding to a particular choice of the interpolating field. The wide spread among these results is therefore
not surprising given that, in general, each model will represent a different such choice. Similarly, a model involving a constant TFF at the singly-virtual vertex in HLbL scattering~\cite{Melnikov:2003xd}
disagrees with the dispersive definition of the pion-pole contribution, so that the resulting number cannot be compared to ours either.

In the end, our central value is remarkably close to early estimates using hadronic models~\cite{Bijnens:1995xf,Hayakawa:1997rq,Knecht:2001qf}, either VMD, LMD+V, or the extended Nambu--Jona-Lasinio model, and falls within the quoted model errors that had been typically estimated at the level of $15\%$. Recent updates in resonance chiral theory~\cite{Guevara:2018rhj} find similar values, however, without an attempt
to quantify the model uncertainty. Our central value is even closer to a calculation of the pion pole using a TFF constructed from rational approximants, with parameters determined from $\pi^0\to\gamma\gamma$ and space-like singly-virtual data~\cite{Masjuan:2017tvw}. The quoted error contains the propagated uncertainties from the data input and estimates of the systematics of the approach by comparing different approximants and varying a parameter that describes doubly-virtual kinematics within a certain range.
In this respect, the main advantages of the dispersive approach concern the fact that also data from the time-like region can be used, as illustrated by the key role of the $e^+e^-\to 3\pi$ data in our analysis; that the sensitivity to the space-like input is significantly reduced in comparison, removing the systematic uncertainty from the asymptotic behavior of the TFF; and that the doubly-virtual dependence is actually predicted within the formalism, eliminating the need for an extrapolation of the singly-virtual input to doubly-virtual kinematics. Further, we have provided an economical way to implement all short-distance constraints, which is not straightforward to achieve in hadronic models, e.g.\ the LMD+V model fails to produce the correct asymptotics for small but finite $q_1^2$ and $q_2^2\to\infty$.
Finally, our result also agrees with a calculation in lattice QCD~\cite{Gerardin:2016cqj}. Currently, an LMD+V ansatz is required to extend the lattice data to the full range of virtualities to perform the $(g-2)_\mu$ integral, but future updates at higher statistics are set to provide a sufficiently fine grid to enable a direct comparison to~\eqref{eq:result_final} in a fully model-independent way.

\section{Conclusions and outlook}
\label{sec:sum}

In this work we presented a comprehensive dispersive reconstruction of the doubly-virtual pion TFF, which determines the residue of the pion-pole contribution to $a_\mu$. 
As a first step, dispersion relations for the pion TFF were derived based on its isospin structure and unitarity relation, wherein the $2\pi$ and $3\pi$ intermediate states 
define the low-lying singularities in the isovector and isoscalar virtualities, respectively. 
As a consequence, the doubly-virtual pion TFF was reconstructed in light of the low-energy theorems for $F_{\pi\gamma\gamma}$ and $F_{3\pi}$, 
the $\pi\pi$ $P$-wave phase shifts from Roy- and Roy-like equations, and experimental input from $e^+e^-\to2\pi,3\pi$. 
Extending previous work, we achieved an improved description of the $e^+e^-\to3\pi$ cross section data after introducing a conformal polynomial to take into account the inelastic effects in the $3\pi$ channel. 
Starting from the unsubtracted dispersion relation~\eqref{eq:un_DR}, the double-spectral representation~\eqref{eq:dsr_disp} was derived afterwards as a convenient 
representation for the evaluation of the pion-pole $(g-2)_\mu$ loop integrals.

Another key advance in this work concerns the consistent matching to constraints from pQCD. To this end,
the LO leading-twist light-cone expansion~\eqref{eq:pQCD} was reformulated in terms of an asymptotic double-spectral density, which leads to an asymptotic contribution~\eqref{eq:asym} 
governing the correct high-energy behavior of the TFF for non-vanishing virtualities. We evaluated the known $\alpha_s$ corrections but found them to be negligible within uncertainties. 
As the final step, we introduced an effective pole term to remedy the normalization of the form factor and account for constraints from space-like singly-virtual data measured in $e^+e^-\to e^+e^-\pi^0$. 
The validity of the dispersive approach was cross-checked by comparing the dispersive prediction for $e^+e^-\to \pi^0\gamma$ based on the time-like singly-virtual TFF 
to cross section data. We found good agreement up to $1\GeV$, with deviations starting to appear in the vicinity of the $\phi$ resonance, right where the phase space 
for inelastic contributions in the $e^+e^-\to 3\pi$ fit was assumed to open.    
We studied the resulting space-like TFF~\eqref{eq:TFF_sl_final} extensively both for singly- and doubly-virtual kinematics, 
in comparison to experimental data, lattice-QCD calculations, and theoretical predictions from pQCD. 

This detailed study of the pion TFF, incorporating all the low-lying singularities and the correct high-energy behavior at $\Order(1/Q^2)$, culminates 
in the first dispersive determination of the pion-pole contribution to the muon $(g-2)_\mu$~\eqref{eq:result_final}, the lowest intermediate state
in a dispersive approach to HLbL scattering.
Our data-driven evaluation produces a central value in line with previous model-dependent estimates, but provides for the first time a determination that fully exploits the constraints
from the fundamental principles of analyticity, unitarity, and crossing symmetry as well as the predictions from pQCD in deriving well-controlled uncertainty estimates.
In fact, despite being already sufficient for a SM prediction of $a_\mu$ at the level of the upcoming experiments,
these uncertainties can be reduced further by virtue of future more precise singly-virtual measurements both in low- and high-energy regimes.

As the largest individual piece, our determination of the pion-pole contribution to $a_\mu$ is a critical step towards a complete data-driven evaluation of HLbL scattering~\cite{Hoferichter:2013ama,Colangelo:2014dfa,Colangelo:2014pva,Colangelo:2015ama,Colangelo:2017qdm,Colangelo:2017fiz}. Moreover, the strategies developed here 
regarding the incorporation of high-energy constraints will facilitate similar studies of the $\eta$ and $\eta'$ TFFs~\cite{Stollenwerk:2011zz,Hanhart:2013vba,Kubis:2015sga,Xiao:2015uva,Kubis:2018bej},
thus paving the way towards a fully data-driven determination of all light pseudoscalar-meson-pole contributions to HLbL scattering in $(g-2)_\mu$.

\acknowledgments
We thank Johan Bijnens, Eric Braaten, Antoine G\'erardin, Tobias Isken, Andrzej Kup\'s\'c,
Andreas Nyffeler, Eduardo de Rafael, Stefan Ropertz, and Peter Stoffer for useful discussions. 
Financial support by
the DFG (CRC 110,
``Symmetries and the Emergence of Structure in QCD''),
the Bonn--Cologne Graduate School of Physics and Astronomy (BCGS),
and the DOE (Grant No.\ DE-FG02-00ER41132)
is gratefully acknowledged.

\appendix

\section{Integral kernels}
\label{app:IK}

The integral kernels $\hat T_1(q_1,q_2;p)$ and $\hat T_2(q_1,q_2;p)$ for~\eqref{eq:tw} read: 
\begin{align}
\hat T_1(q_1,q_2;p) &= -\frac{16}{3} \left((q_1 \cdot q_2)^2-q_1^2 q_2^2\right) m_\mu^2-\frac{16}{3} q_1^2 (p \cdot q_2)^2\notag \\
&+p \cdot q_1\left(\frac{16}{3} p \cdot q_2 q_1 \cdot q_2-\frac{8}{3} q_2^2 q_1 \cdot q_2\right) +p \cdot q_2 \left(8 q_1^2 q_2^2-\frac{16}{3} (q_1 \cdot q_2)^2\right),\notag\\
\hat T_2(q_1,q_2;p) &= -\frac{8}{3} \left((q_1 \cdot q_2)^2-q_1^2 q_2^2\right) m_\mu^2-\frac{8}{3} q_2^2 (p \cdot q_1)^2-\frac{8}{3} q_1^2 (p \cdot q_2)^2 \notag\\
& -\frac{4}{3} q_1^2 p \cdot q_2 \left(q_2^2+q_1 \cdot q_2\right)+p \cdot q_1 \left(\frac{4}{3} \left(q_1^2+q_1 \cdot q_2\right) q_2^2+\frac{16}{3} p \cdot q_2 q_1 \cdot q_2\right).
\end{align}
The kernel functions $ T_1(Q_1,Q_2,\tau)$ and $ T_2(Q_1,Q_2,\tau)$ in~\eqref{eq:km} are given as
\begin{align}
T_1(Q_1,Q_2,\tau) &= \frac{Q_1 \left(\sigma _1^E-1\right) \left(Q_1 \tau  \left(\sigma _1^E+1\right)+4 Q_2 \left(\tau ^2-1\right)\right)-4 \tau  \mmu^2}{Q_1 Q_2 Q_3^2 \mmu^2} \notag\\
&+  X \frac{8 \left(\tau ^2-1\right) \left(2 \mmu^2-Q_2^2\right)}{Q_3^2 \mmu^2} ,\notag\\
T_2(Q_1,Q_2,\tau) &= \frac{1}{2 Q_1 Q_2 Q_3^2 \mmu^2}\bigg[Q_1^2 \tau  \left(\sigma_1^E-1\right) \left(\sigma_1^E+5\right)+Q_2^2 \tau  \left(\sigma_2^E-1\right) \left(\sigma_2^E+5\right)\notag\\
&+4 Q_1 Q_2 \left(\sigma_1^E+\sigma_2^E-2\right)-8 \tau  \mmu^2\bigg] 
+ X \left(\frac{8 \left(\tau ^2-1\right)}{Q_3^2}-\frac{4}{\mmu^2}\right),
\end{align}
where
\begin{align}
X &= \frac{1}{Q_1 Q_2 x} \arctan\left( \frac{z x}{1 - z \tau} \right) , & x &= \sqrt{1 - \tau^2} ,\notag \\
z &= \frac{Q_1 Q_2}{4m_\mu^2} (1-\sigma^E_1)(1-\sigma^E_2) , & \sigma^E_i &= \sqrt{ 1 + \frac{4 m_\mu^2}{Q_i^2} } , \notag\\
Q_3^2 &= Q_1^2 + 2 Q_1 Q_2 \tau + Q_2^2 .
\end{align}

\section{The pion pole in chiral perturbation theory}
\label{app:EFT}

An analysis of HLbL scattering at leading order in ChPT coupled to lepton fields produces the following representation~\cite{RamseyMusolf:2002cy,Knecht:2001qg}\footnote{For the reasons explained in App.~\ref{app:largeNc}, $a_\mu^{\pi^0\text{-pole}}$ does not actually scale with $\Nc^2$. In the following, we therefore set $\Nc=3$ from the start.}  
\beq
\label{amu_EFT}
 a_\mu^{\pi^0\text{-pole, ChPT}}=3\bigg(\frac{\alpha}{\pi}\bigg)^3\bigg(\frac{\mmu}{\Fpi}\bigg)^2\bigg(\frac{1}{4\pi}\bigg)^2
 \bigg\{\log^2\frac{\Lambda}{\mu}+\bigg[\frac{1}{6}\chi(\Lambda)-f(r)+\frac{1}{2}\bigg]\log\frac{\Lambda}{\mu}+C(\Lambda)\bigg\},
\eeq
where
\beq
\label{def_f}
f(r)=\log\frac{\mmu^2}{\mu^2}+\frac{1}{6}r^2\log r-\frac{1}{6}(2r+13)+\frac{1}{3}(2+r)\sqrt{r(4-r)}\arccos\frac{\sqrt{r}}{2}.
\eeq
Here, $r=\mpii^2/\mmu^2$, $\Lambda$ is a UV cutoff, in ChPT to be identified with the scale of chiral symmetry breaking $\Lambda_\chi\sim 4\pi\Fpi$, the IR scale
$\mu$ should be identified with $\mpii$~\cite{Prades:2009tw}, $\chi(\Lambda)$ is a LEC that renormalizes the $1$-loop ChPT expression for $\pi^0\to e^+e^-$, and $C(\Lambda)$ subsumes all terms not enhanced by a logarithm.

The precise definition of $\chi(\Lambda)$ depends on the scheme, which in~\eqref{amu_EFT} is chosen in accordance with~\cite{Savage:1992ac}. Explicitly, conventions can be specified using the reduced amplitude for $P\to \ell^+\ell^-$
\beq
\label{A_def}
A_\ell(q^2)=\frac{2i}{\pi^2 q^2}\int\diff^4 k\frac{k^2q^2-(q\cdot k)^2}{k^2(q-k)^2\big((p-k)^2-\ml^2\big)}\tilde F\big(k^2,(q-k)^2\big),
\eeq
where $q^2=M_P^2$ denotes the mass of the pseudoscalar, $p^2=\ml^2$ the lepton mass, and $\tilde F\big(q_1^2,q_2^2\big)$ the TFF for $P\to \gamma^*\gamma^*$ normalized by the chiral anomaly
\beq
\tilde F(q_1^2,q_2^2)=\frac{F(q_1^2,q_2^2)}{F_{\pi\gamma\gamma}}.
\eeq
For the decay kinematics one has, in addition, $(p-q)^2=\ml^2$, and thus $2p\cdot q=M_P^2$.

At leading order in ChPT $\tilde F(q_1^2,q_2^2)=1$ and the integral in~\eqref{A_def} diverges. This divergence is cured by introducing counterterms based on the Lagrangian~\cite{Savage:1992ac}
\beq
\mathcal{L}=\frac{3i\alpha^2}{32\pi^2}\big(\bar\ell\gamma^\mu\gamma_5\ell\big)\Big\{\chi_1\Tr\big(Q^2\{U^\dagger,\partial_\mu U\}\big)+
\chi_2\Tr\big(Q U^\dagger Q\partial_\mu U - Q \partial_\mu U^\dagger Q U\big)\Big\},
\eeq
where $Q$ is the charge matrix and $U$ contains the meson fields. Altogether, this leads to~\cite{Savage:1992ac}
\begin{align}
\label{A_ChPT}
 \Re A^\text{ChPT}_\ell(q^2)&=3\log\frac{\ml}{\Lambda}-\frac{\chi(\Lambda)}{4}-\frac{7}{2}
 +\frac{1}{\beta_\ell}\bigg[\frac{\pi^2}{12}+\frac{1}{4}\log^2\frac{1-\beta_\ell}{1+\beta_\ell}+\text{Li}_2\bigg(\frac{\beta_\ell-1}{\beta_\ell+1}\bigg)\bigg],
\end{align}
where
\beq
\text{Li}_2(x)=-\int_0^x\diff t\frac{\log(1-t)}{t},\qquad \beta_\ell=\sqrt{1-\frac{4\ml^2}{q^2}},
\eeq
and $\chi(\Lambda)=\chi_1^\text{r}(\Lambda)+\chi_2^\text{r}(\Lambda)$. Note, however, that the choice of scheme is not unique in the literature: another popular choice~\cite{Ametller:1993we} is related by 
$\chi(\Lambda)=\chi^\text{\cite{Savage:1992ac}}(\Lambda)=\chi^\text{\cite{Ametller:1993we}}(\Lambda)-4$.

Since the pion pole as defined in dispersion theory~\cite{Colangelo:2014dfa,Colangelo:2015ama} coincides with the diagrammatic expression~\eqref{eq:tw}, we can start from this expression
to analyze how the ChPT constraints emerge within dispersion relations.
First, we expand the kernel functions in terms of muon propagators as far as possible, using relations of the form
\beq
\int\frac{\diff^4q_2}{(2\pi)^4}\frac{F(q_1^2,q_2^2)F\big((q_1+q_2)^2,0\big)}{(q_2^2-\mpii^2)q_2^2(q_1+q_2)^2} q_2^\mu = 
\int\frac{\diff^4q_2}{(2\pi)^4}\frac{F(q_1^2,q_2^2)F\big((q_1+q_2)^2,0\big)}{(q_2^2-\mpii^2)q_2^2(q_1+q_2)^2} \frac{q_1\cdot q_2}{q_1^2}q_1^\mu,
\eeq
which follow from a standard tensor decomposition. This produces
\begin{align}
\label{rep_Ti}
 a_{\mu,T_1}^{\pi^0\text{-pole, disp}}&=-\frac{32\pi^2}{3\Fpi^2}\bigg(\frac{\alpha}{\pi}\bigg)^3
 \int\frac{\diff^4 q_2}{(2\pi)^4}\frac{\tilde F(q_2^2,0)}{q_2^2(q_2^2-\mpii^2)}\bigg(\frac{2\mmu^2+q_2^2}{(p-q_2)^2-\mmu^2}-1\bigg)\notag\\
 &\times \int\frac{\diff^4q_1}{(2\pi)^4} \frac{q_1^2q_2^2-(q_1\cdot q_2)^2}{q_1^2(q_1+q_2)^2\big((p+q_1)^2-\mmu^2\big)}\tilde F\big(q_1^2,(q_1+q_2)^2\big),\notag\\
  a_{\mu,T_2}^{\pi^0\text{-pole, disp}}&=-\frac{16\pi^2}{3\Fpi^2}\bigg(\frac{\alpha}{\pi}\bigg)^3
\int\frac{\diff^4 q_1}{(2\pi)^4}
\int\frac{\diff^4 q_2}{(2\pi)^4}\frac{\tilde F(q_1^2,q_2^2)\tilde F\big((q_1+q_2)^2,0\big)}{q_1^2q_2^2(q_1+q_2)^2\big((q_1+q_2)^2-\mpii^2\big)}\notag\\
&\times\bigg[\frac{\big(q_1^2+q_1\cdot q_2\big)q_2^2}{(p-q_2)^2-\mmu^2}
+\frac{\big(q_2^2+q_1\cdot q_2\big)q_1^2}{(p+q_1)^2-\mmu^2}\notag\\
&\qquad+\frac{2\mmu^2\big(q_1^2q_2^2-(q_1\cdot q_2)^2\big)-q_1^2q_2^2(q_1+q_2)^2}{\big((p+q_1)^2-\mmu^2\big)\big((p-q_2)^2-\mmu^2\big)}\bigg].
\end{align}
Accordingly, the representation for the $T_1$ term can be expressed as
\beq
\label{disp_T1}
 a_{\mu,T_1}^{\pi^0\text{-pole, disp}}=-\bigg(\frac{\alpha}{\pi}\bigg)^3\frac{1}{3\Fpi^2}\frac{1}{i}\int\frac{\diff^4q_2}{(2\pi)^4}\frac{\tilde F(q_2^2,0)}{q_2^2-\mpii^2}\bigg(\frac{2\mmu^2+q_2^2}{(p-q_2)^2-\mmu^2}-1\bigg)I_\mu(q_2^2),
 \eeq
 where
 \beq
 I_\ell(q^2)=\frac{2i}{\pi^2q^2}\int\diff^4 k\frac{k^2q^2-(q\cdot k)^2}{k^2(q-k)^2\big((p-k)^2-\ml^2\big)}\tilde F\big(k^2,(q-k)^2\big)
 \eeq
has been defined in close analogy to $A_\ell(q^2)$, the difference being that $q^2$ is not restricted to $\mpii^2$.
We checked numerically for a VMD form factor that the representation~\eqref{rep_Ti} reproduces the known result.  

In~\cite{Knecht:2001qf} it was established that the $T_2$ term remains finite even for a pointlike form factor, so that the corresponding integral cannot contribute to any singularities.  
The $\log$-enhanced terms in~\eqref{amu_EFT} all originate from the approximation where the form factors are put equal to unity, at this order in the chiral expansion their structure is not resolved. 
Matching the dispersive representation~\eqref{disp_T1} onto~\eqref{amu_EFT} therefore requires taking the pointlike limit in the appropriate fashion. First, we note that for $I_\mu(q_2^2)$ we cannot use 
the form~\eqref{A_ChPT}, since this relies on the specific kinematics for the pseudoscalar decay. Explicit calculation with Feynman parameters shows that in addition to the $\log$-divergent piece there is a contribution involving $\log(-q_2^2)$, whose coefficient is related to the $\log\Lambda$ term. The corresponding structure is therefore
\beq
\label{Imu}
I_\mu(q_2^2)=-3\log\frac{\Lambda}{\mu}-\frac{\chi(\Lambda)}{4}+\frac{3}{2}\log\bigg(-\frac{q_2^2}{\mu^2}\bigg)+C_\mu,
\eeq
with some constant piece $C_\mu$. The chiral LEC still regulates the divergence since its specific form does not depend on the kinematics. Once the form factor is replaced by its pointlike limit, the same LEC therefore describes the renormalization of the $\pi^0\to\ell^+\ell^-$ vertex (assuming lepton flavor universality).
This argument already shows that for the dispersive formalism to be consistent with the chiral constraints derived in~\cite{RamseyMusolf:2002cy,Knecht:2001qg} it suffices that the form factor used be consistent with the LEC $\chi(\Lambda)$, as extracted from $\pi^0\to e^+e^-$ or $\eta\to\ell^+\ell^-$.

The individual terms in~\eqref{amu_EFT} can then be understood as follows: for the second loop integral we have
\begin{align}
 &\frac{1}{i}\int\frac{\diff^4q_2}{(2\pi)^4}\frac{1}{q_2^2-\mpii^2}\bigg(\frac{2\mmu^2+q_2^2}{(p-q_2)^2-\mmu^2}-1\bigg)\notag\\
 &=\frac{1}{16\pi^2}\bigg(3\mmu^2\log\frac{\Lambda^2}{\mu^2}-2\mmu^2\int_0^1\diff x(1+x)\log\frac{x^2\mmu^2+(1-x)\mpii^2}{\mu^2}\bigg)\notag\\
 &=\frac{3\mmu^2}{16\pi^2}\bigg(\log\frac{\Lambda^2}{\mu^2}-f(r)-\frac{1}{2}\bigg),
\end{align}
with $f(r)$ as given in~\eqref{def_f}. Next, the $\log(-q_2^2)$ piece leads to a term
\begin{align}
\frac{1}{16\pi^4}\bigg(2\mmu^2\int_0^1\diff x(1+x)\frac{1}{i}\int\frac{\diff^4 q_2}{q_2^4}\frac{3}{2}\log\bigg(-\frac{q_2^2}{\mu^2}\bigg)\bigg)
&=\frac{3\mmu^2}{16\pi^4}\frac{3}{2}2\pi^2\int^\frac{\Lambda}{\mu}\frac{\diff x\, x^3}{x^4}\log x^2\notag\\
&=\frac{3\mmu^2}{16\pi^2}3\log^2\frac{\Lambda}{\mu}.
\end{align}
Adding the individual contributions we find
\begin{align}
  a_{\mu,T_1,\text{ div}}^{\pi^0\text{-pole, disp}}&=-\bigg(\frac{\alpha}{\pi}\bigg)^3\frac{1}{3\Fpi^2}
  \frac{3\mmu^2}{16\pi^2}\\
  &\times\bigg[\bigg(-3\log\frac{\Lambda}{\mu}-\frac{\chi(\Lambda)}{4}+C_\mu\bigg)\bigg(2\log\frac{\Lambda}{\mu}-f(r)-\frac{1}{2}\bigg)+3\log^2\frac{\Lambda}{\mu}\bigg]\notag\\
  &=3\bigg(\frac{\alpha}{\pi}\bigg)^3\bigg(\frac{\mmu}{\Fpi}\bigg)^2\bigg(\frac{1}{4\pi}\bigg)^2
 \bigg\{\log^2\frac{\Lambda}{\mu}+\bigg[\frac{1}{6}\chi(\Lambda)-f(r)+\tilde C_\mu\bigg]\log\frac{\Lambda}{\mu}+\ldots\bigg\}.\notag
\end{align}
Taking the pointlike limit of~\eqref{disp_T1} in this way therefore reproduces the basic features of the direct ChPT result~\eqref{amu_EFT}, in particular the coefficient of the double logarithm, the contribution from $\chi(\Lambda)$, and the part of the coefficient of the single logarithm that is non-analytic in the quark mass. 
The analytic contribution, $\tilde C_\mu=1/2$, requires a more careful treatment of the renormalization schemes~\cite{RamseyMusolf:2002cy,Knecht:2001qg} and certainly cannot be expected to emerge from 
a naive cutoff regularization of the loop integrals.

In conclusion, the above discussion demonstrates that dispersion relations for HLbL scattering in the form of~\cite{Colangelo:2014dfa,Colangelo:2015ama} fulfill the low-energy constraints from ChPT. 
Most aspects of~\eqref{amu_EFT} can already be derived from a pointlike form factor alone, so that the corresponding constraints are automatically maintained due to the structure of the loop integrals, which become identical to ChPT once the form factor is set to unity. The only information about the pion TFF beyond its pointlike limit is contained in the LEC $\chi(\Lambda)$, which is needed to renormalize the $\pi^0\to\mu^+\mu^-$ vertex due to the missing form-factor suppression for high momenta. Such a contribution therefore does not arise in a dispersive approach where the full form factor enters, but consistency with the chiral constraint is automatic as long as the employed form factor agrees with experimental constraints from $\pi^0\to e^+e^-$ and/or $\eta\to\ell^+\ell^-$ (the latter if $SU(3)$ symmetry is assumed). This comparison can indeed proceed in terms of $\chi(\Lambda)$: a given representation for the pion TFF can be turned into a prediction for this LEC, which can then be compared to the experimental value as extracted from the decay width. Equivalently, the decay width calculated from the form factor could be directly compared to the experimental result, with the chiral LEC one particular choice how to present the relation between HLbL scattering and the rare meson decays. We stress, however, that the comparison in terms of the TFF directly is actually preferable since it dispenses with the need for the chiral expansion. 

\section{Large-$\boldsymbol{\Nc}$ scaling}
\label{app:largeNc}

If the chiral anomaly $F_{\pi \gamma\gamma}$ were to scale with $\Nc$,  the ChPT expression for the pion pole would acquire on overall factor $\Nc^2$~\cite{RamseyMusolf:2002cy,Knecht:2001qg},
and together with the scaling $\Fpi^2\sim\Nc$ this would reproduce the overall $\Nc$ scaling of the quark-loop contribution to HLbL scattering, see e.g.~\cite{deRafael:1993za,Melnikov:2003xd}. 

However, as pointed out in~\cite{Abbas:1990kd,Gerard:1995bv,Bar:2001qk} this argument is not consistent because to ensure anomaly cancellation in the SM the quark charges need to be rescaled as well. We consider directly the $SU(3)$ case, where
\beq
Q_u=\frac{1}{2}\bigg(1+\frac{1}{\Nc}\bigg),\qquad Q_d=Q_s=-\frac{1}{2}\bigg(1-\frac{1}{\Nc}\bigg).
\eeq
For the decay of $\pi^0\to\gamma\gamma$ as well as the octet and singlet decays of the $\eta$, $\eta'$ system, $\eta_8,\eta_0\to\gamma\gamma$, one finds that the charge factors 
\begin{align}
\label{cancellation}
 (Q_u^2-Q_d^2)\Nc&=1,\notag\\
 \frac{1}{\sqrt{3}}(Q_u^2+Q_d^2-2Q_s^2)\Nc&=\frac{1}{\sqrt{3}},\notag\\
  \sqrt{\frac{2}{3}}(Q_u^2+Q_d^2+Q_s^2)\Nc&=\sqrt{\frac{3}{8}}\Nc-\frac{1}{\sqrt{6}}+\sqrt{\frac{3}{8}}\frac{1}{\Nc},
\end{align}
actually cancel the $\Nc$ scaling except for in the singlet component. Accordingly, a test of $\Nc=3$ either has to rely on $\eta$, $\eta'$ decays, where the mixing adds further complications~\cite{Borasoy:2004ua}, or more complicated decays such as $\eta\to\pi\pi\gamma$~\cite{Bar:2001qk,Borasoy:2004mf}.
Note that for such a test the implicit dependence of $\Fpi$ on $\Nc$ is irrelevant since $\Fpi$ would simply be taken from experiment. 

For the HLbL tensor we consider the corresponding flavor decomposition of the current
\beq
j^\mu=(Q_u-Q_d)j_3^\mu + \frac{1}{\sqrt{3}}(Q_u+Q_d-2Q_s)j_8^\mu + \sqrt{\frac{2}{3}}(Q_u+Q_d+Q_s)j_0^\mu,
\eeq
where
\begin{align}
j_3^\mu&=\frac{1}{2}(\bar u\gamma^\mu u-\bar d\gamma^\mu d),&
j_8^\mu&=\frac{1}{2\sqrt{3}}(\bar u\gamma^\mu u+\bar d\gamma^\mu d-2\bar s\gamma^\mu s),\notag\\
j_0^\mu&=\frac{1}{\sqrt{6}}(\bar u\gamma^\mu u+\bar d\gamma^\mu d+\bar s\gamma^\mu s).
\end{align}
Collecting terms at different orders in $\Nc$ this produces
\beq
j^\mu = j_3^\mu+\frac{1}{\sqrt{3}}j_8^\mu-\frac{1}{\sqrt{6}}j_0^\mu +\sqrt{\frac{3}{2}}\frac{1}{\Nc}j_0^\mu
\equiv j_\text{LO}^\mu + j_\text{NLO}^\mu,
\eeq
where we have named the two currents according to their $\Nc$ scaling,
\beq
\label{currents_Nc}
j_\text{LO}^\mu=\frac{1}{2}(\bar u\gamma^\mu u-\bar d\gamma^\mu d-\bar s\gamma^\mu s),\qquad
j_\text{NLO}^\mu=\frac{1}{2\Nc}(\bar u\gamma^\mu u+\bar d\gamma^\mu d+\bar s\gamma^\mu s).
\eeq
Restricted onto $SU(2)$, these currents correspond to the isovector and isoscalar component, respectively.

The leading $\Nc$ behavior of the quark loop can therefore only occur when each current receives a contribution from $j_\text{LO}^\mu$.
However, since the currents~\eqref{currents_Nc} correspond to charges $Q_\text{LO}=\diag(1,-1,-1)$ and $Q_\text{NLO}=\unity$, both of which fulfill $Q^2=\unity$, this implies that 
$\pi^0$ and $\eta_8$ have to couple to exactly one of them each---otherwise the charge factor $\Tr(Q^2\lambda_a)$, with Gell-Mann matrices $\lambda_a$, $a=3,8$, vanishes---and therefore 
cannot contribute at leading order in $\Nc$, completely in line with the cancellation observed in~\eqref{cancellation}. For the $\pi^0$, this result simply follows from isospin conservation, 
see~\eqref{eq:isod}, which forces exactly one of the currents to be isoscalar.  

We are thus led to the prediction that the $\pi^0$ and $\eta_8$ poles should be suppressed by $1/\Nc^2$ compared to the singlet component $\eta_0$,
in clear contradiction to phenomenology. To obtain a more realistic estimate one needs to include both the chiral scaling and, potentially, $\eta$--$\eta'$ mixing.
Since the mixing disappears in the chiral limit, the effect should scale with $m_s$, in such a way that the overlap of the $\eta$ with the singlet $\eta_0$ should be suppressed by $M_K^2/\Lambda_\chi^2$.
For a typical choice of $\Lambda_\chi$ this $\Nc$-leading but quark-mass-suppressed contribution to the $\eta$ from the $\eta_0$ is therefore not that different from the $\Nc$-suppressed $\eta_8$ itself. 
Taking everything together, the $\eta$ and $\eta'$ poles should be suppressed by
\beq
\frac{M_\eta^2}{\mpii^2}\bigg\{1,\frac{1}{\Nc}\frac{\Lambda_\chi^2}{M_K^2},\frac{1}{\Nc^2}\frac{\Lambda_\chi^4}{M_K^4}\bigg\}\gtrsim 10, \qquad \frac{M_{\eta'}^2}{\mpii^2}\frac{1}{\Nc^2}\sim 6,
\eeq
relative to the $\pi^0$ pole, respectively. While the $\eta'$ contribution comes out correctly, the one from the $\eta$ pole is predicted to be too small by about a factor $3$ 
(depending on the exact choice of $\Lambda_\chi$), and accordingly the hierarchy between $\eta$ and $\eta'$ is reversed. 
Worse, the $1/\Nc$ suppression of the $\pi^0$ pole compounds the mismatch with the pion loop, which has often been considered as leading in a chiral counting but subleading in $\Nc$, see e.g.~\cite{Jegerlehner:2009ry}, but with the corrected $\Nc$ assignments in the charges its contribution would be expected to be enhanced by one power in $\Nc$ and two in the chiral scaling compared to
the $\pi^0$ pole, in spectacular disagreement with phenomenology.
From our perspective, this casts doubt on the viability of the large-$\Nc$ expansion as an organizing principle for HLbL scattering.

A potential way around these conclusions would require considering QCD on its own, not as part of the SM gauge theories. This is essentially done in the original 
literature~\cite{tHooft:1973alw,tHooft:1974pnl,Witten:1979kh}, where it was shown that planar diagrams dominate in the limit $\Nc\to\infty$, $\alpha_s\Nc$ fixed. One could then argue that
the factors of $\Nc$ that originate in the quark charges due to anomaly cancellation do not correspond to this topological expansion 
and should therefore not be counted in this notion of the large-$\Nc$ limit~\cite{deRafael:2002tj}. 
On the other hand, the large-$\Nc$ scaling of~\eqref{currents_Nc} does provide an explanation for the suppression of the isoscalar current in electromagnetic reactions, which raises
the question why the implied hierarchy fails in the context of HLbL scattering. 

\section{Anomalous thresholds and analyticity}
\label{sec:anom_thr}

\begin{figure}[t]
	\centering
	\includegraphics[width=0.4\linewidth]{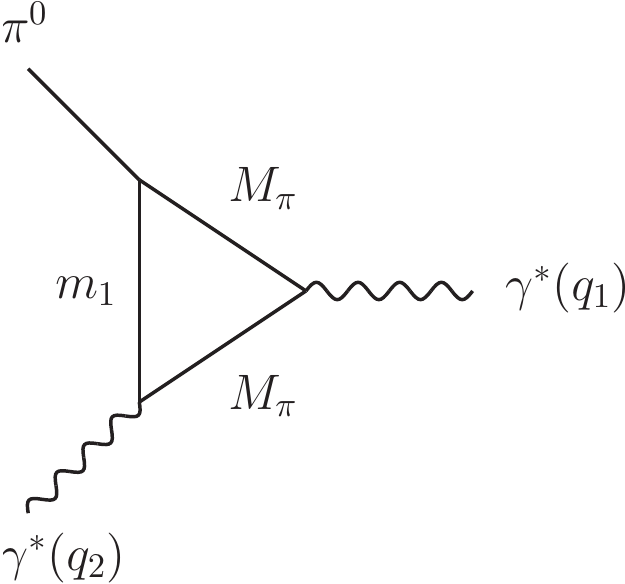}
	\caption{Triangle topology in the $\pi^0\to\gamma^*\gamma^*$ transition. For the physical case $m_1\geq 2\mpi$ no anomalous thresholds occur.}
	\label{fig:pi_triangle}
\end{figure}

The presence of two electromagnetic currents in the $\pi^0\to\gamma^*\gamma^*$ transition together with light pion intermediate states makes it appear likely that anomalous 
thresholds~\cite{Mandelstam:1960zz} require a modification of the integration contours in~\eqref{eq:dsr}, and indeed for similar quantities in the context of HLbL scattering,
e.g.\ the partial waves for $\gamma^*\gamma^*\to\pi\pi$, such complications do arise for time-like virtualities~\cite{Hoferichter:2013ama,Colangelo:2015ama}.
For the pion TFF the crucial analytic properties can be derived from the triangle diagram $C_0$ shown in Fig.~\ref{fig:pi_triangle}, depending on the mass $m_1$~\cite{Lucha:2006vc}.

The key assumption in the derivation of the dispersion relation for $F_{vs}(q_1^2,q_2^2)$ is that the dependence on the isovector virtuality permits a standard dispersive reconstruction. 
The corresponding imaginary part reads ($s=q_1^2$)
\beq
\Im C_0(s)=\frac{\theta(s-4\mpi^2)}{\sqrt{\lambda(s,\mpi^2,q_2^2)}}
\log\frac{s-3\mpi^2-q_2^2+2m_1^2-\sigma_\pi(s)\sqrt{\lambda(s,\mpi^2,q_2^2)}}{s-3\mpi^2-q_2^2+2m_1^2+\sigma_\pi(s)\sqrt{\lambda(s,\mpi^2,q_2^2)}},
\eeq
which defines the critical points
\beq
s_\pm(q_2^2)=\frac{1}{2}\Big\{3\mpi^2+q_2^2-m_1^2\pm\sigma_\pi(m_1^2)\sqrt{\lambda\big(m_1^2,\mpi^2,q_2^2\big)}\Big\}.
\eeq
Anomalous thresholds arise if either point, as a function of $q_2^2$, crosses the unitarity cut and moves onto the first sheet. The trajectory of $s_-(q_2^2)$ indeed comes close at $q_2^2=\mpi^2+2m_1^2$, but since the KT equations are solved for $q_2^2\to q_2^2+i\eps$, the intersection with the real axis occurs at
\beq
s_c=4\mpi^2\bigg(1-\frac{\eps^2}{4m_1^2\big(m_1^2-4\mpi^2\big)}\bigg).
\eeq
In the KT solution the mass $m_1^2$ is replaced by a spectral function whose support starts at $s'=4\mpi^2$, so that the intersection with the unitarity cut is narrowly avoided. 
However, this derivation shows that if there were a lighter state with mass below $2\mpi$, the trajectory would indeed move onto the first sheet and require a modification of the integration contour.

\begin{figure}[t]
\centering
\includegraphics[width=0.25\linewidth]{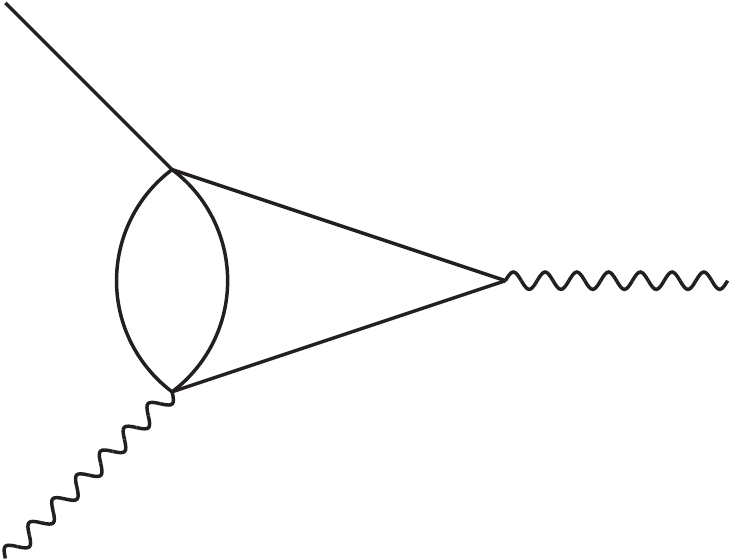}\qquad
\includegraphics[width=0.25\linewidth]{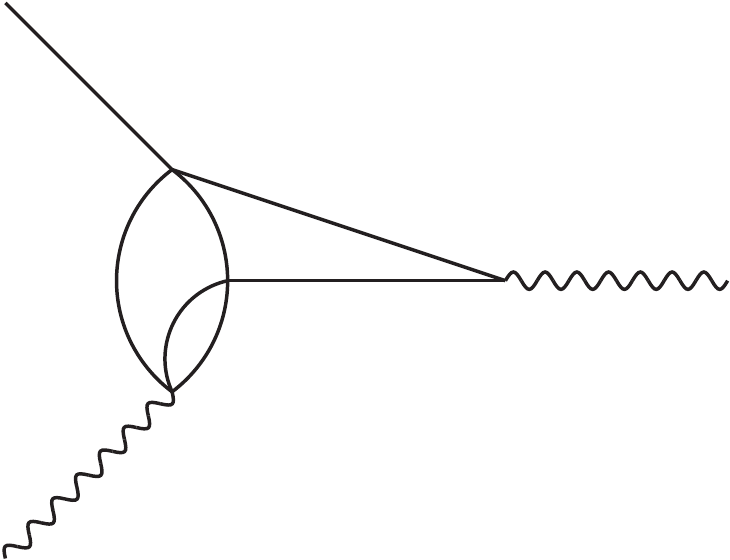}\qquad
\includegraphics[width=0.25\linewidth]{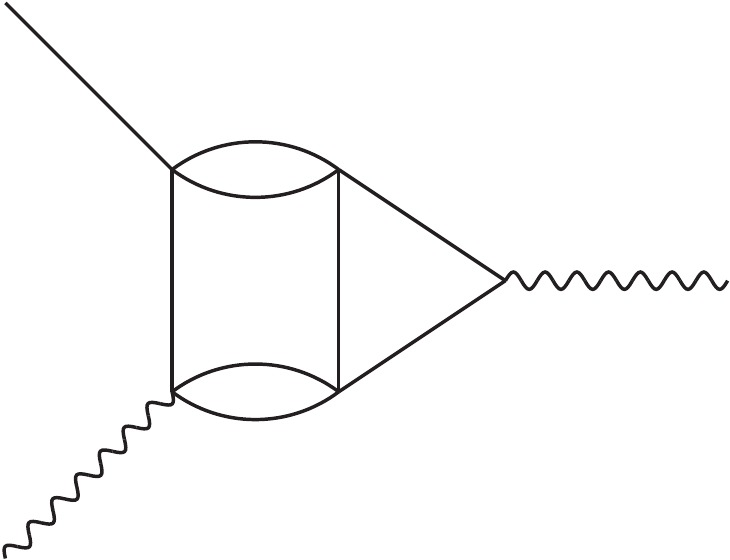}
\caption{Topologies for $\pi^0\to\gamma^*\gamma^*$. The solid lines all refer to pion states, but the analytic properties of these Feynman diagrams are again indicative of the general analytic structure.}
\label{fig:piggloops}
\end{figure}

In general, the occurrence of anomalous thresholds in a dispersion relation in the photon virtuality $q_1^2$ depends crucially on the form of the $\gamma^*\to 3\pi$ amplitude. 
The preceding discussion applies if that amplitude may be described by a dispersion relation in the crossed channel with threshold above $2\mpi$, in particular the first diagram in Fig.~\ref{fig:piggloops}.
Even at two-loop order (in $\gamma^*\to 3\pi$, see second diagram in Fig.~\ref{fig:piggloops}, corresponding to three loops for the TFF) such a representation exists, and even more so a representation free of anomalous thresholds~\cite{Gasser:2011ju}. Indeed, an anomalous threshold in the $\gamma^*\to3\pi$ amplitude would likely trigger an anomalous threshold in the pion TFF itself. 
In this way, the first problematic diagram occurs at three-loop order for the $\gamma^*\to 3\pi$ amplitude (third diagram in Fig.~\ref{fig:piggloops}, corresponding to four loops in the TFF): the $3\pi$ triangle should give rise to anomalous thresholds. However, this diagram involves an additional cut, implying that the corresponding $\gamma^*\to3\pi$ amplitude cannot be decomposed in terms of single-variable functions
anymore. Such contributions involving $4\pi$ cuts cannot be fully accounted for in our dispersive analysis of the $\gamma^*\to3\pi$ amplitude, 
and thus appear within the estimates for higher intermediate states, but not in the dispersive part of the decomposition. 

Apart from anomalous thresholds, it is surprising that a simple dispersion relation for the TFF arises despite the complicated analytic structure of the partial wave $f_1(s,q_2^2)$. To test this assumption numerically, we separated the normalization according to $f_1(s,q^2)=a(q^2)\bar f_1(s,q^2)$ and expressed the form factor in terms of
\beq
F_{vs}(q_1^2,q_2^2)=a(q_2^2)g(q_1^2,q_2^2),\qquad 
g(q_1^2,q_2^2)=\frac{1}{12\pi^2}\int^{\infty}_{4\mpi^2}\diff x\frac{q_{\pi}^3(x)\big(F_\pi^{V}(x)\big)^*\bar f_1(x,q_2^2)}{x^{1/2}(x-q_1^2)}.
\eeq 
The requirement that the single and double dispersion relations be equivalent then implies that
\beq
a(q_2^2)g(q_1^2,q_2^2)=\frac{1}{\pi}\int_{\sthr}^{\infty}\diff y \frac{\Im \big[a(y) g(q_1^2,y)\big]}{y-q_2^2},
\eeq
and since, by construction, $a(q^2_2)$ is analytic the same is true for $F_{vs}(q_1^2,q_2^2)$ as soon as $g(q_1^2,q_2^2)$ is analytic. Taking $q_1^2$ space-like, this statement follows from
{\allowdisplaybreaks
\begin{align}
 \frac{1}{\pi}\int\diff y\frac{\Im F_{vs}(q_1^2,y)}{y-q_2^2-i\eps}
 &=\frac{1}{\pi^2}\int\diff y \frac{\Im a(y)}{y-q_2^2-i\eps}\int \diff y'\frac{\Im g(q_1^2,y')}{y'-y+i\eps} \notag\\
 &\qquad+\frac{1}{\pi^2}\int\diff y' \frac{\Im g(q_1^2,y')}{y'-q_2^2-i\eps}\int \diff y\frac{\Im a(y)}{y-y'-i\eps}\notag\\
 &=\frac{1}{\pi^2}\int\diff y\, \Im a(y)\int \diff y'\, \Im g(q_1^2,y') \notag\\
 &\qquad\times\frac{1}{y'-y+i\eps} \bigg(\frac{1}{y-q_2^2-i\eps}
 -\frac{1}{y'-q_2^2-i\eps}\bigg)\notag\\
 &=\frac{1}{\pi^2}\int\diff y \frac{\Im a(y)}{y-q_2^2-i\eps}\int \diff y' \frac{\Im g(q_1^2,y')}{y'-q_2^2-i\eps}\notag\\
 &=a(q_2^2)g(q_1^2,q_2^2)=F_{vs}(q_1^2,q_2^2),
\end{align}}%
and the general case follows by analytic continuation in $q_1^2$.
From the KT solution we do not have access to $g(q_1^2,q_2^2)$ above $q_2^2=(1.8\GeV)^2$, but we can still check if, with a reasonable high-energy completion of the imaginary part,
the resulting function $g(q_1^2,q_2^2)$ fulfills a dispersion relation. Empirically, we observe that with a continuation according to $1/y^2$
a once-subtracted dispersion relation does reproduce the KT result, providing another check on the consistency of our dispersive formalism for the pion TFF.

\section{Scale estimate from light-cone QCD sum rules}
\label{sec:LV-app}

We start with a dispersive representation of the doubly-virtual pion TFF for space-like momenta
\beq
  \label{eq:startLCSR}
  F_{\pi^0\gamma^*\gamma^*}(q_1^2,q_2^2) = \frac{1}{\pi}\int_0^\infty \diff s \frac{\Im F_{\pi^0\gamma^*\gamma^*}(s,q_2^2)}{s-q_1^2}
\eeq
and split the spectral information into high and low energies~\cite{Khodjamirian:1997tk,Agaev:2010aq}:
\beq
  \label{eq:LCSR-high-low}
  F_{\pi^0\gamma^*\gamma^*}(q_1^2,q_2^2) = \frac{G_V(q_2^2)}{M_V^2-q_1^2} + 
  \frac{1}{\pi}\int_{\sm}^\infty \diff s \, \frac{\Im F_{\pi^0\gamma^*\gamma^*}(s,q_2^2)}{s-q_1^2} .
\eeq
For the low-energy part we use a VMD model~\cite{sakurai-book}: 
\beq
  \label{eq:VMDLCSR}
  \Im F_{\pi^0\gamma^*\gamma^*}(s,q_2^2) \approx G_V(q_2^2) \, \pi \, \delta(s-M_V^2) \quad \text{for} \quad s<\sm  , 
\eeq
with a vector-meson mass $M_V$ and a quantity $G_V$ proportional to the electromagnetic form factor for the transition of the 
vector meson to the pion. 

Duality between hadronic and quark--gluon (``OPE'') degrees of freedom suggests that at high energies, properly energy-averaged 
quantities should agree for both representations~\cite{Khodjamirian:1997tk,Agaev:2010aq}. Therefore one demands 
\beq
  \label{eq:dual1}
  \frac{1}{\pi}\int_{\sm}^\infty \diff s \, \frac{\Im F_{\pi^0\gamma^*\gamma^*}(s,q_2^2)}{s-q_1^2} \approx
\frac{1}{\pi}\int_{\sm}^\infty \diff s \, \frac{\Im F^\text{OPE}_{\pi^0\gamma^*\gamma^*}(s,q_2^2)}{s-q_1^2}
\eeq
for any value of $q_1^2$ (and sufficiently  large $\sm$) and 
\beq
  \label{eq:dual2}
  \frac{1}{\pi}\int_0^\infty \diff s \, \frac{\Im F_{\pi^0\gamma^*\gamma^*}(s,q_2^2)}{s-q_1^2} \approx
  \frac{1}{\pi}\int_0^\infty \diff s \, \frac{\Im F^\text{OPE}_{\pi^0\gamma^*\gamma^*}(s,q_2^2)}{s-q_1^2}
\eeq
for asymptotically large $q_1^2$. Taken together, these relations allow one to determine~\cite{Khodjamirian:1997tk,Agaev:2010aq} 
both parts on the right-hand side of~\eqref{eq:LCSR-high-low}, leading to 
\beq
  \label{eq:detGV}
  G_V(q^2) \approx \frac{1}{\pi}\int_0^{\sm} \diff s \, \Im F^\text{OPE}_{\pi^0\gamma^*\gamma^*}(s,q^2)
\eeq
and
\beq
  F_{\pi^0\gamma^*\gamma^*}(q_1^2,q_2^2) \approx 
  \frac{1}{M_V^2-q_1^2} \, \frac{1}{\pi}\int_0^{\sm} \diff s \, \Im F^\text{OPE}_{\pi^0\gamma^*\gamma^*}(s,q_2^2) 
  + \frac{1}{\pi}\int_{\sm}^\infty \diff s \, \frac{\Im F^\text{OPE}_{\pi^0\gamma^*\gamma^*}(s,q_2^2)}{s-q_1^2} . 
  \label{eq:relpionTFFOPEVMD}
\eeq
The pion TFF is symmetric in its two virtualities whereas the right-hand side of~\eqref{eq:relpionTFFOPEVMD} is not. We symmetrize
the expression by hand and obtain
\begin{align}
  F_{\pi^0\gamma^*\gamma^*}(q_1^2,q_2^2) &\approx
  \frac{1}{2}\Bigg[\frac{1}{M_V^2-q_1^2} \, \frac{1}{\pi}\int_0^{\sm} \diff s \, \Im F^\text{OPE}_{\pi^0\gamma^*\gamma^*}(s,q_2^2) 
  + \frac{1}{\pi}\int_{\sm}^\infty \diff s \, \frac{\Im F^\text{OPE}_{\pi^0\gamma^*\gamma^*}(s,q_2^2)}{s-q_1^2} \notag\\
  &+ \frac{1}{M_V^2-q_2^2} \, \frac{1}{\pi}\int_0^{\sm} \diff s \, \Im F^\text{OPE}_{\pi^0\gamma^*\gamma^*}(q_1^2,s) 
  + \frac{1}{\pi}\int_{\sm}^\infty \diff s \, \frac{\Im F^\text{OPE}_{\pi^0\gamma^*\gamma^*}(q_1^2,s)}{s-q_2^2}\Bigg]. 
  \label{eq:relpionTFFOPEVMDsym}
\end{align}

In~\cite{Khodjamirian:1997tk,Agaev:2010aq}, a Borel transformation has been applied to~\eqref{eq:dual2} and a Borelized
version of~\eqref{eq:relpionTFFOPEVMD} is used for the singly-virtual pion TFF. In the following, 
we use the symmetrized finite-energy sum rule~\eqref{eq:relpionTFFOPEVMDsym} as it is. 
It has the advantage that it contains only two 
non-perturbative parameters, the vector-meson mass $M_V$ and the ``continuum threshold'' $\sm$, i.e.\ the onset of the 
asymptotic regime.

Finally, we need the OPE expression for the spectral information. To this end, we use the asymptotic LO
leading-twist expression~\eqref{eq:pQCD} that relates the pion TFF to the 
pion distribution amplitude~\cite{Lepage:1979zb,Lepage:1980fj,Brodsky:1981rp}.
The final expression for this LCSR VMD approach (LV) is
\begin{align}
  \label{eq:finalLV}
  F^\text{LV}_{\pi^0\gamma^*\gamma^*}(q_1^2,q_2^2) & := 
  \frac{F_\pi}{3} \, \int_0^{x_1} \diff x \, \frac{\phi_\pi(x)}{(1-x)(M_V^2-q_2^2)}
  - \frac{F_\pi}{3} \, \int_{x_1}^1 \diff x \, \frac{\phi_\pi(x)}{x q_1^2 + (1-x)q_2^2}  \notag \\
  & + \frac{F_\pi}{3} \, \int_0^{x_2} \diff x \, \frac{\phi_\pi(x)}{(1-x)(M_V^2-q_1^2)}
  - \frac{F_\pi}{3} \, \int_{x_2}^1 \diff x \, \frac{\phi_\pi(x)}{x q_2^2 + (1-x)q_1^2},
\end{align}
where
\beq
  \label{eq:defxq}
  x_i := \frac{\sm}{\sm-q_i^2}.
\eeq

\begin{figure}[t]
  \centering
      \includegraphics[keepaspectratio,width=0.8\textwidth]{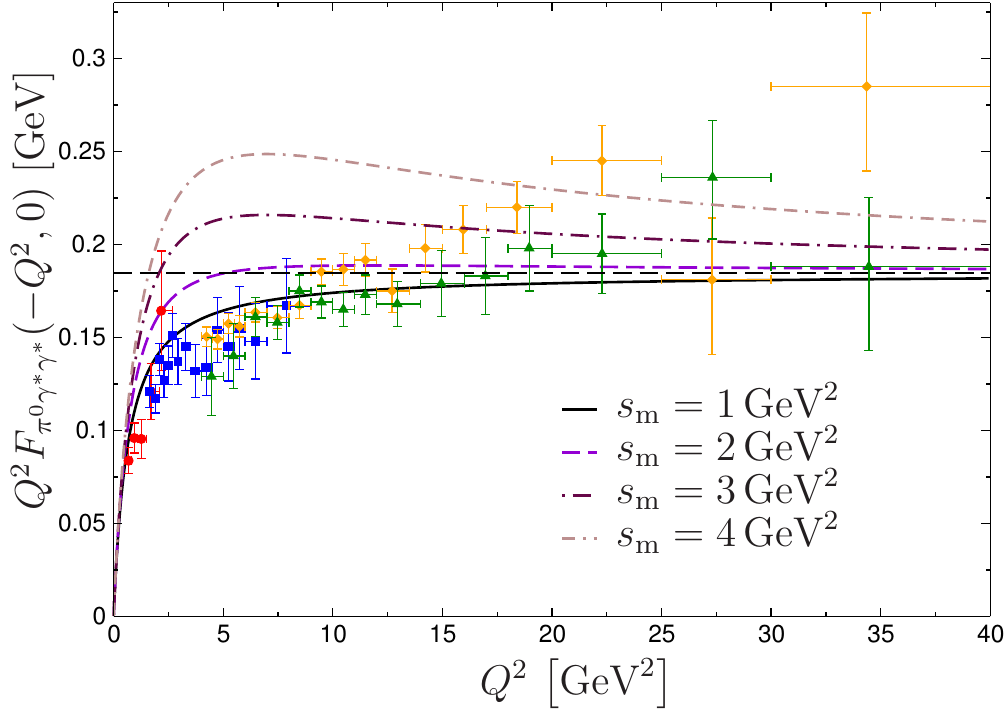}
  \caption{Comparison of~\eqref{eq:finalLV} to singly-virtual pion TFF 
    data~\cite{Behrend:1990sr,Gronberg:1997fj,Aubert:2009mc,Uehara:2012ag}. Color coding for the experimental points as in Fig.~\ref{fig:TFF_spacelike_asym}.}
  \label{fig:LV}
\end{figure}

Expression~\eqref{eq:finalLV} shows very satisfying high- and low-energy limits provided one 
chooses $M_V^2 = 8 \pi^2 F_\pi^2$~\cite{Meissner:1987ge,Marco:1999xz,Leupold:2003zb}. 
In line with the chiral anomaly one obtains 
\beq
  \label{eq:limitsLVlow}
  F^\text{LV}_{\pi^0\gamma^*\gamma^*}(0,0) = \frac{2F_\pi}{3M_V^2} \, \int_0^1\, \frac{\phi_\pi(x)}{1-x} = \frac{2F_\pi}{M_V^2} 
  = \frac{1}{4 \pi^2 F_\pi}  .
\eeq
The Brodsky--Lepage limit~\cite{Lepage:1979zb,Lepage:1980fj,Brodsky:1981rp} is recovered:
\begin{align}
  \label{eq:limitsLVBL}
  F^\text{LV}_{\pi^0\gamma^*\gamma^*}(-Q^2,0) & = 
  \frac{F_\pi}{3} \, \int_0^{x_Q} \diff x \, \frac{\phi_\pi(x)}{(1-x)M_V^2}
  + \frac{F_\pi}{3} \, \int_{x_Q}^1 \diff x \, \frac{\phi_\pi(x)}{x Q^2}  \notag\\
  &+ \frac{F_\pi}{3} \, \int_0^{1} \diff x \, \frac{\phi_\pi(x)}{(1-x)(M_V^2+Q^2)}  \notag \\ 
& = \frac{1}{Q^2} \, \frac{2F_\pi}{3} \, \int_{0}^1 \diff x \, \frac{\phi_\pi(x)}{x} + \Order(1/Q^4) 
  = \frac{2F_\pi }{Q^2} + \Order(1/Q^4)  .
\end{align}
Finally, for large $Q_1^2$, $Q_2^2$ one finds the relation
\beq
  \label{eq:limitsLVOPE}
  F^\text{LV}_{\pi^0\gamma^*\gamma^*}(-Q_1^2,-Q_2^2) =
  \frac{2 F_\pi}{3} \, \int_0^1 \diff x \, \frac{\phi_\pi(x)}{x Q_1^2 + (1-x) Q_2^2} + \Order(1/Q_i^4) ,
\eeq
which is in line with the OPE prediction~\cite{Lepage:1979zb,Lepage:1980fj,Brodsky:1981rp,Manohar:1990hu}. 
More generally, if both virtualities are space-like, \eqref{eq:finalLV} vanishes as soon as one of the two virtualities 
becomes infinitely large, irrespective of the value of the other virtuality. This property is not so easy to achieve for 
hadronic resonance saturation models.

Before we show the results, 
we stress again that the QCD sum rule formula~\eqref{eq:finalLV} containing in particular the VMD model 
for the low-energy part is not meant for a full-fledged quantitative calculation of the pion TFF, 
but for understanding the size of $\sm$. Figure~\ref{fig:LV} shows a comparison of formula~\eqref{eq:finalLV} to the 
data on the singly-virtual pion TFF for different values of $\sm$. 
Obviously, large values of $\sm$ do not agree with the data while a value of $\sm = 1\GeV^2$ provides a consistent picture.


\end{document}